\pdfoutput=1

\documentclass[11pt]{article}

\usepackage[final]{acl}
\usepackage{times}
\usepackage{latexsym}
\usepackage{booktabs}
\usepackage{multirow}
\usepackage{hhline}
\usepackage{comment}
\usepackage{todonotes}
\usepackage{makecell}

\usepackage[T1]{fontenc}

\usepackage[utf8]{inputenc}

\usepackage{microtype}
\usepackage{enumitem}

\usepackage{inconsolata}
\usepackage[explicit]{titlesec}
\usepackage[table]{xcolor}    

\usepackage{graphicx}
\usepackage{listings}
\usepackage{adjustbox}
\usepackage{geometry}

\titlespacing{\paragraph}{%
  0em}{
  0\baselineskip}{
 0\baselineskip}%

\newcommand{\dataset}{{\textsc{StarQA}}}
\newcommand{\size}{362}
\newcommand{\system}{{\textsc{Text2SQLCode}}}
\newcommand{\hybrid}{{\textsc{Hybrid}}}
\AtBeginDocument{\normalfont\selectfont}
\usepackage{listings}
\title{${\Huge \star}$ \dataset{}: A Question Answering Dataset for Complex Analytical Reasoning over Structured Databases}

\author{Mounica Maddela$^1$, Lingjue Xie$^{1}$, Daniel Preo\c{t}iuc-Pietro$^1$, Mausam$^2$\thanks{Most of the work was done when the author was on a sabbatical at Bloomberg.}\\
$^{1}$Bloomberg \\
$^{2}$Yardi School of Artificial Intelligence, Indian Institute of Technology, Delhi \\
{\small {\tt \{mmaddela3,lxie91,dperotiucpie\}@bloomberg.net}, {\tt mausam@cse.iitd.ac.in}}\\
}

\begin{document}
\maketitle

\begin{abstract}
Semantic parsing methods for converting text to SQL queries enable question answering over structured data and can greatly benefit analysts who routinely perform complex analytics on vast data stored in specialized relational databases. 
Although several benchmarks measure the abilities of text to SQL, the complexity of their questions is inherently limited by the level of expressiveness in query languages and none focus explicitly on questions involving complex analytical reasoning which require operations such as calculations over aggregate analytics, time series analysis or scenario understanding.
In this paper, we introduce \dataset{}, the first public human-created dataset of complex analytical reasoning questions and answers on three specialized-domain databases. 
In addition to generating SQL directly using LLMs, we evaluate a novel approach (\system{}) that decomposes the task into a combination of SQL and Python: SQL is responsible for data fetching, and Python more naturally performs reasoning. Our results demonstrate that identifying and combining the abilities of SQL and Python is beneficial compared to using SQL alone, yet the dataset still remains quite challenging for the existing state-of-the-art LLMs. 
%
\end{abstract}

\section{Introduction}

Text to SQL semantic parsing approaches enable intuitive and efficient human interaction with structured databases through natural language queries. They alleviate the need for the user to fully comprehend the underlying DB schema or the query language used to answer their question.  Large language models (LLMs) exhibit very good performance in this task in constrained settings, measured by over 90\% execution match accuracies on popular data sets such as Spider \cite{yu-etal-spider} and WikiSQL \cite{zhong-2017-wikisql}. 

An important user base for these methods can be analysts whose goals are to perform analyses involving complex analytics that require reasoning over large, proprietary datasets, in order to obtain quantitative insights that can drive data-driven decision making. Examples include economic analysts deciding on a country's policies, financial analysts making investment decisions, operations researchers that improve business operations or data-driven sports analysts making match or acquisition strategies. It is estimated that about 52\% of professional developers regularly use SQL, and over one-third must run very complex queries efficiently~\cite{stackoverflow,text2sqlstate}. However, almost all of the 36 benchmark datasets, surveyed in \citet{liu-2024-survey}, contain less than two \texttt{select} clauses (suggesting lack of nested queries and complex set operations) and less than one mathematical, scalar or aggregation function per query. 

In response, we introduce a novel dataset \dataset{} -- \textbf{St}ructured Data \textbf{A}nalytics \& \textbf{R}easoning \textbf{Q}uestion \textbf{A}nswering. It contains \size{} realistic complex analytics and reasoning questions and answers over publicly available databases from three domains -- movies (\texttt{IMDb}), sports (\texttt{ES}), and E-commerce (\texttt{OL}). Each question requires one or more types of reasoning, such as math operations, time-series analysis, nested queries, calculations over aggregate analytics, temporal reasoning, and scenario understanding. Table \ref{t:categories} illustrates examples of questions in \dataset{} and their categories.


The types of reasoning in \dataset{} raise a question about whether SQL syntax alone is the best choice for obtaining the correct answer. Cases that require operations such nested loops, complex conditional logic or data manipulation can lead to very complex SQL syntax and long queries, which can be cumbersome for a model to generate. 

We propose \system{} as a potential solution to decompose the task in multiple steps that interweave SQL for efficient data operations over databases and Python code for operations that are too burdensome to achieve using SQL, such as performing analytics, data wrangling and computations. A challenge of this solution approach is its multi-step nature that can lead to compounding errors. Thus, we explore its hybrid variants, where we apply \system{} only on the subset of difficult reasoning questions as automatically identified through the Text2SQL parser's outputs.

We extensively benchmark recent open-weight and commercial LLMs, including reasoning models, on \dataset{}. Results show that \dataset{} is challenging for all methods and models tested, with the median performance (across six LLMs) of a standalone Text2SQL model reaching only 36.5\% execution accuracy. However, our best performing hybrid approaches (\hybrid{}$_{single}$ and \hybrid{}$_{multi}$) achieve improved performance on this task and significantly outperform Text2SQL, with median performance of 45.4\% and 44.5\% respectively. We also find that most of our questions cannot be solved by existing LLMs with access to web search tools -- reinforcing the unique potential of large-scale, rich and potentially private proprietary structured data. Finally, our analyses highlight the nature of the errors and suggest directions for future research in building stronger reasoning systems over structured databases. We release \dataset{} publicly and freely for research.\footnote{\scriptsize\url{https://zenodo.org/records/17157169}}

\section{Related Work}



Semantic parsing techniques convert natural language inputs into formal representations, such as $\lambda$-calculus for robotics \cite{williams-2018-robotics}, SQL for structured databases \cite{kat-2023-text2sqlsurvey}, SPARQL for KBs \cite{patidar-2024-fusic}, Cypher for graph data \cite{ozsoy-etal-2025-text2cypher}, and other low-resource languages for specific applications \cite{dutta-etal-2024-retrieval}. While our dataset is specific to SQL, our methods and observations could be more generally applicable.

The latest Text2SQL systems obtain nearly $90\%$ performance on popular benchmarks that study constrained settings of the task such as WikiSQL\footnote{\scriptsize\url{https://github.com/salesforce/WikiSQL}} and Spider.\footnote{\scriptsize\url{https://yale-lily.github.io/spider}} Various benchmarks highlight challenges in specific domains such as finance \cite{sen-fiben}, accounting \cite{kumar-etal-2024-booksql} and science \cite{zhang-science}. Other works study generic notions of complexity such as variety in SQL dialects and interaction with real-world enterprise databases \cite{lei-2024-spider2}, complex and large database schemas \cite{li-academic, sen-fiben}, issues with entity linking \cite{wang-etal-2022-improving-text}, and use of external knowledge \cite{li-etal-2021-dual}. After surveying 36 benchmark datasets for Text2SQL, \citet{liu-2024-survey} identify reasoning as a key underexplored aspect of the task. \dataset{} is developed to bridge this gap.

Most related, \texttt{Archer}~\cite{zheng-etal-2024-archer} contains questions on arithmetic, commonsense, and counterfactual reasoning; however, they can all be solved by the basic SQL language and do not require any procedural extensions or scripting languages. This limits the overall complexity of the questions -- they do not require string manipulations, complex cell processing, or statistical functions -- some of the features of \dataset{}.
SPIDER 2.0~\cite{lei-2024-spider2}, on the other hand, combines many levels of complexity in its questions on real-world enterprise databases and is best suited for developing a complete agentic workflow with multiple components. However, it is not an ideal dataset for evaluating analytical reasoning abilities and is not annotated with nature of reasoning required.

The main LLM-based approach for Text2SQL systems \cite{shi-2024-text2sqlsurvey-llms} includes building schema and entity linkers \cite{wang-2025-dbcopilot} whose outputs, along with retrieved exemplars \cite{gurawa-2025-ragtext2sql}, are input to an LLM for SQL generation. CoT prompting \cite{tai-etal-2023-exploring}, 
self-consistency over multiple generations \cite{sun-etal-2023}, self-correction \cite{pourreza-2023-dinsql}, and iterative repair \cite{sawhney-2025-iterative} are also used to improve generation quality.

Our proposed solution is influenced by the rapid progress in code generation systems \cite{zan-etal-2023-large}, and investigates whether SQL and Python can be effectively combined. To the best of our knowledge, there is limited exploration of this synergy in literature. TPTU \cite{ruan-2023-tptu} studies this but with somewhat artificial prompts where, often a \emph{simple} mathematical operation (e.g., log or factorial) is applied over a statistic outputted by an SQL query. Another work \cite{sui-2023-reboost} assesses Text2Python as an alternative to Text2SQL, but finds that former performs worse, primarily because of high generation length requirements. It does not explore their combination.

Our approach is also related to table-augmented generation \cite{biswal-2024-tag} where SQL results are provided in context, for LLMs to generate a final answer. This succeeds only when SQL outputs are short, and processing post-SQL is limited. It does not study combinations with other languages.

\begin{table*}[t!]
\rowcolors{2}{gray!25}{white}
\centering
\renewcommand{\arraystretch}{1.2}
\scriptsize
\begin{tabular}{|p{2.6cm}|p{11.2cm}|c|}
\hline
\rowcolor{gray!50}
\textbf{Category} & \textbf{Example} & \textbf{\#Samples} \\
\hline
Stat./Math. Operations & \textit{Is there a \textbf{statistically significant correlation (use Spearman’s)} between a movie’s runtime and its average rating?} &  85 \\
Analytics over Non-entries in Database & \textit{What is the \textbf{most common uncased prefix (first 3 characters)} across all movie titles released in the 1990s, removing any preceding articles (a, an, the)?} &   20 \\
Nested Queries & \textit{Find the actor/actress with the \textbf{most appearances in movies that are in the top 10\% by rating every decade}.} &   147 \\
String Manipulation & \textit{Which movie title has the highest number of \textbf{distinct characters (ignoring case and spaces)}?} &  46 \\
Calculations over Aggregate Analytics & \textit{Percentage (no decimals) for \textbf{a team to win the league if they are 3 points or more ahead of all the other teams at the end of January}} &  100 \\
Complex Columns & \textit{Identify the most popular \textbf{co-occurring genre pairs} among TV series} &  16 \\
Temporal Reasoning & \textit{Print the number of directors that got an 9.0 rating for their debut in \textbf{1980s}.} &  32 \\
Complex Filtering & \textit{Which director has directed movies in \textbf{at least ten different languages and has an average movie rating above 8.5}?} &  36 \\
Unit Conversions & \textit{How many years would it take to watch every movie in the database?} &  9 \\
Scenario Understanding & \textit{Which teams that won the league would not have won it \textbf{if a win would have been only 2 points}.} & 16  \\
Time Series Analysis & \textit{Teams in the English Premier League that \textbf{lost four matches in a row after winning four in a row}.} &  55 \\
Commonsense Knowledge & \textit{The longest number of months either team name was undefeated in the \textbf{Old Firm} between 2008-2016?} &  18 \\

\hline
\end{tabular}
\caption{\label{t:categories}Reasoning categories in \dataset{}, along with an example and number of questions that exhibit this category. A question may have multiple categories. Output formatting instructions in the questions are omitted for brevity.}
\end{table*}

\section{The \dataset{} Dataset}

We create a new dataset to study complex analytical reasoning question answering over SQL databases. The dataset is built to contain questions that require one or more of the categories described in Table~\ref{t:categories}. The \dataset{} dataset consists of the following: (1) questions, (2) their answers, (3) categories required to answer a question, and (4) reference code used to obtain the gold answer.
The code is only presented as a reference to verify the answer and is not used in any experiments in this paper.

\subsection{Databases}

To create \dataset{}, we use three complex and large databases from the movies, sports, and E-commerce domains. We use only three DBs so that data creation is invested primarily on creating a diverse set of questions with depth in reasoning and analytics, which is the goal of this dataset. This can be achieved only if the annotators understand the data, the domain and DB organization deeply enough to ask and answer interesting questions, limiting the number of databases we can scale to. As constructed, we expect \dataset{} to be useful primarily for development and evaluation,  not for training.

\paragraph{\texttt{IMDb}}. This database\footnote{\scriptsize\url{https://developer.imdb.com/non-commercial-datasets/}} is provided by IMDb and contains historical information about over 10 million movie titles, including TV series, their crews, biographical information about crew members, movie ratings on IMDb and number of votes. In total, the \texttt{IMDb} database has 7 tables, with a table having up to 9 columns and a total of 11 million rows.

\paragraph{\texttt{ES}}. This database\footnote{\scriptsize\url{https://www.kaggle.com/datasets/hugomathien/soccer}}\textsuperscript{,}\footnote{\scriptsize\url{https://www.kaggle.com/datasets/jiezi2004/soccer}} consists of soccer (football) statistics from the top-flight leagues of 11 European countries between 2008 and 2016. In total, the database includes information on over 25,000 matches, 10,000 players, their attributes, match events, line-ups, formations and betting odds. In total, the database has 16 tables, with tables having up to 115 columns and 26,000 rows.

\paragraph{\texttt{OL}}. This database\footnote{\scriptsize\url{https://www.kaggle.com/datasets/olistbr/brazilian-ecommerce}} consists of 100k orders from the \texttt{Olist} - online store in Brazil, from 2016 to 2018. It provides a multi-dimensional view of orders, including information on order status, price, payment, freight performance, customer location, product attributes, and customer reviews. In total, the database has 9 tables, with tables having up to 9 columns and 112,000 rows.

\subsection{Dataset Creation}

\dataset{} is manually created in its entirety by the authors of the paper, who all have training and experience in programming and NLP, intermediate or higher knowledge of SQL, advanced knowledge of Python, obtained advanced knowledge of the databases and their structure, and have a personal interest in the domains represented in the dataset.

Our aim is to build a dataset of questions that are both realistic, akin to those that an analyst working with these databases may ask, and that cover as many of the complex analytical reasoning categories described in Table~\ref{t:categories} as possible.

We first start by selecting, for each database, the categories of questions that would be natural to the domain and the underlying database structure. After all questions were written, these were reviewed by the other authors and edited until they satisfied the following criteria: correctness (the question-answer pair is syntactically and semantically correct), credibility (if the question is one an analyst might ask), specificity (lack of ambiguity in the information that is requested) and clarity of the expected output and its format. 

We construct the \dataset{} dataset to facilitate exact automatic evaluation of the performance of systems. Each question clearly states the expected returned fields, and the precision of the numbers in case a fraction or percentage is expected. The answer to each question is formatted in a canonical format represented by a list of tuples, with each item in the list representing a correct answer to the question. Each question can have no answer, one unique answer tuple or more. Each tuple may have one or more constituents -- no specific order for them is imposed within a tuple, and set-match is used for evaluation. We also aim to limit ambiguities in question formulation wherever possible, such that no model is penalized for picking different interpretations e.g., we replace \textit{Find the actor} with \textit{Find the actor, male or female}. While these requirements may make the question appear less natural, they maximize the ability to accurately benchmark the abilities of a model in answering the questions correctly.



The goal of \dataset{} is to measure analytical reasoning abilities. A common challenge in other Text2SQL datasets that include reasoning questions~\cite{lei-2024-spider2,zheng-etal-2024-archer} is entity linking in a new database and adapting the model to the new database schema. We consider these as orthogonal challenges to our goal. In order for them to have a limited impact in our dataset, we normalize the mentions of specific entities (e.g., actors, league or team names) to a common and unambiguous name.

The answers to all the questions were obtained through executing a combination of SQL and Python code written by the dataset creators that is released with the dataset. The authors spent, after data collection and data preparation, on average between 15 to 30 minutes writing the answer to a single question.

\subsection{Dataset Statistics}

Overall, the \dataset{} dataset consists of \size{} data points, with 100 questions on \texttt{IMDb}, 162 questions on \texttt{ES}, and 100 questions from \texttt{Olist}. Out of these, 55 questions have a single numerical answer, 8 questions have a null answer, and most questions (299) have tuples as an answer. The average number of tuples in an answer is 2.35 (up to 66) and the average number of constituents in a tuple is 2.43 (up to 8). The distribution over the types of analytical reasoning categories and example questions are presented in Table~\ref{t:categories}.

\section{The \system{} Approach}
\label{sec:system}

\begin{figure}[t]
\centering
\includegraphics[width=1.00\columnwidth]{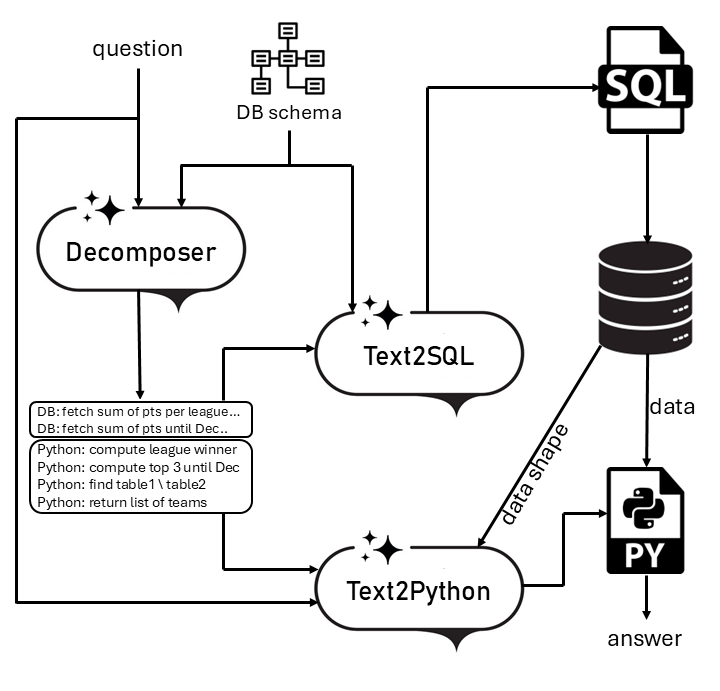}
\vspace*{-2em}
\caption{The architecture of \system{}$_{multi}$ 
}
\label{fig:diagram}
\end{figure}

We propose \system{}, which combines the use of text to SQL semantic parser (Text2SQL) and a Python code generator (Text2Python) for answering the analytical reasoning questions. The underlying insight is that such questions can typically be split into data fetching and processing or computation steps. SQL is required for data fetching, as the data is provided in relational databases. If the required analytical reasoning is complex, a procedural language (such as Python) may be needed due to the limited expressiveness of the SQL language in performing some operations. Consequently, a split may be desirable, where data fetching, and perhaps simple processing, is conducted in SQL and complex processing (if needed) is done in Python. 

We operationalize this by designing a three-step workflow (Figure \ref{fig:diagram}): (1) a decomposer, (2) a Text2SQL model, and (3) a Text2Python model. We implemented two variants of the workflow: \system{}$_{multi}$ and \system{}$_{single}$. \system{}$_{multi}$  uses separate LLM calls for each step with specific prompts and output formats. The LLM outputs for each step are post-processed for the next step. In contrast, \system$_{single}$ uses a single prompt to perform decomposition and generates a complete Python function wrapped around the SQL query. We note that the idea of question decomposition has been explored within Text2SQL \cite{pourreza-2023-dinsql, wang-etal-2025-mac}, though not in combination with Python.

\paragraph{Decomposer:}~
The goal here is to decompose the user question into a series of steps, along with annotation on whether they are Text2SQL steps or Text2Python steps. We provide the DB schema -- tables, column types and descriptions, primary and foreign keys, sample rows per table -- in the prompt. We also provide a handful of positive and negative exemplars (from another domain, for a fair comparison with other approaches). 

We provide these additional guidelines in the prompt: (1) Multiple Text2SQL steps should not be dependent on each other, and should be independently executable; (2) Python may be omitted for simpler questions; (3) Each step should be in natural language and should be prefixed with either ``\texttt{Text2SQL:} " or ``\texttt{Python:} ". We then ask the LLM to perform chain of thought and provide its final output. In Figure \ref{fig:diagram}, the two types of steps (SQL and Python) in the decomposition are shown in two separate boxes.

\paragraph{Text2SQL:}~
We generate an SQL query for each Text2SQL step identified by the decomposer. For the \system{}$_{multi}$ approach, a separate Text2SQL call is made for each identified step. We provide the database schema and the specific text of the step. The generated queries are then executed to fetch data into a dataframe. If the query fails, we reprompt the LLM for corrections, up to three times. For the \system{}$_{single}$, the LLM directly generates SQL queries that are embedded within the final Python function, creating a single-pass solution.
  

\paragraph{Text2Python:}~
This step handles the complex analytical reasoning that goes beyond reasoning abilities. In the \system{}$_{multi}$ approach, this step is executed after all the necessary data has been fetched from the Text2SQL calls. We construct a prompt with the original user question, the decomposition, and the shape of each dataframe generated via the SQL calls from the previous step. The shape includes the names of columns and a few sample rows of data (similar to \citet{maamari-shape}). We prompt the model to generate a single Python function with a specific signature:\\
\texttt{compute\_result(listOfDFs: List[DataFrame]) -> List[Tuple]}.

We then run the generated Python code on the full dataframes retrieved in the Text2SQL step(s).  \system{}$_{single}$  also implements a similar Python function with the DB path as the input:\\
\texttt{compute\_result(db\_path) -> List[Tuple]}.

This function also contains the embedded SQL queries. For both cases, if the code runs successfully, we output the result, else, we reprompt the LLM for correction (maximum of three times). 

We note this framework can be extended by using retrieval when the DB is too large for the schema to fit in the prompt or by using an explicit entity linking step. We eschew these in our experiments for simplicity, as they are not required for \dataset{}.


\subsection{A Text2SQL-Text2SQLCode Hybrid}
Our early experiments indicated that LLMs cannot assess well for which questions to use the decomposition and invoke Python. We thus propose a hybrid approach where we only run \system{} when Text2SQL can not reliably provide a result. We approximate this by using self-consistency \cite{arora-etal-2023-llms} over three runs as a proxy for question difficulty: if two of the three runs provide the same answer, that answer is taken as the prediction; otherwise, the \system{} is invoked and its answer is taken as the prediction. 
This system should be able to gain meaningfully from the strengths of both Text2SQL and \system{} on more complex questions. We implement two hybrid variants \hybrid{}$_{multi}$ and \hybrid{}$_{single}$ corresponding to \system{}$_{multi}$ and \system{}$_{single}$ respectively.


\section{Experimental Setup}

\subsection{System Details}


As baselines, we use the standalone Text2SQL component as described in Section \ref{sec:system}, using the same experimental and task setup. 
Additionally, we also implement a self-consistency \cite{wang-2023-self} approach ($K$Text2SQL + SC) where we output the majority of $K$ answers, or one at random if there is no majority. For our experiments, we use $K = \{3,5\}$. We test all approaches on a variety of current state-of-the-art LLMs including both open- and closed-sourced models, including reasoning models. 
We use the following closed-source (API) models: Claude 3.7, GPT 4.1, GPT o1, GPT o3-mini with their latest versions as of 14 September 2025, and the following open-weight models: DeepSeekV3.1 and Qwen2.5. Out of these, GPT o1 and GPT o3-mini are reasoning models. We use the default temperature for all the models. 


\paragraph{Prompts:} ~We construct one prompt for each database. The prompts are formatted to contain the following information:
\begin{itemize}[noitemsep,topsep=0pt,leftmargin=1em]
    \item A description of the task;
    \item The database schema, including table names, column names, data types and column definitions;
    \item Data samples for each table, such that the LLM is familiar with data types and formats;
    \item A list of all possible values for all fields containing categorical data and their description;
    \item  We instruct that the output should be a list of tuples without any additional names or descriptions, and should not output any additional information, even if relevant. These instructions are added to ensure that the output can be auto-evaluated, as LLMs have a tendency to output additional information.
\end{itemize}

The above metadata is provided in order for the model to obtain as much grounding in the data schema and values as possible. All prompt templates from our experiments are in the supplementary material.

\subsection{Evaluation Metrics}
We test all our approaches on accuracy of execution by running the generated code and comparing the outputs with the reference answer. The output format (list of tuples) is consistent across all questions and is provided as instructions in the prompt. A manual inspection of the output of the model shows this format is widely respected by all models. The execution match metric is insensitive to the order in which the tuples are provided, or the order in which a tuple is constructed. Given these considerations, we are confident that the execution accuracy is an accurate representation of system performance. 

\subsection{Additional Comparisons}

\paragraph{Knowledge:}~
The LLMs' parametric knowledge may contain information about the questions and answers present in \dataset{}, especially if these questions are about information available online. As an additional comparison point to understand the complexity of the \dataset{}, we test the ability of the models to answer these questions directly from their parametric knowledge, by providing explicit instructions in the prompt to the model to not use SQL or code. 

\paragraph{LLMs using Search:}~
We aim to show the performance of state-of-the-art LLM-based agents that can use reasoning, search and other tools in answering questions. These systems can access information and statistics available on websites, or identify articles related to similar questions available online, in order to provide answers to the \dataset{} questions. Albeit, this is not a comparable method to other approaches studied in the paper, as it has access to additional information, we test this in order to understand the value differential that exists in structured data sources. As an exponent of such system, we use Gemini Pro 2.5 (Gemini-2.5-Pro-Preview-05-06) on the week of May 12th, as this tops the Chatbot Arena Leaderboard.\footnote{\url{https://lmarena.ai/}} Given the model does not always adhere to the output format, we additionally process the outputs manually for all questions to obtain its performance score.

\paragraph{\texttt{Archer} Dataset:}~
To demonstrate the generalizability of our \system{} approaches, we also include results on the English validation set of the \texttt{Archer} dataset (AR) \cite{zheng-etal-2024-archer}, an existing text to SQL dataset. This dataset contains 104 questions from 2 databases. 

\section{Results}

\begin{table}[t!]
\centering
\small
\renewcommand{\arraystretch}{1.2}
\resizebox{0.98\columnwidth}{!}{
\begin{tabular}{ccccccc|c}
\hline
\textbf{Model} & \textbf{Method}  & \textbf{\dataset{}} & \textbf{IMDB} & \textbf{ES} & \textbf{OL} &  \textbf{Calls} &   \textbf{AR} \\ \hline
\multirow{3}{*}{\makecell{Claude\\3.7}} 
 & Knowledge & 4.9 & 1.3 & 8.8 & 2.3 & \textbf{1} & 2.6 \\
 & Text2SQL & 35.7 & 29.3 & 43.4 & 29.7 & \textbf{1} &29.5 \\
& Text2SQLCode$_{single}$ & 39.9 & \textbf{50.7} & 34.8 & 37.3 & \textbf{1} &32.0 \\
 & Text2SQLCode$_{multi}$ & 44.0 & 45.3 & 48.8 & 35.0 & 2.8 & \textbf{36.2} \\
 & 3Text2SQL + SC & 37.8 & 30.3 & 46.1 & 32.0 & 3 & 30.1 \\
 & 5Text2SQL + SC & 38.9 & 36.0 & 45.2 & 31.7 & 5 & 29.2 \\
 & Hybrid$_{single}$ & 42.5 & 41.7 & 46.9 & \textbf{36.5} & 3.4 & 32.0 \\
 & Hybrid$_{multi}$ & \textbf{44.2} & 42.3 & \textbf{50.8} & 35.3 & 4.1 & 32.4 \\
 \cline{2-8}
 & Oracle$_{single}$ & 51.5 & 57.3 & 52.5 & 44.0 & - & 48.4 \\
 & Oracle$_{multi}$ & 52.7 & 54.7 & 56.4 & 44.7 & - & 45.8 \\ \hline
\multirow{3}{*}{\makecell{GPT\\ 4.1}} 
 & Knowledge & 6.6 & 0.0 & 13.4 & 2.3 & \textbf{1} & 6.1 \\
 & Text2SQL & 37.2 & 43.0 & 40.5 & 26.0 & \textbf{1} &28.2 \\
 & Text2SQLCode$_{single}$ & 40.4 & 57.7 & 32.3 & \textbf{36.3} & \textbf{1} &30.4 \\
 & Text2SQLCode$_{multi}$ & 40.9 & 53.0 & 36.8 & 35.3 & 2.9 & \underline{\textbf{41.3}} \\
 & 3Text2SQL + SC & 40.5 & 47.7 & 43.2 & 29.0 & 3 & 28.9 \\
 & 5Text2SQL + SC & 39.7 & 44.4 & 43.7 & 28.4 & 5 & 29.5 \\
 & Hybrid$_{single}$ & \textbf{46.4} & \underline{\textbf{60.0}} & 44.7 & 35.7 & 3.2 & 32.7 \\
 & Hybrid$_{multi}$ & 44.6 & 55.0 & \textbf{45.5} & 32.7 & 4.1 & 40.7 \\
 \cline{2-8}
 & Oracle$_{single}$ & 53.8 & 65.7 & 51.2 & 46.0 & - & 38.5 \\
 & Oracle$_{multi}$ & 53.1 & 62.7 & 52.1 & 45.3 & - & 46.2 \\ \hline
\multirow{3}{*}{\makecell{GPT\\o1}}
 & Knowledge & 11.6 & 1.7 & 24.5 & 0.7 & \textbf{1} & 5.5  \\
 & Text2SQL & 39.1 & 41.3 & 43.2 & 30.3 & \textbf{1} &24.7 \\
 & Text2SQLCode$_{single}$ & 48.0 & \textbf{59.0} & 47.9 & 38.0 & \textbf{1} &34.9 \\
 & Text2SQLCode$_{multi}$ & 43.8 & 50.3 & 46.3 & 33.3 & 2.6 & 32.7 \\
 & 3Text2SQL + SC & 41.8 & 44.3 & 46.1 & 32.2 & 3 & 25.0 \\
 & 5Text2SQL + SC & 41.5 & 43.4 & 46.4 & 31.7 & 5 & 25.3 \\
 & Hybrid$_{single}$ & \textbf{48.2} & 50.0 & \textbf{52.7} & \underline{\textbf{38.5}} & 3.3 & \textbf{36.2} \\
 & Hybrid$_{multi}$ & 45.7 & 49.3 & 49.2 & 36.3 & 3.8 & 34.6 \\
 \cline{2-8}
 & Oracle$_{single}$ & 56.8 & 66.1 & 57.8 & 45.7 & -   & 38.1 \\
 & Oracle$_{multi}$ & 52.8 & 60.0 & 54.5 & 43.0 & - & 36.9  \\ \hline
\multirow{3}{*}{\makecell{GPT\\ o3-mini}} 
 & Knowledge & 3.5 & 3.0 & 5.1 & 1.5 & \textbf{1} & 5.8\\
 & Text2SQL & 43.8 & 47.3 & 50.0 & 30.3 & \textbf{1} &26.9 \\
 & Text2SQLCode$_{single}$ & 45.9 & \textbf{59.9} & 44.0 & \textbf{35.0} & \textbf{1} &34.0 \\
 & Text2SQLCode$_{multi}$ & 44.1 & 51.0 & 46.7 & 33.0 & 2.6 & \textbf{36.5} \\
 & 3Text2SQL + SC & 45.2 & 48.3 & 52.1 & 31.0 & 3 & 27.9 \\
 & 5Text2SQL + SC & 46.0 & 50.8 & 51.9 & 31.7 & 5 & 29.2 \\
 & Hybrid$_{single}$ & \underline{\textbf{48.3}} & 54.0 & 53.5 & 34.3 & 3.2 & 32.7 \\
 & Hybrid$_{multi}$ & 47.5 & 53.0 & \underline{\textbf{53.7}} & 32.0 & 3.7 & 35.6 \\\cline{2-8}
 & Oracle$_{single}$ & 55.9 & 66.0 & 57.4 & 43.3 & -  & 37.8 \\
 & Oracle$_{multi}$ & 52.8 & 58.3 & 56.8 & 40.7 & - & 40.1 \\ \hline
\multirow{3}{*}{\makecell{DeepSeek \\ V3.1}}  
 & Knowledge & 3.1 & 1.3 & 4.5 & 2.7 & \textbf{1} & 1.9 \\
 & Text2SQL & 29.9 & 30.0 & 34.2 & 22.7 & \textbf{1} & 26.9 \\
  & Text2SQLCode$_{single}$ & 28.7 & 49.3 & 18.1 & 25.3 & \textbf{1} & 25.3 \\
 & Text2SQLCode$_{multi}$ & 43.6 & 52.5 & 44.4 & 33.3 & 2.81  & \textbf{36.2} \\

 & 3TextSQL + SC & 32.2 & 33.0 & 37.2 & 23.3 & 3 & 28.9 \\
 & 5Text2SQL + SC & 32.6 & 32.4 & 36.9 & 25.7 & 5 & 28.2 \\
 & Hybrid$_{single}$ & 37.3 & 49.0 & 36.2 & 27.3 & 3.5 & 26.3 \\
 & Hybrid$_{multi}$ & \textbf{44.4} & \textbf{53.6} & \textbf{45.1} & \textbf{34.1} &  4.41 & 34.6 \\

 \cline{2-8}
 & Oracle$_{single}$ & 42.5 & 59.0 & 38.9 & 31.7 & - & 36.2 \\
 & Oracle$_{multi}$ & 51.7 & 61.5 & 51.9 & 41.7 & - & 41.3 \\\hline
\multirow{3}{*}{\makecell{Qwen\\ 2.5}} 
 & Knowledge & 0.71 & 0.3 & 0.6 & 1.3 & \textbf{1} & 0 \\
 & Text2SQL & 11.4 & 13.3 & 11.7 & 9.0 & \textbf{1} &12.8 \\
 & Text2SQLCode$_{single}$ & 12.1 & 13.3 & 9.7 & 14.7 & \textbf{1} &13.1 \\
 & Text2SQLCode$_{multi}$ & 15.7 & 22.0 & 11.9 & 15.7 & 3 & 13.3 \\
 & 3Text2SQL + SC & 12.2 & 14.7 & 11.9 & 10.3 & 3 & 14.1 \\
 & 5Text2SQL + SC & 12.2 & 14.2 & 12.2 & 10.2 & 5 & 13.8 \\
 & Hybrid$_{single}$ & 16.7 & 19.3 & 13.2 & 19.7 & 3.7 & 14.1 \\
 & Hybrid$_{multi}$ & \textbf{18.3} & \textbf{24.0} & \textbf{13.6} & \textbf{20.3} & 5.2 & \textbf{14.4} \\ \cline{2-8}
 & Oracle$_{single}$ & 20.6 & 23.0 & 19.1 & 20.7 & - & 23.7 \\
 & Oracle$_{multi}$ & 24.0 & 30.7 & 21.2 & 22.0 &  - & 18.6 \\ \hline
  \hhline{|=|=|=|=|=|=|=|=|}
\makecell{Gemini \\Pro 2.5}  &  Search & 31.2 & 4.0 & 65.4 & 3.0 & - & - \\ \hline
\end{tabular}
}
\caption{\label{t:results} Execution accuracy (average over three runs) and the average \#LLM calls for 6 LLMs with zero-shot prompting on the \dataset{} dataset, its three underlying databases, and the \texttt{Archer} (AR) dataset. Here, Oracle refers to combining the best output of TextSQL and Text2SQLCode post-hoc. Underline represents the best overall result for that dataset, and bold shows the best result for model-dataset pair. The Calls column refers to the average number of LLM calls used by each method.}
\end{table}


Table~\ref{t:results} reports the performance of all six LLMs tested on the \dataset{} and \texttt{Archer} datasets using execution accuracy, each score is an average of three runs.

\textbf{Overall performance on \dataset{} is moderate.} The best result obtained on the overall dataset is 48.3\%. This highlights that the \dataset{} represents a challenging dataset for current state-of-the-art LLMs, especially because there are no challenges regarding entity linking and adapting to a new DB schema present like in other datasets~\cite{lei-2024-spider2,zheng-etal-2024-archer}. The best performing model overall is GPT o3-mini, but the gaps between models (Claude 3.7, GPT 4.1, GPT o1, GPT o3-mini) are narrow, within 4\% overall. Results on the three sections of the dataset show other models performing best (GPT 4.1 for \texttt{IMDb}, GPT-o3-mini for \texttt{ES}, and GPT-o1 for \texttt{OL}), with no clear pattern of what the superior model is. Interestingly, despite reasoning models claiming better performance on coding tasks such as this, the two reasoning models tested (GPT o1 and GPT o3-mini) do not perform substantially better overall on \dataset{} than non-reasoning models. 

Overall, a clear pattern is that the closed-source models outperform open-source models we tested. This is especially evident when using Text2SQL alone. In this setup, the GPT o3-mini model performs substantially better (+4.7\% overall) than any other model.

\textbf{\system{} performs better than Text2SQL alone.} For all the models, the performance of the \system{} methods (both single and multi-step versions) are higher than the Text2SQL method alone, with the overall accuracy improvements ranging up to 14.5\% on \dataset{} for DeepSeek and  13.1\% on \texttt{Archer} for GPT-4.1. Overall, these results highlight that decomposing the task into SQL and Python code execution is a promising direction to improve the performance of the model. We see consistent improvements not only on our own dataset but also on \texttt{Archer}, which demonstrates the generalizability of our approach. We highlight that each \system{}$_{multi}$ run requires multiple LLM calls, between 2 and 3, depending on whether Python code is generated.

\textbf{Self-consistency for Text2SQL improves performance.} Self-consistency over three Text2SQL executions improves performance of the system, in line with past work. Performance improvements are on average 2.1\% and range between 0.8\% and 3.3\% percent. However, we do not always see improvements between three and five executions. In fact, self-consistency with five executions shows performance degradation on \dataset{} for GPT-o1 and GPT-4.1. This shows that simply increasing the number of executions is not enough to improve the overall performance. Note that the improvements come at a cost trade-off, as three or five LLM calls are necessary. Performance of self-consistency, even with Text2SQL alone, is, in many cases, comparable to using \system{} and comes at a similar number of LLM calls. 

\textbf{Different \system{} approaches perform well on different domains.}  Hybrid methods are best for overall accuracy on the \dataset{} and \texttt{ES}, but performance on other domains and \texttt{Archer} is mixed. For 3 of 6 models, \system{}$_{single}$ is the most accurate on \texttt{IMDb}, while \system{}$_{multi}$ is most accurate on \texttt{Archer} for 4 out of 6 models. These experimental results show that the \system{} decomposition should be deployed selectively on questions. An additional indicator is the Oracle method performance, which takes post-hoc the best prediction from the Text2SQL and the \system{} methods. This achieves better performance than any of the two systems by substantial margins from 3\%-9\%. These suggest that, while the model can produce the correct answer, it is hindered by its ability to estimate confidence in its predictions. We study this further in the next section.

\textbf{Methods not using structured data do not perform well on \dataset.} We conduct two experiments where the structured database is not used, in order to measure the value this data brings to the task. As expected, using the model's internal knowledge to answer the questions leads to a low overall performance below 10\%, with the exception of GPT o1 at 11.6\%. Even if the questions are what analysts would ask and thus could be present in the original data used to train the model, usually simple additional constraints (e.g.,  time range filtering) is enough to incapacitate the model to produce the right answer. 

Next, we check the results of the Gemini Pro 2.5 system. The system heavily relies on live searches over the web (usually more than 10 for our questions) to retrieve relevant articles and perform reasoning using the retrieved data in order to find the correct answer. Given that the questions in the \texttt{EuroSoccer} dataset are realistic records that people interested in this sport may ask themselves, as well as summary statistics (e.g., league tables) being readily available online, the performance of the system on the \texttt{EuroSoccer} section is good and better than all models that use structured data (65.4\%). Still, there are still substantial gaps in performance, especially on specific categories such as time series analysis. On the other hand, on the \texttt{IMDb} and \texttt{Olist} sections of the dataset, the performance is very low. This is caused by the nature and difficulty of the questions, which do not ask for well-known records, and because the size of the search space is much vaster than for \texttt{EuroSoccer}, which is only dealing with 11 leagues for 8 seasons.

Both these experiments highlight both the value that is stored in the structured data and that the dataset itself is a challenging dataset in general. Due to the type of the questions and the fine-grained data they require, we expect that this dataset could also represent a valuable general benchmark of LLMs both when using or not using the structured data for grounding.

\subsection{Analysis on Use of Python Code}


\begin{table}[t!]
\centering
\small
\renewcommand{\arraystretch}{1.2}
\resizebox{1.0\columnwidth}{!}{
\begin{tabular}{|cc|ccc|}
\hline
\textbf{Model} & \textbf{Method} & \multicolumn{3}{c|}{\textbf{Pct Questions w/ Python execution}} \\
& & ~~~~\textbf{Full}~~~~ & \textbf{\makecell{T2SQL\\correct}} & \textbf{\makecell{T2SQL\\incorrect}}\\ \hline
\multirow{2}{*}{\makecell{Claude \\ 3.7}} & \system{}$_{multi}$ & 85.2 & -4.8 & $+$2.5 \\
 & \hybrid{}$_{multi}$ & 34.2 & -18.8 & $+$9.7 \\ \hline
\multirow{2}{*}{\makecell{GPT \\ 4.1}} & \system{}$_{multi}$ & 87.6 & -4.7 & $+$2.5  \\
 & \hybrid{}$_{multi}$ & 38.0 & -17.7 & $+$9.4 \\ \hline
\multirow{2}{*}{\makecell{GPT \\ o1}} & \system{}$_{multi}$ & 53.2 & -8.4 & $+$4.1 \\
 & \hybrid{}$_{multi}$ & 23.0 & -13.2 & $+$6.5  \\ \hline
\multirow{2}{*}{\makecell{GPT \\ o3-mini}}  & \system{}$_{multi}$ & 54.9 & -9.4 & $+$5.7 \\
 & \hybrid{}$_{multi}$ & 19.6 & -12.9 & $+$7.8 \\ \hline
\multirow{2}{*}{\makecell{DeepSeek \\ V3.1}}  & \system{}$_{multi}$ & 79.3 & -3.6 & $+$1.6 \\
 &  \hybrid{}$_{multi}$ & 41.2 & -23.1 & $+$10.1 \\ \hline
\multirow{2}{*}{\makecell{Qwen \\ 2.5}} & \system{}$_{multi}$ & 97.8 & -3.0 & $+$0.3\\
 & \hybrid{}$_{multi}$ & 68.5 & -39.2 & $+$4.6 \\
 \hline
\end{tabular}
}
\caption{\label{t:routing_analysis} Post-hoc analysis showing the percentage of questions routed to the Python executor for three sets: full data, the subset where Text2SQL answer is correct and incorrect, respectively. For latter two, we report the relative value w.r.t. the full set value. }
\end{table}

We further investigate the question routing performance of \system{} and \hybrid{}, as results indicated that \hybrid{} performs substantially better likely because of more judicious use of Python. Recall that \system{} has the ability to not use Python if it does not deem it necessary. We measure if this is indeed used for the more difficult questions or it is overused for easier questions, leading to potential loss of performance due to issues in decomposition. To understand this, Table \ref{t:routing_analysis} (column -- `Full') reports the percentage of questions that use Python in \system{}$_{multi}$. Interestingly, we find that all non-reasoning models use Python for 80\% or more questions, hinting at its overuse, given the Text2SQL performance is mostly in the 30-40\% range.

We then compute the same value, but for the two subsets of questions that Text2SQL gets correct or not. While the model can not know this at test time, an ideal model would be aware of its Text2SQL capabilities, and only invoke Python for questions that Text2SQL can not get right. We find that in \system{} there is only $<$ 9\% relative change in either direction for the two subsets, meaning that \system{}'s routing of questions to Python is almost oblivious of the actual accuracy of Text2SQL on the question. This result holds across all six LLMs tested. These results substantiate why in several cases \system{} underperforms Text2SQL (see Table \ref{t:results}), as \system{} is a pipeline and errors can compound more than for Text2SQL, which needs to output a single SQL statement. This is perhaps expected, given that LLMs are generally not well calibrated \cite{arora-etal-2023-llms,stengel-eskin-van-durme-2023-calibrated,liu2025calibratingllmstexttosqlparsing}.

The \hybrid{} system, on the other hand, uses Text2SQL system's own consistency as its proxy for difficulty. The same experiment performed on \hybrid{} shows that there is a up to 39\% relative deviation in routing percentage on the two subsets, leading the system to gain from the complementary strengths of the two systems.

\subsection{Error Analysis}

\paragraph{Text2SQL vs \system{}:}~
Our approaches excel in questions requiring: (1) Advanced string manipulation and regular expressions, (2) Conditional aggregation using derived categories, (3) A clear split between database access (SQL) and logic-heavy processing (Python), and (4) Large-scale joins and filtering in SQL, combined with dynamic grouping, per-row aggregation, and in-memory calculations in Python. We present two examples below.

Example 1: \textit{For each genre with at least 100,000 known movies, compute the percentage of movies rated above 8.0 and rated below 5.0.}

This highlights the limits of pure Text2SQL with semi-structured data (e.g., comma-separated genres in IMDB). SQLite lacks native string-splitting, forcing fragile workarounds. All GPT-4.1 Text2SQL runs failed with runtime errors. \system{} approaches succeeded by using SQL for bulk retrieval and Python for splitting and aggregation — leveraging each language’s strengths for robust, correct results.

Example 2: \textit{Find the TV shows where the absolute rating difference between the first and last episode is the highest. Each season of the TV show can be considered as a separate show.}

This example shows the efficiency of \system{} for complex per-group logic. This question caused pure Text2SQL runs with GPT-o3-mini to stall for 20+ minutes due to SQL inefficiencies. Our methods split the work: SQL fetched episode data, while Python grouped seasons, found boundary episodes, and computed differences. This division leveraged each language’s strengths, produced the correct output, and completed in under three minutes—demonstrating how decomposition scales and performs better than SQL alone.

\paragraph{\system{} Errors:}~
We manually inspect the errors made by \system{}$_{multi}$  and \system{}$_{single}$ systems. We find that the majority of errors are logical mistakes in implementing code/SQL for the user question. Examples include missing clauses (question asks for movies, and query forgets to filter on \texttt{titletype=movie}),  inaccurate null processing (question requires aggregation per season, and code considers \texttt{null} season as a separate season), and missed steps (question requires a final count and answer outputs a list, or question requires a final aggregate, whereas answer outputs it per season). Another error type occurs due to inaccurate decomposition -- for example, the question asks for a player's name, but the model misses retaining this information in its Text2SQL prompt. So, the code lacks name data, and can at best output player IDs, leading to execution errors. Other less frequent errors include answers with additional information, or in wrong format/units, and Text2SQL prompts dependent on each other. In most cases, we find that broadly the solution contains some of the correct elements of the query, but includes subtle errors, loses context or performs reasoning errors which ultimately lead to an incorrect answer.

We present additional analysis in Appendix \ref{app:additional_error_analysis}.

\section{Conclusions}

We introduce \dataset{}, the first public human-created dataset for QA over structured data aimed at measuring the abilities of models to perform complex analytical reasoning. In dataset construction, we focused on several specific categories of reasoning such as statistical operations and analysis, time-series analysis, complex conditional logic, multi-criteria filtering, using commonsense knowledge and performing scenario understanding. Through our experiments, we demonstrated that this dataset is challenging for current state-of-the-art LLMs, with the best performance reaching up to 48.3\%. We experimented with generating SQL code alone, and also proposed to decompose the solution into using SQL and Python code, where the code can be simpler and more expressive and showed that the latter approach produces generally better results, especially when invoked only if necessary.

Future work can further explore improving the performance of models on this data set by leveraging the idea of task decomposition and potentially combining this into agentic workflows, that can plan, examine and correct their mistakes and invoke other tools. The \dataset{} itself could be extended to include even more types of analytical reasoning and cover more domains, 
in order to enable study and methods for cross-domain training.

\section*{Limitations}

Our dataset and methods are evaluated on three underlying databases that we consider expose several challenges related to complex analytical reasoning. We acknowledge these do not represent the entire complexity of all databases, although we consider our dataset as a solid first step to explore other data bases. Further, we limit our study to queries and dataset in English, hence we did not test the generalizability to other languages and multi-lingual models, and leave this to future work. Our prompts for testing LLMs are constructed with best prompting strategies from the literature, however there is more scope to perform prompt optimization and tuning in order to achieve the best results. 
All of our experiments are relevant for SQL databases, but, in principle, the ideas herein should be equally suited to complex reasoning questions on data sources such as a knowledge graph or a semi-structured table. 

\section*{Acknowledgements}
We acknowledge Ella Hoffman-Coyle, and Shuyi Wang for their early discussions on the project. We also thank Danna Zheng for sharing their evaluation code with us. Most of the work was done when Mausam was on a full-time sabbatical at Bloomberg. For the remaining work, Mausam acknowledges support from IBM and Verisk grants to IIT Delhi.


\bibliography{anthology,custom}

\clearpage

\appendix

\onecolumn
\section{\dataset{} Categorization Results}

\begin{table*}[ht]
\centering
\small
\renewcommand{\arraystretch}{1.2}
\resizebox{1.0\columnwidth}{!}{
\begin{tabular}{ccccccccccccc}
\hline
\textbf{Model} & \textbf{Method}  & \textbf{Stats} & \textbf{Non-entries in DB} & \textbf{Nested Joins} & \textbf{Strings} & \textbf{Agg Analytics} & \textbf{Complex Column} & \textbf{Temporal Reasoning} & \textbf{Complex Filtering} & \textbf{Time Series} & \textbf{Scenario} & \textbf{Common sense} \\ \hline
\multirow{3}{*}{\makecell{Claude\\3.7}} & Text2SQL & 42.9 & 15.1 & 38.2 & 41.0 & 27.9 & 25.7 & 20.7 & 38.9 & 23.9 & 41.1 & 17.2 \\
 & Text2SQLCode$_{single}$ & 41.6 & \textbf{18.4} & 35.1 & \textbf{47.8} & 29.5 & 22.3 & 22.0 & 41.3 & \textbf{36.8} & 35.6 & 14.9 \\
 & Text2SQLCode$_{multi}$ & 44.3 & 16.1 & 41.3 & 47.5 & \textbf{37.1} & \textbf{35.0} & 37.9 & 40.8 & 35.6 & 36.6 & \textbf{18.7} \\
 & Hybrid$_{single}$ & 43.3 & 16.7 & 41.6 & 46.1 & 28.7 & 22.7 & 40.3 & \textbf{47.4} & 34.4 & \textbf{43.1} & 12.4 \\
 & Hybrid$_{multi}$ & \textbf{43.8} & 14.4 & \textbf{43.5} & 44.4 & 32.0 & 25.4 & \textbf{43.6} & 46.4 & 35.6 & 38.5 & 16.1 \\ \hline
 
 \multirow{3}{*}{\makecell{GPT\\4.1}} & Text2SQL & 29.3 & 19.6 & 34.6 & 39.3 & 36.1 & 14.2 & 26.1 & 34.9 & 36.4 & 41.7 & 12.4 \\
 & Text2SQLCode$_{single}$ & 35.9 & 24.7 & 35.6 & 52.6 & 36.2 & \textbf{36.5} & 29.5 & 37.0 & 46.1 & 37.9 & \textbf{22.4} \\
 & Text2SQLCode$_{multi}$ & \textbf{38.6} & 21.3 & 37.7 & 44.4 & 39.2 & 24.4 & 31.1 & 40.9 & 33.5 & 36.7 & 11.2 \\
 & Hybrid$_{single}$ & 34.9 & \textbf{25.3} & 41.8 & \textbf{49.1} & \textbf{41.2} & 28.1 & \textbf{33.3} & 42.9 & \textbf{56.6} & \textbf{56.9} & 18.7 \\
 & Hybrid$_{multi}$ & 34.4 & 22.4 & \textbf{42.3} & 47.6 & 40.9 & 20.3 & 33.7 & \textbf{43.7} & 44.9 & 50.0 & 13.7 \\ \hline
 
 \multirow{3}{*}{\makecell{GPT\\o1}}  & Text2SQL & 34.0 & 15.6 & 35.1 & 42.3 & 35.5 & 15.8 & 21.7 & 44.4 & 47.5 & 39.4 & 23.6 \\
 & Text2SQLCode$_{single}$ & \textbf{49.1} & \textbf{24.2} & \textbf{45.0} & \textbf{65.1} & 40.1 & \textbf{38.0} & \textbf{38.3} & 49.8 & 50.4 & \textbf{42.7} & 26.1 \\
 & Text2SQLCode$_{multi}$ & 45.0 & 17.9 & 35.4 & 50.9 & 34.4 & 32.8 & 18.7 & 42.8 & 34.6 & 32.0 & 31.0 \\
 & Hybrid$_{single}$ & 44.7 & 19.0 & 44.9 & 56.0 & \textbf{43.0} & 27.8 & 30.9 & \textbf{52.2} & \textbf{58.0} & 42.6 & \textbf{32.4} \\
 & Hybrid$_{multi}$ & 43.5 & 19.0 & 41.9 & 48.2 & 41.0 & 25.0 & 26.7 & 46.3 & 54.5 & 40.8 & 31.0 \\ \hline

 \multirow{3}{*}{\makecell{GPT\\o3-mini}}  & Text2SQL & 39.6 & 17.3 & 35.3 & 46.1 & 34.6 & 36.2 & 33.1 & 48.2 & 50.7 & 30.6 & 26.1 \\
 & Text2SQLCode$_{single}$ & \textbf{43.6} & \textbf{27.6} & \textbf{41.0} & \textbf{55.7} & \textbf{37.0} & \textbf{38.3} & 31.7 & 50.5 & 55.6 & 26.9 & 26.1 \\
 & Text2SQLCode$_{multi}$ & 42.1 & 22.4 & 34.8 & 50.3 & 34.5 & 30.5 & 24.7 & 51.0 & 44.3 & \textbf{32.0} & 24.9 \\
 & Hybrid$_{single}$ & 41.9 & 20.1 & 38.4 & 48.6 & 32.8 & 38.0 & \textbf{37.4} & \textbf{55.8} & \textbf{55.7} & 30.2 & 29.8 \\
 & Hybrid$_{multi}$ & 42.1 & 19.0 & 37.7 & 48.0 & 33.6 & 46.4 & 34.6 & 56.0 & 51.2 & 29.7 & \textbf{31.0} \\ \hline
 
  \multirow{3}{*}{\makecell{Deepseek\\V3.1}} 
  & Text2SQL & 26.2 & 15.5 & 24.6 & 37.3 & 21.2 & 24.5 & 22.3 & 31.6 & 22.8 & 25.0 & 8.7 \\
 & Text2SQLCode$_{single}$  & 30.2 & \textbf{27.0} & 22.5 & \textbf{47.3} & 20.5 & 15.3 & 21.4 & 26.9 & 24.3 & 22.7 & 9.9  \\
 & Text2SQLCode$_{multi}$ &  \textbf{45.6} & 19.0 & \textbf{39.6} & 46.6 & \textbf{36.0} & \textbf{47.8} & \textbf{35.7} & 40.5 & \textbf{40.7} & 36.9 & \textbf{16.8} \\
 & Hybrid$_{single}$ &  35.9 & 27.0 & 31.2 & 47.2 & 26.0 & 38.0 & 26.8 & 39.9 & 31.1 & 27.3 & 11.2 \\
 & Hybrid$_{multi}$ & 45.4 & 20.7 & 39.2 & 45.5 & 35.0 & 40.3 & 29.8 & \textbf{43.4} & 40.0 & \textbf{42.6} & 14.9 \\ \hline
\end{tabular}
}
\caption{\label{t:categories_results} Execution accuracy (average over three runs) over different reasoning categories of \dataset{}.}
\end{table*}

 Table \ref{t:categories_results} shows our approaches consistently outperform Text2SQL on questions involving statistical operations, non-entries in DB, string operations, nested joins, and complex filtering. While a few models show a slight degradation for other categories, our models generally perform better. This analysis is suggestive due to the small number of questions in some categories.

\section{Additional Error Analysis}
\label{app:additional_error_analysis}
\paragraph{Text2SQL: Syntactic Error Diversity and Debuggability:}~
Models varied significantly in the types and consistency of their SQL-related errors. Some, like Qwen-2.5 and DeepSeek-v3.1, produced a broad range of failure modes—including unsupported functions, misused columns, and unstable output formatting—indicating weaker alignment with the SQL dialect and greater fragility. In contrast, models like GPT-o3-mini and GPT-o1 generated more uniform and predictable errors, making them easier to debug and more reliable in execution. Models with lower error diversity tended to exhibit stronger syntactic control and were better suited for integration in systems that rely on stable SQL generation.

\paragraph{Text2SQL-SC: Self-Consistency and Majority-Voting Behavior:}~
Models also differed in how consistently they generated predictions across multiple runs. GPT-o3-mini, GPT-o1, and GPT-4.1 produced highly stable outputs, with correct predictions aligning across runs—making them well-suited for majority-voting strategies. In contrast, Qwen-2.5 repeatedly generated the same incorrect outputs, preventing fallback mechanisms (like Text2SQLCode) from activating and leading to consistently poor results. This shows that self-consistency only adds value when the underlying predictions are reliable.

While \hybrid{} approaches often enhance the robustness of \system{}, they are not universally better. Problems arise when a model: (1) consistently produces incorrect SQL predictions, causing self-consistency voting to lock in bad outputs, and (2) fails again during Python execution, making recovery impossible even if fallback is triggered.

This was the case with Qwen-2.5, which combined high runtime failure, poor prediction diversity, and low correctness—rendering hybridization ineffective. In contrast, GPT-o3-mini and GPT-o1 demonstrate how hybridization can succeed: their SQL outputs are both reliable and diverse, enabling fallback to Python when necessary and ensuring correctness when it is not.

\clearpage

\section{Example Text2SQL Prompt for \texttt{Eurosoccer}}
\UseRawInputEncoding
\lstset{
  breaklines=true,
  breakatwhitespace=true,
  columns=flexible,
  keepspaces=true,
  basicstyle=\ttfamily\small,
  showstringspaces=false,
  frame=single
}

\begin{lstlisting}
You are a helpful assistant to a database engineer. The user has provided a complex instruction. Your job is to write a valid SQL for it.

This is how the database was created.
CREATE TABLE "Player_Attributes" (
	`id`	INTEGER PRIMARY KEY AUTOINCREMENT,
	`player_fifa_api_id`	INTEGER,
	`player_api_id`	INTEGER,
	`date`	TEXT,
	`overall_rating`	INTEGER,
	`potential`	INTEGER,
	`preferred_foot`	TEXT,
	`attacking_work_rate`	TEXT,
	`defensive_work_rate`	TEXT,
	`crossing`	INTEGER,
	`finishing`	INTEGER,
	`heading_accuracy`	INTEGER,
	`short_passing`	INTEGER,
	`volleys`	INTEGER,
	`dribbling`	INTEGER,
	`curve`	INTEGER,
	`free_kick_accuracy`	INTEGER,
	`long_passing`	INTEGER,
	`ball_control`	INTEGER,
	`acceleration`	INTEGER,
	`sprint_speed`	INTEGER,
	`agility`	INTEGER,
	`reactions`	INTEGER,
	`balance`	INTEGER,
	`shot_power`	INTEGER,
	`jumping`	INTEGER,
	`stamina`	INTEGER,
	`strength`	INTEGER,
	`long_shots`	INTEGER,
	`aggression`	INTEGER,
	`interceptions`	INTEGER,
	`positioning`	INTEGER,
	`vision`	INTEGER,
	`penalties`	INTEGER,
	`marking`	INTEGER,
	`standing_tackle`	INTEGER,
	`sliding_tackle`	INTEGER,
	`gk_diving`	INTEGER,
	`gk_handling`	INTEGER,
	`gk_kicking`	INTEGER,
	`gk_positioning`	INTEGER,
	`gk_reflexes`	INTEGER,
	FOREIGN KEY(`player_fifa_api_id`) REFERENCES `Player`(`player_fifa_api_id`),
	FOREIGN KEY(`player_api_id`) REFERENCES `Player`(`player_api_id`)
)
/* 
3 example rows:
SELECT * FROM Player_Attributes LIMIT 3;
id player_fifa_api_id player_api_id                date overall_rating potential preferred_foot attacking_work_rate defensive_work_rate crossing finishing heading_accuracy short_passing volleys dribbling curve free_kick_accuracy long_passing ball_control acceleration sprint_speed agility reactions balance shot_power jumping stamina strength long_shots aggression interceptions positioning vision penalties marking standing_tackle sliding_tackle gk_diving gk_handling gk_kicking gk_positioning gk_reflexes 
 1             218353        505942 2016-02-18 00:00:00             67        71          right              medium              medium       49        44               71            61      44        51    45                 39           64           49           60           64      59        47      65         55      58      54       76         35         71            70          45     54        48      65              69             69         6          11         10              8           8 
 2             218353        505942 2015-11-19 00:00:00             67        71          right              medium              medium       49        44               71            61      44        51    45                 39           64           49           60           64      59        47      65         55      58      54       76         35         71            70          45     54        48      65              69             69         6          11         10              8           8 
 3             218353        505942 2015-09-21 00:00:00             62        66          right              medium              medium       49        44               71            61      44        51    45                 39           64           49           60           64      59        47      65         55      58      54       76         35         63            41          45     54        48      65              66             69         6          11         10              8           8 
*/

CREATE TABLE `Player` (
	`id`	INTEGER PRIMARY KEY AUTOINCREMENT,
	`player_api_id`	INTEGER UNIQUE,
	`player_name`	TEXT,
	`player_fifa_api_id`	INTEGER UNIQUE,
	`birthday`	TEXT,
	`height`	INTEGER,
	`weight`	INTEGER
)
/* 
3 example rows:
SELECT * FROM Player LIMIT 3;
id player_api_id        player_name player_fifa_api_id            birthday height weight 
 1        505942 Aaron Appindangoye             218353 1992-02-29 00:00:00 182.88    187 
 2        155782    Aaron Cresswell             189615 1989-12-15 00:00:00 170.18    146 
 3        162549        Aaron Doran             186170 1991-05-13 00:00:00 170.18    163 
*/

CREATE TABLE `Match` (
	`id`	INTEGER PRIMARY KEY AUTOINCREMENT,
	`country_id`	INTEGER,
	`league_id`	INTEGER,
	`season`	TEXT,
	`stage`	INTEGER,
	`date`	TEXT,
	`match_api_id`	INTEGER UNIQUE,
	`home_team_api_id`	INTEGER,
	`away_team_api_id`	INTEGER,
	`home_team_goal`	INTEGER,
	`away_team_goal`	INTEGER,
	`home_player_X1`	INTEGER,
	`home_player_X2`	INTEGER,
	`home_player_X3`	INTEGER,
	`home_player_X4`	INTEGER,
	`home_player_X5`	INTEGER,
	`home_player_X6`	INTEGER,
	`home_player_X7`	INTEGER,
	`home_player_X8`	INTEGER,
	`home_player_X9`	INTEGER,
	`home_player_X10`	INTEGER,
	`home_player_X11`	INTEGER,
	`away_player_X1`	INTEGER,
	`away_player_X2`	INTEGER,
	`away_player_X3`	INTEGER,
	`away_player_X4`	INTEGER,
	`away_player_X5`	INTEGER,
	`away_player_X6`	INTEGER,
	`away_player_X7`	INTEGER,
	`away_player_X8`	INTEGER,
	`away_player_X9`	INTEGER,
	`away_player_X10`	INTEGER,
	`away_player_X11`	INTEGER,
	`home_player_Y1`	INTEGER,
	`home_player_Y2`	INTEGER,
	`home_player_Y3`	INTEGER,
	`home_player_Y4`	INTEGER,
	`home_player_Y5`	INTEGER,
	`home_player_Y6`	INTEGER,
	`home_player_Y7`	INTEGER,
	`home_player_Y8`	INTEGER,
	`home_player_Y9`	INTEGER,
	`home_player_Y10`	INTEGER,
	`home_player_Y11`	INTEGER,
	`away_player_Y1`	INTEGER,
	`away_player_Y2`	INTEGER,
	`away_player_Y3`	INTEGER,
	`away_player_Y4`	INTEGER,
	`away_player_Y5`	INTEGER,
	`away_player_Y6`	INTEGER,
	`away_player_Y7`	INTEGER,
	`away_player_Y8`	INTEGER,
	`away_player_Y9`	INTEGER,
	`away_player_Y10`	INTEGER,
	`away_player_Y11`	INTEGER,
	`home_player_1`	INTEGER,
	`home_player_2`	INTEGER,
	`home_player_3`	INTEGER,
	`home_player_4`	INTEGER,
	`home_player_5`	INTEGER,
	`home_player_6`	INTEGER,
	`home_player_7`	INTEGER,
	`home_player_8`	INTEGER,
	`home_player_9`	INTEGER,
	`home_player_10`	INTEGER,
	`home_player_11`	INTEGER,
	`away_player_1`	INTEGER,
	`away_player_2`	INTEGER,
	`away_player_3`	INTEGER,
	`away_player_4`	INTEGER,
	`away_player_5`	INTEGER,
	`away_player_6`	INTEGER,
	`away_player_7`	INTEGER,
	`away_player_8`	INTEGER,
	`away_player_9`	INTEGER,
	`away_player_10`	INTEGER,
	`away_player_11`	INTEGER,
	`B365H`	NUMERIC,
	`B365D`	NUMERIC,
	`B365A`	NUMERIC,
	`BWH`	NUMERIC,
	`BWD`	NUMERIC,
	`BWA`	NUMERIC,
	`IWH`	NUMERIC,
	`IWD`	NUMERIC,
	`IWA`	NUMERIC,
	`LBH`	NUMERIC,
	`LBD`	NUMERIC,
	`LBA`	NUMERIC,
	`PSH`	NUMERIC,
	`PSD`	NUMERIC,
	`PSA`	NUMERIC,
	`WHH`	NUMERIC,
	`WHD`	NUMERIC,
	`WHA`	NUMERIC,
	`SJH`	NUMERIC,
	`SJD`	NUMERIC,
	`SJA`	NUMERIC,
	`VCH`	NUMERIC,
	`VCD`	NUMERIC,
	`VCA`	NUMERIC,
	`GBH`	NUMERIC,
	`GBD`	NUMERIC,
	`GBA`	NUMERIC,
	`BSH`	NUMERIC,
	`BSD`	NUMERIC,
	`BSA`	NUMERIC,
	FOREIGN KEY(`country_id`) REFERENCES `country`(`id`),
	FOREIGN KEY(`league_id`) REFERENCES `League`(`id`),
	FOREIGN KEY(`home_team_api_id`) REFERENCES `Team`(`team_api_id`),
	FOREIGN KEY(`away_team_api_id`) REFERENCES `Team`(`team_api_id`),
	FOREIGN KEY(`home_player_1`) REFERENCES `Player`(`player_api_id`),
	FOREIGN KEY(`home_player_2`) REFERENCES `Player`(`player_api_id`),
	FOREIGN KEY(`home_player_3`) REFERENCES `Player`(`player_api_id`),
	FOREIGN KEY(`home_player_4`) REFERENCES `Player`(`player_api_id`),
	FOREIGN KEY(`home_player_5`) REFERENCES `Player`(`player_api_id`),
	FOREIGN KEY(`home_player_6`) REFERENCES `Player`(`player_api_id`),
	FOREIGN KEY(`home_player_7`) REFERENCES `Player`(`player_api_id`),
	FOREIGN KEY(`home_player_8`) REFERENCES `Player`(`player_api_id`),
	FOREIGN KEY(`home_player_9`) REFERENCES `Player`(`player_api_id`),
	FOREIGN KEY(`home_player_10`) REFERENCES `Player`(`player_api_id`),
	FOREIGN KEY(`home_player_11`) REFERENCES `Player`(`player_api_id`),
	FOREIGN KEY(`away_player_1`) REFERENCES `Player`(`player_api_id`),
	FOREIGN KEY(`away_player_2`) REFERENCES `Player`(`player_api_id`),
	FOREIGN KEY(`away_player_3`) REFERENCES `Player`(`player_api_id`),
	FOREIGN KEY(`away_player_4`) REFERENCES `Player`(`player_api_id`),
	FOREIGN KEY(`away_player_5`) REFERENCES `Player`(`player_api_id`),
	FOREIGN KEY(`away_player_6`) REFERENCES `Player`(`player_api_id`),
	FOREIGN KEY(`away_player_7`) REFERENCES `Player`(`player_api_id`),
	FOREIGN KEY(`away_player_8`) REFERENCES `Player`(`player_api_id`),
	FOREIGN KEY(`away_player_9`) REFERENCES `Player`(`player_api_id`),
	FOREIGN KEY(`away_player_10`) REFERENCES `Player`(`player_api_id`),
	FOREIGN KEY(`away_player_11`) REFERENCES `Player`(`player_api_id`)
)
/* 
3 example rows:
SELECT * FROM Match LIMIT 3;
id country_id league_id    season stage                date match_api_id home_team_api_id away_team_api_id home_team_goal away_team_goal home_player_X1 home_player_X2 home_player_X3 home_player_X4 home_player_X5 home_player_X6 home_player_X7 home_player_X8 home_player_X9 home_player_X10 home_player_X11 away_player_X1 away_player_X2 away_player_X3 away_player_X4 away_player_X5 away_player_X6 away_player_X7 away_player_X8 away_player_X9 away_player_X10 away_player_X11 home_player_Y1 home_player_Y2 home_player_Y3 home_player_Y4 home_player_Y5 home_player_Y6 home_player_Y7 home_player_Y8 home_player_Y9 home_player_Y10 home_player_Y11 away_player_Y1 away_player_Y2 away_player_Y3 away_player_Y4 away_player_Y5 away_player_Y6 away_player_Y7 away_player_Y8 away_player_Y9 away_player_Y10 away_player_Y11 home_player_1 home_player_2 home_player_3 home_player_4 home_player_5 home_player_6 home_player_7 home_player_8 home_player_9 home_player_10 home_player_11 away_player_1 away_player_2 away_player_3 away_player_4 away_player_5 away_player_6 away_player_7 away_player_8 away_player_9 away_player_10 away_player_11 B365H B365D B365A  BWH  BWD  BWA  IWH IWD IWA LBH LBD  LBA  PSH  PSD  PSA  WHH  WHD  WHA  SJH SJD SJA  VCH  VCD  VCA  GBH  GBD  GBA  BSH  BSD  BSA 
 1          1         1 2008/2009     1 2008-08-17 00:00:00       492473             9987             9993              1              1           None           None           None           None           None           None           None           None           None            None            None           None           None           None           None           None           None           None           None           None            None            None           None           None           None           None           None           None           None           None           None            None            None           None           None           None           None           None           None           None           None           None            None            None          None          None          None          None          None          None          None          None          None           None           None          None          None          None          None          None          None          None          None          None           None           None  1.73   3.4     5 1.75 3.35  4.2 1.85 3.2 3.5 1.8 3.3 3.75 None None None  1.7  3.3 4.33  1.9 3.3   4 1.65  3.4  4.5 1.78 3.25    4 1.73  3.4  4.2 
 2          1         1 2008/2009     1 2008-08-16 00:00:00       492474            10000             9994              0              0           None           None           None           None           None           None           None           None           None            None            None           None           None           None           None           None           None           None           None           None            None            None           None           None           None           None           None           None           None           None           None            None            None           None           None           None           None           None           None           None           None           None            None            None          None          None          None          None          None          None          None          None          None           None           None          None          None          None          None          None          None          None          None          None           None           None  1.95   3.2   3.6  1.8  3.3 3.95  1.9 3.2 3.5 1.9 3.2  3.5 None None None 1.83  3.3  3.6 1.95 3.3 3.8    2 3.25 3.25 1.85 3.25 3.75 1.91 3.25  3.6 
 3          1         1 2008/2009     1 2008-08-16 00:00:00       492475             9984             8635              0              3           None           None           None           None           None           None           None           None           None            None            None           None           None           None           None           None           None           None           None           None            None            None           None           None           None           None           None           None           None           None           None            None            None           None           None           None           None           None           None           None           None           None            None            None          None          None          None          None          None          None          None          None          None           None           None          None          None          None          None          None          None          None          None          None           None           None  2.38   3.3  2.75  2.4  3.3 2.55  2.6 3.1 2.3 2.5 3.2  2.5 None None None  2.5 3.25  2.4 2.63 3.3 2.5 2.35 3.25 2.65  2.5  3.2  2.5  2.3  3.2 2.75 
*/

CREATE TABLE `League` (
	`id`	INTEGER PRIMARY KEY AUTOINCREMENT,
	`country_id`	INTEGER,
	`name`	TEXT UNIQUE,
	FOREIGN KEY(`country_id`) REFERENCES `country`(`id`)
)
/* 
3 example rows:
SELECT * FROM League LIMIT 3;
  id country_id                   name 
   1          1 Belgium Jupiler League 
1729       1729 England Premier League 
4769       4769         France Ligue 1 
*/

CREATE TABLE `Country` (
	`id`	INTEGER PRIMARY KEY AUTOINCREMENT,
	`name`	TEXT UNIQUE
)
/* 
3 example rows:
SELECT * FROM Country LIMIT 3;
  id    name 
   1 Belgium 
1729 England 
4769  France 
*/

CREATE TABLE "Team" (
	`id`	INTEGER PRIMARY KEY AUTOINCREMENT,
	`team_api_id`	INTEGER UNIQUE,
	`team_fifa_api_id`	INTEGER,
	`team_long_name`	TEXT,
	`team_short_name`	TEXT
)
/* 
3 example rows:
SELECT * FROM Team LIMIT 3;
id team_api_id team_fifa_api_id   team_long_name team_short_name 
 1        9987              673         KRC Genk             GEN 
 2        9993              675     Beerschot AC             BAC 
 3       10000            15005 SV Zulte-Waregem             ZUL 
*/

CREATE TABLE `Team_Attributes` (
	`id`	INTEGER PRIMARY KEY AUTOINCREMENT,
	`team_fifa_api_id`	INTEGER,
	`team_api_id`	INTEGER,
	`date`	TEXT,
	`buildUpPlaySpeed`	INTEGER,
	`buildUpPlaySpeedClass`	TEXT,
	`buildUpPlayDribbling`	INTEGER,
	`buildUpPlayDribblingClass`	TEXT,
	`buildUpPlayPassing`	INTEGER,
	`buildUpPlayPassingClass`	TEXT,
	`buildUpPlayPositioningClass`	TEXT,
	`chanceCreationPassing`	INTEGER,
	`chanceCreationPassingClass`	TEXT,
	`chanceCreationCrossing`	INTEGER,
	`chanceCreationCrossingClass`	TEXT,
	`chanceCreationShooting`	INTEGER,
	`chanceCreationShootingClass`	TEXT,
	`chanceCreationPositioningClass`	TEXT,
	`defencePressure`	INTEGER,
	`defencePressureClass`	TEXT,
	`defenceAggression`	INTEGER,
	`defenceAggressionClass`	TEXT,
	`defenceTeamWidth`	INTEGER,
	`defenceTeamWidthClass`	TEXT,
	`defenceDefenderLineClass`	TEXT,
	FOREIGN KEY(`team_fifa_api_id`) REFERENCES `Team`(`team_fifa_api_id`),
	FOREIGN KEY(`team_api_id`) REFERENCES `Team`(`team_api_id`)
)
/* 
3 example rows:
SELECT * FROM Team_Attributes LIMIT 3;
id team_fifa_api_id team_api_id                date buildUpPlaySpeed buildUpPlaySpeedClass buildUpPlayDribbling buildUpPlayDribblingClass buildUpPlayPassing buildUpPlayPassingClass buildUpPlayPositioningClass chanceCreationPassing chanceCreationPassingClass chanceCreationCrossing chanceCreationCrossingClass chanceCreationShooting chanceCreationShootingClass chanceCreationPositioningClass defencePressure defencePressureClass defenceAggression defenceAggressionClass defenceTeamWidth defenceTeamWidthClass defenceDefenderLineClass 
 1              434        9930 2010-02-22 00:00:00               60              Balanced                 None                    Little                 50                   Mixed                   Organised                    60                     Normal                     65                      Normal                     55                      Normal                      Organised              50               Medium                55                  Press               45                Normal                    Cover 
 2              434        9930 2014-09-19 00:00:00               52              Balanced                   48                    Normal                 56                   Mixed                   Organised                    54                     Normal                     63                      Normal                     64                      Normal                      Organised              47               Medium                44                  Press               54                Normal                    Cover 
 3              434        9930 2015-09-10 00:00:00               47              Balanced                   41                    Normal                 54                   Mixed                   Organised                    54                     Normal                     63                      Normal                     64                      Normal                      Organised              47               Medium                44                  Press               54                Normal                    Cover 
*/

CREATE TABLE goals (
                id INTEGER PRIMARY KEY,
                match_id INTEGER NOT NULL,
                team INTEGER NOT NULL,
                player1 INTEGER NOT NULL,
                player2 INTEGER NOT NULL,
                goal_type TEXT,
                subtype TEXT,
                elapsed INTEGER NOT NULL,
                elapsed_plus INTEGER,
                n INTEGER NOT NULL,
                sortorder INTEGER NOT NULL,
                FOREIGN KEY (match_id) REFERENCES match (id)
                FOREIGN KEY (team) REFERENCES team (team_api_id)
                FOREIGN KEY (player1) REFERENCES player (player_api_id)
                FOREIGN KEY (player2) REFERENCES player (player_api_id)
            )
/* 
3 example rows:
SELECT * FROM goals LIMIT 3;
    id match_id team player1 player2 goal_type subtype elapsed elapsed_plus  n sortorder 
375301     7809 9823   30872                 n              12              21         0 
375302     7809 9823   32118                 p              16              22         0 
375303     7809 9790   33101                 n              25              20         0 
*/

CREATE TABLE cards (
                id INTEGER PRIMARY KEY,
                match_id INTEGER NOT NULL,
                team INTEGER NOT NULL,
                player1 INTEGER NOT NULL,
                card_type TEXT,
                subtype TEXT,
                elapsed INTEGER NOT NULL,
                elapsed_plus INTEGER,
                n INTEGER NOT NULL,
                sortorder INTEGER NOT NULL,
                FOREIGN KEY (match_id) REFERENCES match (id)
                FOREIGN KEY (team) REFERENCES team (team_api_id)
                FOREIGN KEY (player1) REFERENCES player (player_api_id)
            )
/* 
3 example rows:
SELECT * FROM cards LIMIT 3;
    id match_id team player1 card_type subtype elapsed elapsed_plus  n sortorder 
375310     7809 9790   30749         y              19              26         0 
375311     7809 9823   38244         y              64              28         0 
375312     7809 9790   33101         y              44              27         0 
*/

CREATE TABLE fouls (
                id INTEGER PRIMARY KEY,
                match_id INTEGER NOT NULL,
                team INTEGER NOT NULL,
                player1 INTEGER NOT NULL,
                player2 INTEGER NOT NULL,
                subtype TEXT,
                elapsed INTEGER NOT NULL,
                elapsed_plus INTEGER,
                injury_time INTEGER,
                n INTEGER NOT NULL,
                sortorder INTEGER NOT NULL,
                FOREIGN KEY (match_id) REFERENCES match (id)
                FOREIGN KEY (team) REFERENCES team (team_api_id)
                FOREIGN KEY (player1) REFERENCES player (player_api_id)
                FOREIGN KEY (player2) REFERENCES player (player_api_id)
            )
/* 
3 example rows:
SELECT * FROM fouls LIMIT 3;
    id match_id team player1 player2 subtype elapsed elapsed_plus injury_time   n sortorder 
375531     1730 8659   36832   26181               2                          227         0 
375541     1730 9825   31291   23257               3                          230         5 
375551     1730 9825   30935   31088 pushing       5                          232         0 
*/

CREATE TABLE corners (
                id INTEGER PRIMARY KEY,
                match_id INTEGER NOT NULL,
                team INTEGER NOT NULL,
                player1 INTEGER NOT NULL,
                subtype TEXT,
                elapsed INTEGER NOT NULL,
                elapsed_plus INTEGER,
                n INTEGER NOT NULL,
                sortorder INTEGER NOT NULL,
                FOREIGN KEY (match_id) REFERENCES match (id)
                FOREIGN KEY (team) REFERENCES team (team_api_id)
                FOREIGN KEY (player1) REFERENCES player (player_api_id)
            )
/* 
3 example rows:
SELECT * FROM corners LIMIT 3;
    id match_id team player1 subtype elapsed elapsed_plus   n sortorder 
375535     1730 9825   39297   short       3              228         0 
375538     1730 9825   39297   cross       3              229         3 
375562     1730 9825   30960   short       9              238         2 
*/

Here is a basic description of some of the less-clear column names used in the database:
- home_player_ids: For a given n (1-11), the ID of the player playing for the home-team is in home_player_n, and his coordinates are X=home_player_Xn and Y=home_player_Yn. Same for away team. Note that X coordinates are on the goalline ranging from 1-9, and the Y coordinates are along the length of the field ranging 1-11.
  In this encoding, all goalkeepers have coordinate (1,1). If the second field player of the hometeam has coordinates (2,3), (i.e.: home_playerX2=2 and home_player_Y2=3), we can infer that he plays on the left side of the field (X=2) as a defender (Y=3). He must thus be a left defender in this match. We can assume that Y coordinates between 2 and 5 are defenders, Y coordinates between 6 and 8 are midfielders and Y coordinates 9-11 are forwards.
- All columns ending with H are the odds for the Home team in the match. Likewise, columns ending with D are odds for Draw, & A are for Away team. The other characters in the column names represent the name of the betting website. (B365=Bet365, LB=LadBrokes, WH=William Hill, SJ=Stan James, BW=BWin. Other websites are not known)
- All dates are in the format yyyy-mm-dd hh:mm:ss. Generally hh:mm:ss are 00:00:00 for most values in these columns
- The eleven leagues covered in the DB are: Belgium Jupiler League, England Premier League, France Ligue 1, Germany 1. Bundesliga, Italy Serie A, Netherlands Eredivisie, Poland Ekstraklasa, Portugal Liga ZON Sagres, Scotland Premier League, Spain LIGA BBVA, Switzerland Super League
- Many textual ratings in player_attributes, attaching_work_rate and defensive_work_rate are from {high, medium, low}. The preferred_foot are from {left, right}
- The heights are in cms and weights are in lbs.
- The class ratings in team_attributes take values as follows:
    buildUpPlaySpeedClass: {Balanced, Fast, Slow}
    buildUpPlayDribblingClass: {Little, Normal, Lots}
    buildUpPlayPassingClass: {Mixed, Long, Short}
    buildUpPlayPositioningClass / chanceCreationPositioningClass: {Organised, Free Form}
    chanceCreationPassingClass: {Normal, Risky, Safe}
    chanceCreationCrossingClass / chanceCreationShootingClass: {Normal, Lots, Little}
    defencePressureClass: {Deep, Medium, High}
    defenceAggressionClass: {Press, Double, Contain}
    defenceTeamWidthClass: {Normal, Wide, Narrow}
    defenceDefenderLineClass: {Cover, Offside Trap}
- For the events (goals, fouls, cards, corners) tables:
    Elapsed represents the minute number when goal happened. Elapsed_plus indicates that this happened in extra time. E.g., if it happens in 46th minute in first half, elapsed_plus is 1
    Player1 is the primary actor (player who scores or who fouls), and player2 is the secondary actor (e.g., who assists, or who is fouled)
    Team id is player1's team ID
    injury_time is time of break for the injury
    card_type is {y, r, y2}
    Goal_type takes values as: {n - normal goal, p - penalty, dg - goal disallowed, o - own goal, npm - penalty saved by the goalkeeper, psm - missed penalty, rp - retake penalty}
    card subtype takes values as: {advantage, diving, emergency_brake, hands, kicked_ball_away, pushing, Removing Shirt, serious_fouls, shirt_pull, stall_time, Unsportsmanlike Cond, verbal_abuse, violence}
    corner subtype takes values as: {cross, cross_left, cross_right, short, short_left, short_right}
    goal subtype takes values as: {backheel, bicycle_kick, crossbar, deflected, direct_freekick, distance, header, indirect freekick, lob, loose_ball, missed, post, saved, saved_back_into_play, shot, tap_in, volley}
    foul subtype takes values as: {advantage, dangerous_play, diving, from_behind, goalkeeper_hands, hands, obstruction, penalty, pull, pushing, serious_foul, shirt_pull, trip}


-- Using valid SQLite, answer the following questions for the tables provided above. Write the SQL query in a separate line enclosed in ```sql ```
-- Make sure your code ONLY outputs the EXACT columns desired in the user instruction. It should NOT output any additional information, even if relevant. It should NOT round up any numerical information, unless especially required for the question.
-- Team, match and league should be outputted as name and not ID. Season is outputted in "YYYY/YYYY" format (Eg: "2011/2012"). Date should be outputted in "DD Month YYYY" format (Eg: "1 January 2025" and not "1 1 2025"). Match information should be outputted as "home_team_name - away_team_name". Score should be outputted as "home_team_score-away_team_score".
-- Seasons between Year1 and Year2 do not include season ending in Year1 or starting in Year2. Eg: only two seasons exist between 2013 and 2015: 2013/2014 and 2014/2015. On the other matches between 2013 and 2015 include matches in second half of 2012/2013 and first half of 2015/2016 seasons.
-- Finally, if multiple entities match question's criteria, output all of them, even if question asks for one.
-- User question: The longest number of months either team name was undefeated in the Old Firm between 2008-2016? Team name and number of months.
The final SQL is: Let's think step by step.
\end{lstlisting}

\section{Example Text2SQLCode$_{single}$ Prompt for \texttt{Eurosoccer}}
\begin{lstlisting}
You are a highly capable AI assistant specializing in database engineering and code generation. 
Your task is to process a user's natural language query in the following steps:
1) Decompose the query into SQL and Python steps
2) Write python code that uses SQL for data fetching operations and python for computing the final answer from the fetched data.


--- Decomposition Stage ---

We have access to a Text2SQL model, which takes as input a textual instruction and converts it into an SQL query. But, it may or may not be possible to just write a single SQL query for the complex user instruction. This often happens when there is a nested for loop or several complicated conditionals or arithmetic operations or string manipulation needed to express user intent programmatically. This may also happen if we are unsure of SQL syntax for some function, but can execute the same in Python code.
Your job is to generate a step-by-step breakdown for satisfying this instruction. Clearly annotate which step will be done by a Python interpreter and which step by Text2SQL query engine. 
If the instruction is simple enough to be solved by Text2SQL, the decomposition should be only one step. If you require multiple Text2SQL queries, make them independent of each other.

Examples: 

User question: Return the most common payment method used for transactions.
Let's think step by step. We have a table of transactions that has payment_method as a column. We can simply count the frequency of each payment method and take top 1. This question can be handled by SQL easily, no need for Python.
Decomposition:
Text2SQL: Return the most common payment method used for transactions. Output the payment method.

User question: return the date of the monthly high price of AAPL stock between Jan 1st, 2024 and June 30, 2024
Lets's think step by step. This question is complex since we need to first group the time-series data by month, and then compute the max separately for each group. We can fetch raw data for AAPL between Jan 1st and June 30 using SQL and compute aggregates using Python.
Decomposition:
Text2SQL: get daily price of AAPL stock between Jan 1st, 2024 and June 30, 2024
Python: for each month (January-June), find the highest price within that month. and its corresponding date.
Python: for each month, output the corresponding date

--- Code Stage --

Now write python code that contains execution of SQL queries for the Text2SQL steps and python code for computing the final result.

The schema for the database is as follows:
CREATE TABLE "Player_Attributes" (
	`id`	INTEGER PRIMARY KEY AUTOINCREMENT,
	`player_fifa_api_id`	INTEGER,
	`player_api_id`	INTEGER,
	`date`	TEXT,
	`overall_rating`	INTEGER,
	`potential`	INTEGER,
	`preferred_foot`	TEXT,
	`attacking_work_rate`	TEXT,
	`defensive_work_rate`	TEXT,
	`crossing`	INTEGER,
	`finishing`	INTEGER,
	`heading_accuracy`	INTEGER,
	`short_passing`	INTEGER,
	`volleys`	INTEGER,
	`dribbling`	INTEGER,
	`curve`	INTEGER,
	`free_kick_accuracy`	INTEGER,
	`long_passing`	INTEGER,
	`ball_control`	INTEGER,
	`acceleration`	INTEGER,
	`sprint_speed`	INTEGER,
	`agility`	INTEGER,
	`reactions`	INTEGER,
	`balance`	INTEGER,
	`shot_power`	INTEGER,
	`jumping`	INTEGER,
	`stamina`	INTEGER,
	`strength`	INTEGER,
	`long_shots`	INTEGER,
	`aggression`	INTEGER,
	`interceptions`	INTEGER,
	`positioning`	INTEGER,
	`vision`	INTEGER,
	`penalties`	INTEGER,
	`marking`	INTEGER,
	`standing_tackle`	INTEGER,
	`sliding_tackle`	INTEGER,
	`gk_diving`	INTEGER,
	`gk_handling`	INTEGER,
	`gk_kicking`	INTEGER,
	`gk_positioning`	INTEGER,
	`gk_reflexes`	INTEGER,
	FOREIGN KEY(`player_fifa_api_id`) REFERENCES `Player`(`player_fifa_api_id`),
	FOREIGN KEY(`player_api_id`) REFERENCES `Player`(`player_api_id`)
)
/* 
3 example rows:
SELECT * FROM Player_Attributes LIMIT 3;
id player_fifa_api_id player_api_id                date overall_rating potential preferred_foot attacking_work_rate defensive_work_rate crossing finishing heading_accuracy short_passing volleys dribbling curve free_kick_accuracy long_passing ball_control acceleration sprint_speed agility reactions balance shot_power jumping stamina strength long_shots aggression interceptions positioning vision penalties marking standing_tackle sliding_tackle gk_diving gk_handling gk_kicking gk_positioning gk_reflexes 
 1             218353        505942 2016-02-18 00:00:00             67        71          right              medium              medium       49        44               71            61      44        51    45                 39           64           49           60           64      59        47      65         55      58      54       76         35         71            70          45     54        48      65              69             69         6          11         10              8           8 
 2             218353        505942 2015-11-19 00:00:00             67        71          right              medium              medium       49        44               71            61      44        51    45                 39           64           49           60           64      59        47      65         55      58      54       76         35         71            70          45     54        48      65              69             69         6          11         10              8           8 
 3             218353        505942 2015-09-21 00:00:00             62        66          right              medium              medium       49        44               71            61      44        51    45                 39           64           49           60           64      59        47      65         55      58      54       76         35         63            41          45     54        48      65              66             69         6          11         10              8           8 
*/

CREATE TABLE `Player` (
	`id`	INTEGER PRIMARY KEY AUTOINCREMENT,
	`player_api_id`	INTEGER UNIQUE,
	`player_name`	TEXT,
	`player_fifa_api_id`	INTEGER UNIQUE,
	`birthday`	TEXT,
	`height`	INTEGER,
	`weight`	INTEGER
)
/* 
3 example rows:
SELECT * FROM Player LIMIT 3;
id player_api_id        player_name player_fifa_api_id            birthday height weight 
 1        505942 Aaron Appindangoye             218353 1992-02-29 00:00:00 182.88    187 
 2        155782    Aaron Cresswell             189615 1989-12-15 00:00:00 170.18    146 
 3        162549        Aaron Doran             186170 1991-05-13 00:00:00 170.18    163 
*/

CREATE TABLE `Match` (
	`id`	INTEGER PRIMARY KEY AUTOINCREMENT,
	`country_id`	INTEGER,
	`league_id`	INTEGER,
	`season`	TEXT,
	`stage`	INTEGER,
	`date`	TEXT,
	`match_api_id`	INTEGER UNIQUE,
	`home_team_api_id`	INTEGER,
	`away_team_api_id`	INTEGER,
	`home_team_goal`	INTEGER,
	`away_team_goal`	INTEGER,
	`home_player_X1`	INTEGER,
	`home_player_X2`	INTEGER,
	`home_player_X3`	INTEGER,
	`home_player_X4`	INTEGER,
	`home_player_X5`	INTEGER,
	`home_player_X6`	INTEGER,
	`home_player_X7`	INTEGER,
	`home_player_X8`	INTEGER,
	`home_player_X9`	INTEGER,
	`home_player_X10`	INTEGER,
	`home_player_X11`	INTEGER,
	`away_player_X1`	INTEGER,
	`away_player_X2`	INTEGER,
	`away_player_X3`	INTEGER,
	`away_player_X4`	INTEGER,
	`away_player_X5`	INTEGER,
	`away_player_X6`	INTEGER,
	`away_player_X7`	INTEGER,
	`away_player_X8`	INTEGER,
	`away_player_X9`	INTEGER,
	`away_player_X10`	INTEGER,
	`away_player_X11`	INTEGER,
	`home_player_Y1`	INTEGER,
	`home_player_Y2`	INTEGER,
	`home_player_Y3`	INTEGER,
	`home_player_Y4`	INTEGER,
	`home_player_Y5`	INTEGER,
	`home_player_Y6`	INTEGER,
	`home_player_Y7`	INTEGER,
	`home_player_Y8`	INTEGER,
	`home_player_Y9`	INTEGER,
	`home_player_Y10`	INTEGER,
	`home_player_Y11`	INTEGER,
	`away_player_Y1`	INTEGER,
	`away_player_Y2`	INTEGER,
	`away_player_Y3`	INTEGER,
	`away_player_Y4`	INTEGER,
	`away_player_Y5`	INTEGER,
	`away_player_Y6`	INTEGER,
	`away_player_Y7`	INTEGER,
	`away_player_Y8`	INTEGER,
	`away_player_Y9`	INTEGER,
	`away_player_Y10`	INTEGER,
	`away_player_Y11`	INTEGER,
	`home_player_1`	INTEGER,
	`home_player_2`	INTEGER,
	`home_player_3`	INTEGER,
	`home_player_4`	INTEGER,
	`home_player_5`	INTEGER,
	`home_player_6`	INTEGER,
	`home_player_7`	INTEGER,
	`home_player_8`	INTEGER,
	`home_player_9`	INTEGER,
	`home_player_10`	INTEGER,
	`home_player_11`	INTEGER,
	`away_player_1`	INTEGER,
	`away_player_2`	INTEGER,
	`away_player_3`	INTEGER,
	`away_player_4`	INTEGER,
	`away_player_5`	INTEGER,
	`away_player_6`	INTEGER,
	`away_player_7`	INTEGER,
	`away_player_8`	INTEGER,
	`away_player_9`	INTEGER,
	`away_player_10`	INTEGER,
	`away_player_11`	INTEGER,
	`B365H`	NUMERIC,
	`B365D`	NUMERIC,
	`B365A`	NUMERIC,
	`BWH`	NUMERIC,
	`BWD`	NUMERIC,
	`BWA`	NUMERIC,
	`IWH`	NUMERIC,
	`IWD`	NUMERIC,
	`IWA`	NUMERIC,
	`LBH`	NUMERIC,
	`LBD`	NUMERIC,
	`LBA`	NUMERIC,
	`PSH`	NUMERIC,
	`PSD`	NUMERIC,
	`PSA`	NUMERIC,
	`WHH`	NUMERIC,
	`WHD`	NUMERIC,
	`WHA`	NUMERIC,
	`SJH`	NUMERIC,
	`SJD`	NUMERIC,
	`SJA`	NUMERIC,
	`VCH`	NUMERIC,
	`VCD`	NUMERIC,
	`VCA`	NUMERIC,
	`GBH`	NUMERIC,
	`GBD`	NUMERIC,
	`GBA`	NUMERIC,
	`BSH`	NUMERIC,
	`BSD`	NUMERIC,
	`BSA`	NUMERIC,
	FOREIGN KEY(`country_id`) REFERENCES `country`(`id`),
	FOREIGN KEY(`league_id`) REFERENCES `League`(`id`),
	FOREIGN KEY(`home_team_api_id`) REFERENCES `Team`(`team_api_id`),
	FOREIGN KEY(`away_team_api_id`) REFERENCES `Team`(`team_api_id`),
	FOREIGN KEY(`home_player_1`) REFERENCES `Player`(`player_api_id`),
	FOREIGN KEY(`home_player_2`) REFERENCES `Player`(`player_api_id`),
	FOREIGN KEY(`home_player_3`) REFERENCES `Player`(`player_api_id`),
	FOREIGN KEY(`home_player_4`) REFERENCES `Player`(`player_api_id`),
	FOREIGN KEY(`home_player_5`) REFERENCES `Player`(`player_api_id`),
	FOREIGN KEY(`home_player_6`) REFERENCES `Player`(`player_api_id`),
	FOREIGN KEY(`home_player_7`) REFERENCES `Player`(`player_api_id`),
	FOREIGN KEY(`home_player_8`) REFERENCES `Player`(`player_api_id`),
	FOREIGN KEY(`home_player_9`) REFERENCES `Player`(`player_api_id`),
	FOREIGN KEY(`home_player_10`) REFERENCES `Player`(`player_api_id`),
	FOREIGN KEY(`home_player_11`) REFERENCES `Player`(`player_api_id`),
	FOREIGN KEY(`away_player_1`) REFERENCES `Player`(`player_api_id`),
	FOREIGN KEY(`away_player_2`) REFERENCES `Player`(`player_api_id`),
	FOREIGN KEY(`away_player_3`) REFERENCES `Player`(`player_api_id`),
	FOREIGN KEY(`away_player_4`) REFERENCES `Player`(`player_api_id`),
	FOREIGN KEY(`away_player_5`) REFERENCES `Player`(`player_api_id`),
	FOREIGN KEY(`away_player_6`) REFERENCES `Player`(`player_api_id`),
	FOREIGN KEY(`away_player_7`) REFERENCES `Player`(`player_api_id`),
	FOREIGN KEY(`away_player_8`) REFERENCES `Player`(`player_api_id`),
	FOREIGN KEY(`away_player_9`) REFERENCES `Player`(`player_api_id`),
	FOREIGN KEY(`away_player_10`) REFERENCES `Player`(`player_api_id`),
	FOREIGN KEY(`away_player_11`) REFERENCES `Player`(`player_api_id`)
)
/* 
3 example rows:
SELECT * FROM Match LIMIT 3;
id country_id league_id    season stage                date match_api_id home_team_api_id away_team_api_id home_team_goal away_team_goal home_player_X1 home_player_X2 home_player_X3 home_player_X4 home_player_X5 home_player_X6 home_player_X7 home_player_X8 home_player_X9 home_player_X10 home_player_X11 away_player_X1 away_player_X2 away_player_X3 away_player_X4 away_player_X5 away_player_X6 away_player_X7 away_player_X8 away_player_X9 away_player_X10 away_player_X11 home_player_Y1 home_player_Y2 home_player_Y3 home_player_Y4 home_player_Y5 home_player_Y6 home_player_Y7 home_player_Y8 home_player_Y9 home_player_Y10 home_player_Y11 away_player_Y1 away_player_Y2 away_player_Y3 away_player_Y4 away_player_Y5 away_player_Y6 away_player_Y7 away_player_Y8 away_player_Y9 away_player_Y10 away_player_Y11 home_player_1 home_player_2 home_player_3 home_player_4 home_player_5 home_player_6 home_player_7 home_player_8 home_player_9 home_player_10 home_player_11 away_player_1 away_player_2 away_player_3 away_player_4 away_player_5 away_player_6 away_player_7 away_player_8 away_player_9 away_player_10 away_player_11 B365H B365D B365A  BWH  BWD  BWA  IWH IWD IWA LBH LBD  LBA  PSH  PSD  PSA  WHH  WHD  WHA  SJH SJD SJA  VCH  VCD  VCA  GBH  GBD  GBA  BSH  BSD  BSA 
 1          1         1 2008/2009     1 2008-08-17 00:00:00       492473             9987             9993              1              1           None           None           None           None           None           None           None           None           None            None            None           None           None           None           None           None           None           None           None           None            None            None           None           None           None           None           None           None           None           None           None            None            None           None           None           None           None           None           None           None           None           None            None            None          None          None          None          None          None          None          None          None          None           None           None          None          None          None          None          None          None          None          None          None           None           None  1.73   3.4     5 1.75 3.35  4.2 1.85 3.2 3.5 1.8 3.3 3.75 None None None  1.7  3.3 4.33  1.9 3.3   4 1.65  3.4  4.5 1.78 3.25    4 1.73  3.4  4.2 
 2          1         1 2008/2009     1 2008-08-16 00:00:00       492474            10000             9994              0              0           None           None           None           None           None           None           None           None           None            None            None           None           None           None           None           None           None           None           None           None            None            None           None           None           None           None           None           None           None           None           None            None            None           None           None           None           None           None           None           None           None           None            None            None          None          None          None          None          None          None          None          None          None           None           None          None          None          None          None          None          None          None          None          None           None           None  1.95   3.2   3.6  1.8  3.3 3.95  1.9 3.2 3.5 1.9 3.2  3.5 None None None 1.83  3.3  3.6 1.95 3.3 3.8    2 3.25 3.25 1.85 3.25 3.75 1.91 3.25  3.6 
 3          1         1 2008/2009     1 2008-08-16 00:00:00       492475             9984             8635              0              3           None           None           None           None           None           None           None           None           None            None            None           None           None           None           None           None           None           None           None           None            None            None           None           None           None           None           None           None           None           None           None            None            None           None           None           None           None           None           None           None           None           None            None            None          None          None          None          None          None          None          None          None          None           None           None          None          None          None          None          None          None          None          None          None           None           None  2.38   3.3  2.75  2.4  3.3 2.55  2.6 3.1 2.3 2.5 3.2  2.5 None None None  2.5 3.25  2.4 2.63 3.3 2.5 2.35 3.25 2.65  2.5  3.2  2.5  2.3  3.2 2.75 
*/

CREATE TABLE `League` (
	`id`	INTEGER PRIMARY KEY AUTOINCREMENT,
	`country_id`	INTEGER,
	`name`	TEXT UNIQUE,
	FOREIGN KEY(`country_id`) REFERENCES `country`(`id`)
)
/* 
3 example rows:
SELECT * FROM League LIMIT 3;
  id country_id                   name 
   1          1 Belgium Jupiler League 
1729       1729 England Premier League 
4769       4769         France Ligue 1 
*/

CREATE TABLE `Country` (
	`id`	INTEGER PRIMARY KEY AUTOINCREMENT,
	`name`	TEXT UNIQUE
)
/* 
3 example rows:
SELECT * FROM Country LIMIT 3;
  id    name 
   1 Belgium 
1729 England 
4769  France 
*/

CREATE TABLE "Team" (
	`id`	INTEGER PRIMARY KEY AUTOINCREMENT,
	`team_api_id`	INTEGER UNIQUE,
	`team_fifa_api_id`	INTEGER,
	`team_long_name`	TEXT,
	`team_short_name`	TEXT
)
/* 
3 example rows:
SELECT * FROM Team LIMIT 3;
id team_api_id team_fifa_api_id   team_long_name team_short_name 
 1        9987              673         KRC Genk             GEN 
 2        9993              675     Beerschot AC             BAC 
 3       10000            15005 SV Zulte-Waregem             ZUL 
*/

CREATE TABLE `Team_Attributes` (
	`id`	INTEGER PRIMARY KEY AUTOINCREMENT,
	`team_fifa_api_id`	INTEGER,
	`team_api_id`	INTEGER,
	`date`	TEXT,
	`buildUpPlaySpeed`	INTEGER,
	`buildUpPlaySpeedClass`	TEXT,
	`buildUpPlayDribbling`	INTEGER,
	`buildUpPlayDribblingClass`	TEXT,
	`buildUpPlayPassing`	INTEGER,
	`buildUpPlayPassingClass`	TEXT,
	`buildUpPlayPositioningClass`	TEXT,
	`chanceCreationPassing`	INTEGER,
	`chanceCreationPassingClass`	TEXT,
	`chanceCreationCrossing`	INTEGER,
	`chanceCreationCrossingClass`	TEXT,
	`chanceCreationShooting`	INTEGER,
	`chanceCreationShootingClass`	TEXT,
	`chanceCreationPositioningClass`	TEXT,
	`defencePressure`	INTEGER,
	`defencePressureClass`	TEXT,
	`defenceAggression`	INTEGER,
	`defenceAggressionClass`	TEXT,
	`defenceTeamWidth`	INTEGER,
	`defenceTeamWidthClass`	TEXT,
	`defenceDefenderLineClass`	TEXT,
	FOREIGN KEY(`team_fifa_api_id`) REFERENCES `Team`(`team_fifa_api_id`),
	FOREIGN KEY(`team_api_id`) REFERENCES `Team`(`team_api_id`)
)
/* 
3 example rows:
SELECT * FROM Team_Attributes LIMIT 3;
id team_fifa_api_id team_api_id                date buildUpPlaySpeed buildUpPlaySpeedClass buildUpPlayDribbling buildUpPlayDribblingClass buildUpPlayPassing buildUpPlayPassingClass buildUpPlayPositioningClass chanceCreationPassing chanceCreationPassingClass chanceCreationCrossing chanceCreationCrossingClass chanceCreationShooting chanceCreationShootingClass chanceCreationPositioningClass defencePressure defencePressureClass defenceAggression defenceAggressionClass defenceTeamWidth defenceTeamWidthClass defenceDefenderLineClass 
 1              434        9930 2010-02-22 00:00:00               60              Balanced                 None                    Little                 50                   Mixed                   Organised                    60                     Normal                     65                      Normal                     55                      Normal                      Organised              50               Medium                55                  Press               45                Normal                    Cover 
 2              434        9930 2014-09-19 00:00:00               52              Balanced                   48                    Normal                 56                   Mixed                   Organised                    54                     Normal                     63                      Normal                     64                      Normal                      Organised              47               Medium                44                  Press               54                Normal                    Cover 
 3              434        9930 2015-09-10 00:00:00               47              Balanced                   41                    Normal                 54                   Mixed                   Organised                    54                     Normal                     63                      Normal                     64                      Normal                      Organised              47               Medium                44                  Press               54                Normal                    Cover 
*/

CREATE TABLE goals (
                id INTEGER PRIMARY KEY,
                match_id INTEGER NOT NULL,
                team INTEGER NOT NULL,
                player1 INTEGER NOT NULL,
                player2 INTEGER NOT NULL,
                goal_type TEXT,
                subtype TEXT,
                elapsed INTEGER NOT NULL,
                elapsed_plus INTEGER,
                n INTEGER NOT NULL,
                sortorder INTEGER NOT NULL,
                FOREIGN KEY (match_id) REFERENCES match (id)
                FOREIGN KEY (team) REFERENCES team (team_api_id)
                FOREIGN KEY (player1) REFERENCES player (player_api_id)
                FOREIGN KEY (player2) REFERENCES player (player_api_id)
            )
/* 
3 example rows:
SELECT * FROM goals LIMIT 3;
    id match_id team player1 player2 goal_type subtype elapsed elapsed_plus  n sortorder 
375301     7809 9823   30872                 n              12              21         0 
375302     7809 9823   32118                 p              16              22         0 
375303     7809 9790   33101                 n              25              20         0 
*/

CREATE TABLE cards (
                id INTEGER PRIMARY KEY,
                match_id INTEGER NOT NULL,
                team INTEGER NOT NULL,
                player1 INTEGER NOT NULL,
                card_type TEXT,
                subtype TEXT,
                elapsed INTEGER NOT NULL,
                elapsed_plus INTEGER,
                n INTEGER NOT NULL,
                sortorder INTEGER NOT NULL,
                FOREIGN KEY (match_id) REFERENCES match (id)
                FOREIGN KEY (team) REFERENCES team (team_api_id)
                FOREIGN KEY (player1) REFERENCES player (player_api_id)
            )
/* 
3 example rows:
SELECT * FROM cards LIMIT 3;
    id match_id team player1 card_type subtype elapsed elapsed_plus  n sortorder 
375310     7809 9790   30749         y              19              26         0 
375311     7809 9823   38244         y              64              28         0 
375312     7809 9790   33101         y              44              27         0 
*/

CREATE TABLE fouls (
                id INTEGER PRIMARY KEY,
                match_id INTEGER NOT NULL,
                team INTEGER NOT NULL,
                player1 INTEGER NOT NULL,
                player2 INTEGER NOT NULL,
                subtype TEXT,
                elapsed INTEGER NOT NULL,
                elapsed_plus INTEGER,
                injury_time INTEGER,
                n INTEGER NOT NULL,
                sortorder INTEGER NOT NULL,
                FOREIGN KEY (match_id) REFERENCES match (id)
                FOREIGN KEY (team) REFERENCES team (team_api_id)
                FOREIGN KEY (player1) REFERENCES player (player_api_id)
                FOREIGN KEY (player2) REFERENCES player (player_api_id)
            )
/* 
3 example rows:
SELECT * FROM fouls LIMIT 3;
    id match_id team player1 player2 subtype elapsed elapsed_plus injury_time   n sortorder 
375531     1730 8659   36832   26181               2                          227         0 
375541     1730 9825   31291   23257               3                          230         5 
375551     1730 9825   30935   31088 pushing       5                          232         0 
*/

CREATE TABLE corners (
                id INTEGER PRIMARY KEY,
                match_id INTEGER NOT NULL,
                team INTEGER NOT NULL,
                player1 INTEGER NOT NULL,
                subtype TEXT,
                elapsed INTEGER NOT NULL,
                elapsed_plus INTEGER,
                n INTEGER NOT NULL,
                sortorder INTEGER NOT NULL,
                FOREIGN KEY (match_id) REFERENCES match (id)
                FOREIGN KEY (team) REFERENCES team (team_api_id)
                FOREIGN KEY (player1) REFERENCES player (player_api_id)
            )
/* 
3 example rows:
SELECT * FROM corners LIMIT 3;
    id match_id team player1 subtype elapsed elapsed_plus   n sortorder 
375535     1730 9825   39297   short       3              228         0 
375538     1730 9825   39297   cross       3              229         3 
375562     1730 9825   30960   short       9              238         2 
*/

Here is a basic description of some of the less-clear column names used in the database:
- home_player_ids: For a given n (1-11), the ID of the player playing for the home-team is in home_player_n, and his coordinates are X=home_player_Xn and Y=home_player_Yn. Same for away team. Note that X coordinates are on the goalline ranging from 1-9, and the Y coordinates are along the length of the field ranging 1-11.
  In this encoding, all goalkeepers have coordinate (1,1). If the second field player of the hometeam has coordinates (2,3), (i.e.: home_playerX2=2 and home_player_Y2=3), we can infer that he plays on the left side of the field (X=2) as a defender (Y=3). He must thus be a left defender in this match. We can assume that Y coordinates between 2 and 5 are defenders, Y coordinates between 6 and 8 are midfielders and Y coordinates 9-11 are forwards.
- All columns ending with H are the odds for the Home team in the match. Likewise, columns ending with D are odds for Draw, & A are for Away team. The other characters in the column names represent the name of the betting website. (B365=Bet365, LB=LadBrokes, WH=William Hill, SJ=Stan James, BW=BWin. Other websites are not known)
- All dates are in the format yyyy-mm-dd hh:mm:ss. Generally hh:mm:ss are 00:00:00 for most values in these columns
- The eleven leagues covered in the DB are: Belgium Jupiler League, England Premier League, France Ligue 1, Germany 1. Bundesliga, Italy Serie A, Netherlands Eredivisie, Poland Ekstraklasa, Portugal Liga ZON Sagres, Scotland Premier League, Spain LIGA BBVA, Switzerland Super League
- Many textual ratings in player_attributes, attaching_work_rate and defensive_work_rate are from {high, medium, low}. The preferred_foot are from {left, right}
- The heights are in cms and weights are in lbs.
- The class ratings in team_attributes take values as follows:
    buildUpPlaySpeedClass: {Balanced, Fast, Slow}
    buildUpPlayDribblingClass: {Little, Normal, Lots}
    buildUpPlayPassingClass: {Mixed, Long, Short}
    buildUpPlayPositioningClass / chanceCreationPositioningClass: {Organised, Free Form}
    chanceCreationPassingClass: {Normal, Risky, Safe}
    chanceCreationCrossingClass / chanceCreationShootingClass: {Normal, Lots, Little}
    defencePressureClass: {Deep, Medium, High}
    defenceAggressionClass: {Press, Double, Contain}
    defenceTeamWidthClass: {Normal, Wide, Narrow}
    defenceDefenderLineClass: {Cover, Offside Trap}
- For the events (goals, fouls, cards, corners) tables:
    Elapsed represents the minute number when goal happened. Elapsed_plus indicates that this happened in extra time. E.g., if it happens in 46th minute in first half, elapsed_plus is 1
    Player1 is the primary actor (player who scores or who fouls), and player2 is the secondary actor (e.g., who assists, or who is fouled)
    Team id is player1's team ID
    injury_time is time of break for the injury
    card_type is {y, r, y2}
    Goal_type takes values as: {n - normal goal, p - penalty, dg - goal disallowed, o - own goal, npm - penalty saved by the goalkeeper, psm - missed penalty, rp - retake penalty}
    card subtype takes values as: {advantage, diving, emergency_brake, hands, kicked_ball_away, pushing, Removing Shirt, serious_fouls, shirt_pull, stall_time, Unsportsmanlike Cond, verbal_abuse, violence}
    corner subtype takes values as: {cross, cross_left, cross_right, short, short_left, short_right}
    goal subtype takes values as: {backheel, bicycle_kick, crossbar, deflected, direct_freekick, distance, header, indirect freekick, lob, loose_ball, missed, post, saved, saved_back_into_play, shot, tap_in, volley}
    foul subtype takes values as: {advantage, dangerous_play, diving, from_behind, goalkeeper_hands, hands, obstruction, penalty, pull, pushing, serious_foul, shirt_pull, trip}

The query is as follows:
The longest number of months either team name was undefeated in the Old Firm between 2008-2016? Team name and number of months

Write a python code that takes the database_path as an input and prints a list of tuples with the desired information. Its type signature is:
def compute_result(database_path: str]) -> List[Tuple]:

The Python code must use sqlite3 to execute the SQL queries, and the outputs can be stored as dataframes.
Please use standard Python libraries for transformations on the dataframe. 
Don't write anything apart from the Python function; use Python comments if needed.
Make sure your code is robust, for instance, some values in the dataframe may be missing -- the code should handle that without crashing.
Assume indentation level as 0 in your code. Your code will be called as print(compute_result(database_path)) to show the output to the user.
If the user question asks for one instance matching a criteria, your code should output all such instances.
Important: make sure your code ONLY outputs the EXACT columns desired in the user instruction. It should NOT output any additional information, even if relevant. It should NOT round up any numerical information, unless especially required for the question.
Important: the output should be returned as a list of tuples without any column names or descriptions. 
Write the final python code enclosed in ```python ```. Provide the full and final code in a single code block. Do not provide any intermediate steps, refinements, or additional explanations.

\end{lstlisting}

\section{Example Text2SQLCode$_{multi}$ Prompt for \texttt{Eurosoccer}}
\subsection{Decomposition}
\begin{lstlisting}
You are a helpful assistant to a database engineer. The user has provided a complex instruction. As a first step, we wish to break it into a series of steps.
You current job is to take this user instruction and create a step-by-step execution plan for achieving it. The raw data is fetched from an SQL database.

We have access to a Text2SQL model, which takes as input a textual instruction and converts it into an SQL query. But, it may or may not be possible to just write a single SQL query for the complex user instruction. This often happens when there is a nested for loop or several complicated conditionals or arithmetic operations or string manipulation needed to express user intent programmatically. This may also happen if we are unsure of SQL syntax for some function, but can execute the same in Python code.
Your job is to generate a step-by-step breakdown for satisfying this instruction. Clearly annotate which step will be done by a Python interpreter and which step by Text2SQL query engine.
If the instruction is simple enough to be solved by Text2SQL, the decomposition should be only one step. If you require multiple Text2SQL queries, make them independent of each other.

Positive example:
User question: Return the most common payment method used for transactions.
Let's think step by step. We have a table of transactions that has payment_method as a column. We can simply count the frequency of each payment method and take top 1. This question can be handled by SQL easily, no need for Python.
Decomposition:
Text2SQL: Return the most common payment method used for transactions. Output the payment method.

Positive example:
User question: return the date of the monthly high price of AAPL stock between Jan 1st, 2024 and June 30, 2024
Lets's think step by step. This question is complex since we need to first group the time-series data by month, and then compute the max separately for each group. We can fetch raw data for AAPL between Jan 1st and June 30 using SQL and compute aggregates using Python.
Decomposition:
Text2SQL: get daily price of AAPL stock between Jan 1st, 2024 and June 30, 2024
Python: for each month (January-June), find the highest price within that month. and its corresponding date.
Python: for each month, output the corresponding date

Negative example:
User question: How many products are sold by Zara at a price higher than 5?
Bad Decomposition:
Text2SQL: get list of all products that are sold by Zara
Text2SQL: get prices of all products identified in the previous step
Python: find the products have prices higher than 5
Python: count the number of such products
Python: output the calculated count
This is not a valid decomposition because the second Text2SQL query depends on the output of the previous Text2SQL query. However, all Text2SQL queries must be independently executable and not dependent on each other.

Negative example:
User question: return the date of the monthly high price of AAPL stock between Jan 1st, 2024 and June 30, 2024
Lets's think step by step. This question is complex since we need to first group the time-series data by month, and then compute the max separately for each group. We can fetch raw data for AAPL between Jan 1st and June 30 using SQL and compute aggregates using Python.
Decomposition:
Text2SQL: get daily price of AAPL stock between Jan 1st, 2024 and June 30, 2024
Python: for each month (January-June), find the highest price within that month. and its corresponding date.
Python: for each month, output the highest price and the corresponding date
This is not a valid decomposition because it outputs additional information (the highest price in a month), which is not requested in the original user question. Only the requested information should be outputted.


There is a similar problem, you need to help with. This is how the database was created.
CREATE TABLE "Player_Attributes" (
	`id`	INTEGER PRIMARY KEY AUTOINCREMENT,
	`player_fifa_api_id`	INTEGER,
	`player_api_id`	INTEGER,
	`date`	TEXT,
	`overall_rating`	INTEGER,
	`potential`	INTEGER,
	`preferred_foot`	TEXT,
	`attacking_work_rate`	TEXT,
	`defensive_work_rate`	TEXT,
	`crossing`	INTEGER,
	`finishing`	INTEGER,
	`heading_accuracy`	INTEGER,
	`short_passing`	INTEGER,
	`volleys`	INTEGER,
	`dribbling`	INTEGER,
	`curve`	INTEGER,
	`free_kick_accuracy`	INTEGER,
	`long_passing`	INTEGER,
	`ball_control`	INTEGER,
	`acceleration`	INTEGER,
	`sprint_speed`	INTEGER,
	`agility`	INTEGER,
	`reactions`	INTEGER,
	`balance`	INTEGER,
	`shot_power`	INTEGER,
	`jumping`	INTEGER,
	`stamina`	INTEGER,
	`strength`	INTEGER,
	`long_shots`	INTEGER,
	`aggression`	INTEGER,
	`interceptions`	INTEGER,
	`positioning`	INTEGER,
	`vision`	INTEGER,
	`penalties`	INTEGER,
	`marking`	INTEGER,
	`standing_tackle`	INTEGER,
	`sliding_tackle`	INTEGER,
	`gk_diving`	INTEGER,
	`gk_handling`	INTEGER,
	`gk_kicking`	INTEGER,
	`gk_positioning`	INTEGER,
	`gk_reflexes`	INTEGER,
	FOREIGN KEY(`player_fifa_api_id`) REFERENCES `Player`(`player_fifa_api_id`),
	FOREIGN KEY(`player_api_id`) REFERENCES `Player`(`player_api_id`)
)

CREATE TABLE `Player` (
	`id`	INTEGER PRIMARY KEY AUTOINCREMENT,
	`player_api_id`	INTEGER UNIQUE,
	`player_name`	TEXT,
	`player_fifa_api_id`	INTEGER UNIQUE,
	`birthday`	TEXT,
	`height`	INTEGER,
	`weight`	INTEGER
)

CREATE TABLE `Match` (
	`id`	INTEGER PRIMARY KEY AUTOINCREMENT,
	`country_id`	INTEGER,
	`league_id`	INTEGER,
	`season`	TEXT,
	`stage`	INTEGER,
	`date`	TEXT,
	`match_api_id`	INTEGER UNIQUE,
	`home_team_api_id`	INTEGER,
	`away_team_api_id`	INTEGER,
	`home_team_goal`	INTEGER,
	`away_team_goal`	INTEGER,
	`home_player_X1`	INTEGER,
	`home_player_X2`	INTEGER,
	`home_player_X3`	INTEGER,
	`home_player_X4`	INTEGER,
	`home_player_X5`	INTEGER,
	`home_player_X6`	INTEGER,
	`home_player_X7`	INTEGER,
	`home_player_X8`	INTEGER,
	`home_player_X9`	INTEGER,
	`home_player_X10`	INTEGER,
	`home_player_X11`	INTEGER,
	`away_player_X1`	INTEGER,
	`away_player_X2`	INTEGER,
	`away_player_X3`	INTEGER,
	`away_player_X4`	INTEGER,
	`away_player_X5`	INTEGER,
	`away_player_X6`	INTEGER,
	`away_player_X7`	INTEGER,
	`away_player_X8`	INTEGER,
	`away_player_X9`	INTEGER,
	`away_player_X10`	INTEGER,
	`away_player_X11`	INTEGER,
	`home_player_Y1`	INTEGER,
	`home_player_Y2`	INTEGER,
	`home_player_Y3`	INTEGER,
	`home_player_Y4`	INTEGER,
	`home_player_Y5`	INTEGER,
	`home_player_Y6`	INTEGER,
	`home_player_Y7`	INTEGER,
	`home_player_Y8`	INTEGER,
	`home_player_Y9`	INTEGER,
	`home_player_Y10`	INTEGER,
	`home_player_Y11`	INTEGER,
	`away_player_Y1`	INTEGER,
	`away_player_Y2`	INTEGER,
	`away_player_Y3`	INTEGER,
	`away_player_Y4`	INTEGER,
	`away_player_Y5`	INTEGER,
	`away_player_Y6`	INTEGER,
	`away_player_Y7`	INTEGER,
	`away_player_Y8`	INTEGER,
	`away_player_Y9`	INTEGER,
	`away_player_Y10`	INTEGER,
	`away_player_Y11`	INTEGER,
	`home_player_1`	INTEGER,
	`home_player_2`	INTEGER,
	`home_player_3`	INTEGER,
	`home_player_4`	INTEGER,
	`home_player_5`	INTEGER,
	`home_player_6`	INTEGER,
	`home_player_7`	INTEGER,
	`home_player_8`	INTEGER,
	`home_player_9`	INTEGER,
	`home_player_10`	INTEGER,
	`home_player_11`	INTEGER,
	`away_player_1`	INTEGER,
	`away_player_2`	INTEGER,
	`away_player_3`	INTEGER,
	`away_player_4`	INTEGER,
	`away_player_5`	INTEGER,
	`away_player_6`	INTEGER,
	`away_player_7`	INTEGER,
	`away_player_8`	INTEGER,
	`away_player_9`	INTEGER,
	`away_player_10`	INTEGER,
	`away_player_11`	INTEGER,
	`B365H`	NUMERIC,
	`B365D`	NUMERIC,
	`B365A`	NUMERIC,
	`BWH`	NUMERIC,
	`BWD`	NUMERIC,
	`BWA`	NUMERIC,
	`IWH`	NUMERIC,
	`IWD`	NUMERIC,
	`IWA`	NUMERIC,
	`LBH`	NUMERIC,
	`LBD`	NUMERIC,
	`LBA`	NUMERIC,
	`PSH`	NUMERIC,
	`PSD`	NUMERIC,
	`PSA`	NUMERIC,
	`WHH`	NUMERIC,
	`WHD`	NUMERIC,
	`WHA`	NUMERIC,
	`SJH`	NUMERIC,
	`SJD`	NUMERIC,
	`SJA`	NUMERIC,
	`VCH`	NUMERIC,
	`VCD`	NUMERIC,
	`VCA`	NUMERIC,
	`GBH`	NUMERIC,
	`GBD`	NUMERIC,
	`GBA`	NUMERIC,
	`BSH`	NUMERIC,
	`BSD`	NUMERIC,
	`BSA`	NUMERIC,
	FOREIGN KEY(`country_id`) REFERENCES `country`(`id`),
	FOREIGN KEY(`league_id`) REFERENCES `League`(`id`),
	FOREIGN KEY(`home_team_api_id`) REFERENCES `Team`(`team_api_id`),
	FOREIGN KEY(`away_team_api_id`) REFERENCES `Team`(`team_api_id`),
	FOREIGN KEY(`home_player_1`) REFERENCES `Player`(`player_api_id`),
	FOREIGN KEY(`home_player_2`) REFERENCES `Player`(`player_api_id`),
	FOREIGN KEY(`home_player_3`) REFERENCES `Player`(`player_api_id`),
	FOREIGN KEY(`home_player_4`) REFERENCES `Player`(`player_api_id`),
	FOREIGN KEY(`home_player_5`) REFERENCES `Player`(`player_api_id`),
	FOREIGN KEY(`home_player_6`) REFERENCES `Player`(`player_api_id`),
	FOREIGN KEY(`home_player_7`) REFERENCES `Player`(`player_api_id`),
	FOREIGN KEY(`home_player_8`) REFERENCES `Player`(`player_api_id`),
	FOREIGN KEY(`home_player_9`) REFERENCES `Player`(`player_api_id`),
	FOREIGN KEY(`home_player_10`) REFERENCES `Player`(`player_api_id`),
	FOREIGN KEY(`home_player_11`) REFERENCES `Player`(`player_api_id`),
	FOREIGN KEY(`away_player_1`) REFERENCES `Player`(`player_api_id`),
	FOREIGN KEY(`away_player_2`) REFERENCES `Player`(`player_api_id`),
	FOREIGN KEY(`away_player_3`) REFERENCES `Player`(`player_api_id`),
	FOREIGN KEY(`away_player_4`) REFERENCES `Player`(`player_api_id`),
	FOREIGN KEY(`away_player_5`) REFERENCES `Player`(`player_api_id`),
	FOREIGN KEY(`away_player_6`) REFERENCES `Player`(`player_api_id`),
	FOREIGN KEY(`away_player_7`) REFERENCES `Player`(`player_api_id`),
	FOREIGN KEY(`away_player_8`) REFERENCES `Player`(`player_api_id`),
	FOREIGN KEY(`away_player_9`) REFERENCES `Player`(`player_api_id`),
	FOREIGN KEY(`away_player_10`) REFERENCES `Player`(`player_api_id`),
	FOREIGN KEY(`away_player_11`) REFERENCES `Player`(`player_api_id`)
)

CREATE TABLE `League` (
	`id`	INTEGER PRIMARY KEY AUTOINCREMENT,
	`country_id`	INTEGER,
	`name`	TEXT UNIQUE,
	FOREIGN KEY(`country_id`) REFERENCES `country`(`id`)
)

CREATE TABLE `Country` (
	`id`	INTEGER PRIMARY KEY AUTOINCREMENT,
	`name`	TEXT UNIQUE
)

CREATE TABLE "Team" (
	`id`	INTEGER PRIMARY KEY AUTOINCREMENT,
	`team_api_id`	INTEGER UNIQUE,
	`team_fifa_api_id`	INTEGER,
	`team_long_name`	TEXT,
	`team_short_name`	TEXT
)

CREATE TABLE `Team_Attributes` (
	`id`	INTEGER PRIMARY KEY AUTOINCREMENT,
	`team_fifa_api_id`	INTEGER,
	`team_api_id`	INTEGER,
	`date`	TEXT,
	`buildUpPlaySpeed`	INTEGER,
	`buildUpPlaySpeedClass`	TEXT,
	`buildUpPlayDribbling`	INTEGER,
	`buildUpPlayDribblingClass`	TEXT,
	`buildUpPlayPassing`	INTEGER,
	`buildUpPlayPassingClass`	TEXT,
	`buildUpPlayPositioningClass`	TEXT,
	`chanceCreationPassing`	INTEGER,
	`chanceCreationPassingClass`	TEXT,
	`chanceCreationCrossing`	INTEGER,
	`chanceCreationCrossingClass`	TEXT,
	`chanceCreationShooting`	INTEGER,
	`chanceCreationShootingClass`	TEXT,
	`chanceCreationPositioningClass`	TEXT,
	`defencePressure`	INTEGER,
	`defencePressureClass`	TEXT,
	`defenceAggression`	INTEGER,
	`defenceAggressionClass`	TEXT,
	`defenceTeamWidth`	INTEGER,
	`defenceTeamWidthClass`	TEXT,
	`defenceDefenderLineClass`	TEXT,
	FOREIGN KEY(`team_fifa_api_id`) REFERENCES `Team`(`team_fifa_api_id`),
	FOREIGN KEY(`team_api_id`) REFERENCES `Team`(`team_api_id`)
)

CREATE TABLE goals (
                id INTEGER PRIMARY KEY,
                match_id INTEGER NOT NULL,
                team INTEGER NOT NULL,
                player1 INTEGER NOT NULL,
                player2 INTEGER NOT NULL,
                goal_type TEXT,
                subtype TEXT,
                elapsed INTEGER NOT NULL,
                elapsed_plus INTEGER,
                n INTEGER NOT NULL,
                sortorder INTEGER NOT NULL,
                FOREIGN KEY (match_id) REFERENCES match (id)
                FOREIGN KEY (team) REFERENCES team (team_api_id)
                FOREIGN KEY (player1) REFERENCES player (player_api_id)
                FOREIGN KEY (player2) REFERENCES player (player_api_id)
            )

CREATE TABLE cards (
                id INTEGER PRIMARY KEY,
                match_id INTEGER NOT NULL,
                team INTEGER NOT NULL,
                player1 INTEGER NOT NULL,
                card_type TEXT,
                subtype TEXT,
                elapsed INTEGER NOT NULL,
                elapsed_plus INTEGER,
                n INTEGER NOT NULL,
                sortorder INTEGER NOT NULL,
                FOREIGN KEY (match_id) REFERENCES match (id)
                FOREIGN KEY (team) REFERENCES team (team_api_id)
                FOREIGN KEY (player1) REFERENCES player (player_api_id)
            )

CREATE TABLE fouls (
                id INTEGER PRIMARY KEY,
                match_id INTEGER NOT NULL,
                team INTEGER NOT NULL,
                player1 INTEGER NOT NULL,
                player2 INTEGER NOT NULL,
                subtype TEXT,
                elapsed INTEGER NOT NULL,
                elapsed_plus INTEGER,
                injury_time INTEGER,
                n INTEGER NOT NULL,
                sortorder INTEGER NOT NULL,
                FOREIGN KEY (match_id) REFERENCES match (id)
                FOREIGN KEY (team) REFERENCES team (team_api_id)
                FOREIGN KEY (player1) REFERENCES player (player_api_id)
                FOREIGN KEY (player2) REFERENCES player (player_api_id)
            )

CREATE TABLE corners (
                id INTEGER PRIMARY KEY,
                match_id INTEGER NOT NULL,
                team INTEGER NOT NULL,
                player1 INTEGER NOT NULL,
                subtype TEXT,
                elapsed INTEGER NOT NULL,
                elapsed_plus INTEGER,
                n INTEGER NOT NULL,
                sortorder INTEGER NOT NULL,
                FOREIGN KEY (match_id) REFERENCES match (id)
                FOREIGN KEY (team) REFERENCES team (team_api_id)
                FOREIGN KEY (player1) REFERENCES player (player_api_id)
            )

Here is a basic description of some of the less-clear column names used in the database:
- home_player_ids: For a given n (1-11), the ID of the player playing for the home-team is in home_player_n, and his coordinates are X=home_player_Xn and Y=home_player_Yn. Same for away team. Note that X coordinates are on the goalline ranging from 1-9, and the Y coordinates are along the length of the field ranging 1-11.
  In this encoding, all goalkeepers have coordinate (1,1). If the second field player of the hometeam has coordinates (2,3), (i.e.: home_playerX2=2 and home_player_Y2=3), we can infer that he plays on the left side of the field (X=2) as a defender (Y=3). He must thus be a left defender in this match. We can assume that Y coordinates between 2 and 5 are defenders, Y coordinates between 6 and 8 are midfielders and Y coordinates 9-11 are forwards.
- All columns ending with H are the odds for the Home team in the match. Likewise, columns ending with D are odds for Draw, & A are for Away team. The other characters in the column names represent the name of the betting website. (B365=Bet365, LB=LadBrokes, WH=William Hill, SJ=Stan James, BW=BWin. Other websites are not known)
- All dates are in the format yyyy-mm-dd hh:mm:ss. Generally hh:mm:ss are 00:00:00 for most values in these columns
- The eleven leagues covered in the DB are: Belgium Jupiler League, England Premier League, France Ligue 1, Germany 1. Bundesliga, Italy Serie A, Netherlands Eredivisie, Poland Ekstraklasa, Portugal Liga ZON Sagres, Scotland Premier League, Spain LIGA BBVA, Switzerland Super League
- Many textual ratings in player_attributes, attaching_work_rate and defensive_work_rate are from high, medium, low. The preferred_foot are from left, right
- The heights are in cms and weights are in lbs.
- The class ratings in team_attributes take values as follows:
    buildUpPlaySpeedClass: Balanced, Fast, Slow
    buildUpPlayDribblingClass: Little, Normal, Lots
    buildUpPlayPassingClass: Mixed, Long, Short
    buildUpPlayPositioningClass / chanceCreationPositioningClass: Organised, Free Form
    chanceCreationPassingClass: Normal, Risky, Safe
    chanceCreationCrossingClass / chanceCreationShootingClass: Normal, Lots, Little
    defencePressureClass: Deep, Medium, High
    defenceAggressionClass: Press, Double, Contain
    defenceTeamWidthClass: Normal, Wide, Narrow
    defenceDefenderLineClass: Cover, Offside Trap
- For the events (goals, fouls, cards, corners) tables:
    Elapsed represents the minute number when goal happened. Elapsed_plus indicates that this happened in extra time. E.g., if it happens in 46th minute in first half, elapsed_plus is 1
    Player1 is the primary actor (player who scores or who fouls), and player2 is the secondary actor (e.g., who assists, or who is fouled)
    Team id is player1's team ID
    injury_time is time of break for the injury
    card_type is y, r, y2
    Goal_type takes values as: n - normal goal, p - penalty, dg - goal disallowed, o - own goal, npm - penalty saved by the goalkeeper, psm - missed penalty, rp - retake penalty
    card subtype takes values as: advantage, diving, emergency_brake, hands, kicked_ball_away, pushing, Removing Shirt, serious_fouls, shirt_pull, stall_time, Unsportsmanlike Cond, verbal_abuse, violence
    corner subtype takes values as: cross, cross_left, cross_right, short, short_left, short_right
    goal subtype takes values as: backheel, bicycle_kick, crossbar, deflected, direct_freekick, distance, header, indirect freekick, lob, loose_ball, missed, post, saved, saved_back_into_play, shot, tap_in, volley
    foul subtype takes values as: advantage, dangerous_play, diving, from_behind, goalkeeper_hands, hands, obstruction, penalty, pull, pushing, serious_foul, shirt_pull, trip

User question: The longest number of months either team name was undefeated in the Old Firm between 2008-2016? Team name and number of months

For this example above, write a sequence of steps, each in a different line. Write only one or at most two steps for Text2SQL. And keep each step short for ease of processing.
Note that for Text2SQL steps, you should NOT write an SQL query directly. Instead, you should write a short prompt for a downstream Text2SQL system. You can also copy relevant text from user question.
Important: Note that for Text2SQL steps, include in the step all information desired in the question. For example, if name is asked, include name in the Text2SQL step, or else the Python code will not be able to return it.
Finally, recall that if the instruction can be written as a single SQL easily, do not use Python at all and just copy the original question in Text2SQL.

Seasons between Year1 and Year2 do not include season ending in Year1 or starting in Year2. Eg: only two seasons exist between 2013 and 2015: 2013/2014 and 2014/2015. On the other matches between 2013 and 2015 include matches in second half of 2012/2013 and first half of 2015/2016 seasons.
Make sure your decomposition ONLY outputs the EXACT columns desired in the user instruction. It should NOT output any additional information, even if relevant.
Team, match and league should be outputted as name and not ID. Season is outputted in "YYYY/YYYY" format (Eg: "2011/2012"). Date should be outputted in "DD Month YYYY" format (Eg: "1 January 2025" and not "1 1 2025"). Match information should be outputted as "home_team_name - away_team_name". Score should be outputted as "home_team_score-away_team_score".

After your chain-of-thought deliberation, start a new line with the word "Decomposition:".
After that begin each step with either "Text2SQL: " or "Python: ".
Do not write any other extra lines.
Let's think step by step.
\end{lstlisting}

\subsection{Python}
\begin{lstlisting}
You are a helpful assistant to a database engineer, helping a user find the best code for their instruction. The user has provided a complex instruction and an LLM has generated a decomposition for it into atomic steps.
Some of those steps require fetching data from an SQL database. That data has already been fetched and is available in Pandas dataframes.

You current job is to take this user instruction and its suggested decomposition, along with the Pandas dataframes and write a Python code that produces the final answer as desired in the user instruction.

Your code should be directly executable by a Python interpreter (you may assume that the dataframes are available to the interpreter already). So, any explanation or your thought process should be written in Python comments.

User instruction: The longest number of months either team name was undefeated in the Old Firm between 2008-2016? Team name and number of months
Decomposition:
Text2SQL: Get all matches between Celtic and Rangers between 2008 and 2016, including match date, home team, away team, and scores. Output the match information as "home_team_name away_team_name" and "home_team_score-away_team_score".
Python: For each match, determine which team (if any) was undefeated. Track the longest streak of undefeated months for each team. Output the team name and the number of months for the longest streak.


The list (listOfDFs) has 1 dataframe corresponding to the Text2SQL query in the decomposition above.
listOfDFs[0] has the columns as: [match_date    object
match_info    object
score         object]
Some sample data in listOfDFs[0] is follows:
   match_date        match_info  score
0        None  Celtic - Rangers  2 - 4
1        None  Rangers - Celtic  0 - 1
2        None  Celtic - Rangers  0 - 0
..        ...               ...    ...
13       None  Celtic - Rangers  1 - 0
14       None  Rangers - Celtic  3 - 2
15       None  Celtic - Rangers  3 - 0


Write a Python function (compute_result) that takes as input the list of dataframes (listOfDFs) as described above and returns a list of tuples with the desired information. Its type signature is:
def compute_result(listOfDFs: List[DataFrame]) -> List[Tuple]:

The Python function must only use standard Python libraries.
Don't write anything apart from the Python function; use Python comments if needed.
Make sure your code is robust, for instance, some values in the dataframe may be missing -- the code should handle that without crashing.
Assume indentation level as 0 in your code. Your code will be called as print(compute_results(listOfDFs)) to show the output to the user.
Important: make sure your code ONLY outputs the EXACT columns desired in the user instruction. It should NOT output any additional information, even if relevant. It should NOT round up any numerical information, unless especially required for the question.
Important: the output should be returned as a list of tuples without any column names or descriptions.
Important: Team, match and league should be outputted as name and not ID. Date should be outputted in "DD Month YYYY" format (Eg: 1 January 2025 and not 1 1 2025). Match information should be outputted as "home_team_name - away_team_name". Score should be outputted as "home_team_score-away_team_score".
Finally, if multiple entities match question's criteria, your code should output all of them, even if question asks for one.
Write the python code enclosed in ```python ```

\end{lstlisting}

\section{Example Text2SQL Prompt for \texttt{IMDb}}
\begin{lstlisting}
You are a helpful assistant to a database engineer. The user has provided a complex instruction. Your job is to write a valid SQL for it.

This is how the database was created.
CREATE TABLE name_basics (
                nconst TEXT PRIMARY KEY,
                primaryName TEXT,
                birthYear INTEGER,
                deathYear INTEGER,
                primaryProfession TEXT,
                knownForTitles TEXT
            )
nconst (string) - alphanumeric unique identifier of the name/person
primaryName (string)– name by which the person is most often credited
birthYear – in YYYY format
deathYear – in YYYY format if applicable, else '\N'
primaryProfession (array of strings)– the top-3 professions of the person. One or more of the following: ['accountant', 'actor', 'actress', ..., 'writer']
knownForTitles (array of tconsts) – titles the person is known for
/* 
3 example rows:
SELECT * FROM name_basics LIMIT 3;
   nconst     primaryName birthYear deathYear                  primaryProfession                          knownForTitles 
nm0000001    Fred Astaire      1899      1987       actor,miscellaneous,producer tt0072308,tt0050419,tt0027125,tt0031983 
nm0000002   Lauren Bacall      1924      2014 actress,soundtrack,archive_footage tt0037382,tt0075213,tt0117057,tt0038355 
nm0000003 Brigitte Bardot      1934        \N  actress,music_department,producer tt0057345,tt0049189,tt0056404,tt0054452 
*/

CREATE TABLE title_basics (
                tconst TEXT PRIMARY KEY,
                titleType TEXT,
                primaryTitle TEXT,
                originalTitle TEXT,
                isAdult INTEGER,
                startYear INTEGER,
                endYear INTEGER,
                runtimeMinutes INTEGER,
                genres TEXT
            )
tconst (string) - alphanumeric unique identifier of the title
titleType (string) – the type/format of the title. Unique values:['short', 'movie', 'tvShort', 'tvMovie', 'tvEpisode', 'tvSeries', 'tvMiniSeries', 'tvSpecial', 'video', 'videoGame', 'tvPilot']
primaryTitle (string) – the more popular title / the title used by the filmmakers on promotional materials at the point of release
originalTitle (string) - original title, in the original language
isAdult (boolean) - 0: non-adult title; 1: adult title
startYear (YYYY) – represents the release year of a title. In the case of TV Series, it is the series start year
endYear (YYYY) – TV Series end year. '\N' for all other title types
runtimeMinutes – primary runtime of the title, in minutes
genres (string array) – includes up to three genres associated with the title. One or more of the following:['Action', 'Adult', 'Adventure', 'Animation', 'Biography', 'Comedy', 'Crime', 'Documentary', 'Drama', 'Family', 'Fantasy', 'Film-Noir', 'Game-Show', 'History', 'Horror', 'Music', 'Musical', 'Mystery', 'News', 'Reality-TV', 'Romance', 'Sci-Fi', 'Short', 'Sport', 'Talk-Show', 'Thriller', 'War', 'Western']
/* 
3 example rows:
SELECT * FROM title_basics LIMIT 3;
   tconst titleType           primaryTitle          originalTitle isAdult startYear endYear runtimeMinutes                   genres 
tt0000001     short             Carmencita             Carmencita       0      1894      \N              1        Documentary,Short 
tt0000002     short Le clown et ses chiens Le clown et ses chiens       0      1892      \N              5          Animation,Short 
tt0000003     short           Poor Pierrot         Pauvre Pierrot       0      1892      \N              5 Animation,Comedy,Romance 
*/

CREATE TABLE title_akas (
                titleid TEXT,
                ordering INTEGER,
                title TEXT,
                region TEXT,
                language TEXT,
                types TEXT,
                attributes TEXT,
                isOriginalTitle INTEGER,
                FOREIGN KEY (titleid) REFERENCES title_basics (tconst)
            )
titleId (string) - a tconst, an alphanumeric unique identifier of the title
ordering (integer) – a number to uniquely identify rows for a given titleId
title (string) – the localized title
region (string) - the region for this version of the title. Unique values:['DE', 'US', 'HU', 'GR', 'RU', 'UA', 'JP', 'RO', 'FR', 'GB', 'CA', 'PT', 'MX', 'AU', 'IT', 'ES', 'FI', 'UY', 'AR', 'PL', 'BG', 'RS', 'BR', 'TR', 'SK', 'XWW', 'DK', 'XEU', 'CZ', 'SE', 'NZ', 'KZ', 'NO', 'XYU', 'AT', 'VE', 'CSHH', 'SI', 'SUHH', 'IN', 'NL', 'LT', 'HR', 'TW', 'CN', 'CO', 'IR', 'SG', 'BE', 'EC', 'IE', 'VN', 'PH', 'DZ', 'CH', 'XWG', 'BF', 'HK', 'XSA', 'EE', 'IS', 'PR', 'DDDE', 'IL', 'EG', 'XKO', 'CL', 'JM', 'KR', 'PE', 'BY', 'GE', 'BA', 'DO', 'TH', 'AE', 'ZA', 'PA', 'LV', 'TJ', 'XSI', 'MY', 'UZ', 'AZ', 'ID', 'PK', 'BD', 'CU', 'AL', 'BO', 'XAS', 'NG', 'YUCS', 'GT', 'PY', 'SV', 'CR', 'KP', 'BUMM', 'MM', 'XPI', 'BJ', 'CM', 'KG', 'MA', 'GL', 'MN', 'LI', 'LU', 'MZ', 'MK', 'BM', 'MD', 'ME', 'LB', 'IQ', 'TM', 'TN', 'HT', 'AM', 'CI', 'LK', 'NP', 'QA', 'SY', 'TO', 'CG', 'SN', 'GH', 'JO', 'NE', 'GN', 'VDVN', 'TD', 'SO', 'SD', 'MC', 'TT', 'GA', 'BS', 'LY', 'AO', 'KH', 'MR', 'AF', 'MG', 'ML', 'GY', 'CY', 'ET', 'GU', 'SR', 'MT', 'TG', 'PG', 'MU', 'BI', 'CF', 'NI', 'ZW', 'ZM', 'GW', 'DJ', 'RW', 'TZ', 'GI', 'LA', 'SC', 'GP', 'XAU', 'FO', 'PS', 'ZRCD', 'MO', 'AW', 'KW', 'CV', 'SL', 'SM', 'CD', 'BT', 'LS', 'HN', 'KE', 'MQ', 'AD', 'ER', 'SA', 'CSXX', 'IM', 'XKV', 'BH', 'BB', 'BZ', 'UG', 'AG', 'NU', 'OM', 'BW', 'LR', 'AN', 'MV', 'YE', 'GM', 'KY', 'NC', 'DM', 'TL', 'MP', 'VA', 'GQ', 'FJ', 'SZ', 'TC', 'RE', 'EH', 'PF', 'VG', 'LC', 'MW', 'BN', 'ST', 'KM', 'FM', 'AI', 'GD', 'VI', 'SB', 'GF', 'AQ', 'MH', 'CW', 'WS', 'VC', 'AS', 'XNA', 'MS', 'VU', 'SH', 'KI', 'KN', 'CC', 'TV', 'CK', 'PW', 'NR', 'JE']
language (string) - the language of the title. Unique values:['ja', 'en', 'sv', 'bg', 'tr', 'ru', 'es', 'sr', 'cs', 'fr', 'hi', 'cmn', 'sk', 'fa', 'ca', 'qbn', 'nl', 'pt', 'uz', 'uk', 'qbp', 'ar', 'rn', 'bs', 'ga', 'de', 'yue', 'th', 'yi', 'ka', 'hr', 'sl', 'he', 'it', 'tg', 'kk', 'da', 'el', 'fi', 'be', 'gsw', 'eu', 'gl', 'az', 'ms', 'pl', 'id', 'mr', 'qbo', 'mi', 'la', 'ta', 'lt', 'lv', 'af', 'hy', 'ur', 'bn', 'te', 'ro', 'kn', 'ml', 'mk', 'tl', 'cy', 'et', 'gd', 'qal', 'gu', 'lb', 'zu', 'xh', 'eka', 'ko', 'tk', 'ky', 'wo', 'zh', 'hu', 'no', 'is', 'sq', 'vi', 'pa', 'sd', 'ps', 'ku', 'roa', 'hil', 'tn', 'rm', 'su', 'jv', 'st', 'prs', 'jsl', 'fro', 'haw', 'mn', 'am', 'ne', 'qac', 'lo', 'my', 'myv', 'br', 'iu', 'cr']
types (array) - Enumerated set of attributes for this alternative title. One or more of the following: ['original', 'imdbDisplay', 'alternative', 'festival', 'dvd', 'working', 'tv', 'video', 'imdbDisplay tv', 'alternative tv', 'imdbDisplay working', 'imdbDisplay festival', 'working tv', 'imdbDisplay video', 'dvd alternative', 'tv video', 'imdbDisplay dvd', 'working video', 'working festival', 'dvd video', 'alternative festival', 'alternative video', 'working alternative']
attributes (array) - Additional terms to describe this alternative title, not enumerated
isOriginalTitle (boolean) – 0: not original title; 1: original title
/* 
3 example rows:
SELECT * FROM title_akas LIMIT 3;
  titleid ordering      title region language       types    attributes isOriginalTitle 
tt0000001        1 Carmencita     \N       \N    original            \N               1 
tt0000001        2 Carmencita     DE       \N          \N literal title               0 
tt0000001        3 Carmencita     US       \N imdbDisplay            \N               0 
*/

CREATE TABLE title_episode (
                tconst TEXT,
                parentTconst TEXT,
                seasonNumber INTEGER,
                episodeNumber INTEGER,
                FOREIGN KEY (tconst) REFERENCES title_basics (tconst),
                FOREIGN KEY (parentTconst) REFERENCES title_basics (tconst)
            )
tconst (string) - alphanumeric identifier of episode
parentTconst (string) - alphanumeric identifier of the parent TV Series
seasonNumber (integer) – season number the episode belongs to
episodeNumber (integer) – episode number of the tconst in the TV series
/* 
3 example rows:
SELECT * FROM title_episode LIMIT 3;
   tconst parentTconst seasonNumber episodeNumber 
tt0031458   tt32857063           \N            \N 
tt0041951    tt0041038            1             9 
tt0042816    tt0989125            1            17 
*/

CREATE TABLE title_crew (
                tconst TEXT,
                directors TEXT,
                writers TEXT,
                FOREIGN KEY (tconst) REFERENCES title_basics (tconst)
            )
tconst (string) - alphanumeric unique identifier of the title
directors (array of nconsts) - director(s) of the given title
writers (array of nconsts) – writer(s) of the given title
/* 
3 example rows:
SELECT * FROM title_crew LIMIT 3;
   tconst directors   writers 
tt0000001 nm0005690        \N 
tt0000002 nm0721526        \N 
tt0000003 nm0721526 nm0721526 
*/

CREATE TABLE title_principals (
                tconst TEXT,
                ordering INTEGER,
                nconst TEXT,
                category TEXT,
                job TEXT,
                characters TEXT,
                FOREIGN KEY (tconst) REFERENCES title_basics (tconst),
                FOREIGN KEY (nconst) REFERENCES name_basics (nconst)
            )
tconst (string) - alphanumeric unique identifier of the title
ordering (integer) – a number to uniquely identify rows for a given titleId
nconst (string) - alphanumeric unique identifier of the name/person
category (string) - the category of job that person was in. Unique values: ['self', 'director', 'producer', 'cinematographer', 'composer', 'writer', 'editor', 'actor', 'actress', 'production_designer', 'archive_footage', 'casting_director', 'archive_sound']
job (string) - the specific job title if applicable, else '\N'
characters (string) - the name of the character played if applicable, else '\N'
/* 
3 example rows:
SELECT * FROM title_principals LIMIT 3;
   tconst ordering    nconst category      job characters 
tt0000001        1 nm1588970     self       \N   ["Self"] 
tt0000001        2 nm0005690 director       \N         \N 
tt0000001        3 nm0005690 producer producer         \N 
*/

CREATE TABLE title_ratings (
                tconst TEXT,
                averageRating REAL,
                numVotes INTEGER,
                FOREIGN KEY (tconst) REFERENCES title_basics (tconst)
            )
tconst (string) - alphanumeric unique identifier of the title
averageRating – weighted average of all the individual user ratings
numVotes - number of votes the title has received
/* 
3 example rows:
SELECT * FROM title_ratings LIMIT 3;
   tconst averageRating numVotes 
tt0000001           5.7     2133 
tt0000002           5.5      289 
tt0000003           6.4     2169 
*/

Important: Null values in any field are represented in the data as '\N' (and not as null).

-- Using valid SQLite, answer the following questions for the tables provided above. Write the SQL query in a separate line enclosed in ```sql ```
-- Find the most 'consistent' TV series (titleType = tvSeries) that contains atleast 10 seasons and where the difference between the highest and lowest-rated episodes is minimal. Each season of the TV show can be considered as a separate show. Return the TV show title, season number(int), and the difference between the highest and lowest-rated episodes.
The final SQL is: Let's think step by step.

\end{lstlisting}

\section{Example Text2SQLCode$_{single}$ Prompt for \texttt{IMDb}}
\begin{lstlisting}
You are a highly capable AI assistant specializing in database engineering and code generation. 
Your task is to process a user's natural language query in the following steps:
1) Decompose the query into SQL and Python steps
2) Write python code that uses SQL for data fetching operations and python for computing the final answer from the fetched data.


--- Decomposition Stage ---

We have access to a Text2SQL model, which takes as input a textual instruction and converts it into an SQL query. But, it may or may not be possible to just write a single SQL query for the complex user instruction. This often happens when there is a nested for loop or several complicated conditionals or arithmetic operations or string manipulation needed to express user intent programmatically. This may also happen if we are unsure of SQL syntax for some function, but can execute the same in Python code.
Your job is to generate a step-by-step breakdown for satisfying this instruction. Clearly annotate which step will be done by a Python interpreter and which step by Text2SQL query engine. 
If the instruction is simple enough to be solved by Text2SQL, the decomposition should be only one step. If you require multiple Text2SQL queries, make them independent of each other.

Examples: 

User question: Return the most common payment method used for transactions.
Let's think step by step. We have a table of transactions that has payment_method as a column. We can simply count the frequency of each payment method and take top 1. This question can be handled by SQL easily, no need for Python.
Decomposition:
Text2SQL: Return the most common payment method used for transactions. Output the payment method.

User question: return the date of the monthly high price of AAPL stock between Jan 1st, 2024 and June 30, 2024
Lets's think step by step. This question is complex since we need to first group the time-series data by month, and then compute the max separately for each group. We can fetch raw data for AAPL between Jan 1st and June 30 using SQL and compute aggregates using Python.
Decomposition:
Text2SQL: get daily price of AAPL stock between Jan 1st, 2024 and June 30, 2024
Python: for each month (January-June), find the highest price within that month. and its corresponding date.
Python: for each month, output the corresponding date

--- Code Stage --

Now write python code that contains execution of SQL queries for the Text2SQL steps and python code for computing the final result.

The schema for the database is as follows:
CREATE TABLE name_basics (
                nconst TEXT PRIMARY KEY,
                primaryName TEXT,
                birthYear INTEGER,
                deathYear INTEGER,
                primaryProfession TEXT,
                knownForTitles TEXT
            )
nconst (string) - alphanumeric unique identifier of the name/person
primaryName (string)– name by which the person is most often credited
birthYear – in YYYY format
deathYear – in YYYY format if applicable, else '\N'
primaryProfession (array of strings)– the top-3 professions of the person. One or more of the following: ['accountant', 'actor', 'actress', ..., 'writer']
knownForTitles (array of tconsts) – titles the person is known for
/* 
3 example rows:
SELECT * FROM name_basics LIMIT 3;
   nconst     primaryName birthYear deathYear                  primaryProfession                          knownForTitles 
nm0000001    Fred Astaire      1899      1987       actor,miscellaneous,producer tt0072308,tt0050419,tt0027125,tt0031983 
nm0000002   Lauren Bacall      1924      2014 actress,soundtrack,archive_footage tt0037382,tt0075213,tt0117057,tt0038355 
nm0000003 Brigitte Bardot      1934        \N  actress,music_department,producer tt0057345,tt0049189,tt0056404,tt0054452 
*/

CREATE TABLE title_basics (
                tconst TEXT PRIMARY KEY,
                titleType TEXT,
                primaryTitle TEXT,
                originalTitle TEXT,
                isAdult INTEGER,
                startYear INTEGER,
                endYear INTEGER,
                runtimeMinutes INTEGER,
                genres TEXT
            )
tconst (string) - alphanumeric unique identifier of the title
titleType (string) – the type/format of the title. Unique values:['short', 'movie', 'tvShort', 'tvMovie', 'tvEpisode', 'tvSeries', 'tvMiniSeries', 'tvSpecial', 'video', 'videoGame', 'tvPilot']
primaryTitle (string) – the more popular title / the title used by the filmmakers on promotional materials at the point of release
originalTitle (string) - original title, in the original language
isAdult (boolean) - 0: non-adult title; 1: adult title
startYear (YYYY) – represents the release year of a title. In the case of TV Series, it is the series start year
endYear (YYYY) – TV Series end year. '\N' for all other title types
runtimeMinutes – primary runtime of the title, in minutes
genres (string array) – includes up to three genres associated with the title. One or more of the following:['Action', 'Adult', 'Adventure', 'Animation', 'Biography', 'Comedy', 'Crime', 'Documentary', 'Drama', 'Family', 'Fantasy', 'Film-Noir', 'Game-Show', 'History', 'Horror', 'Music', 'Musical', 'Mystery', 'News', 'Reality-TV', 'Romance', 'Sci-Fi', 'Short', 'Sport', 'Talk-Show', 'Thriller', 'War', 'Western']
/* 
3 example rows:
SELECT * FROM title_basics LIMIT 3;
   tconst titleType           primaryTitle          originalTitle isAdult startYear endYear runtimeMinutes                   genres 
tt0000001     short             Carmencita             Carmencita       0      1894      \N              1        Documentary,Short 
tt0000002     short Le clown et ses chiens Le clown et ses chiens       0      1892      \N              5          Animation,Short 
tt0000003     short           Poor Pierrot         Pauvre Pierrot       0      1892      \N              5 Animation,Comedy,Romance 
*/

CREATE TABLE title_akas (
                titleid TEXT,
                ordering INTEGER,
                title TEXT,
                region TEXT,
                language TEXT,
                types TEXT,
                attributes TEXT,
                isOriginalTitle INTEGER,
                FOREIGN KEY (titleid) REFERENCES title_basics (tconst)
            )
titleId (string) - a tconst, an alphanumeric unique identifier of the title
ordering (integer) – a number to uniquely identify rows for a given titleId
title (string) – the localized title
region (string) - the region for this version of the title. Unique values:['DE', 'US', 'HU', 'GR', 'RU', 'UA', 'JP', 'RO', 'FR', 'GB', 'CA', 'PT', 'MX', 'AU', 'IT', 'ES', 'FI', 'UY', 'AR', 'PL', 'BG', 'RS', 'BR', 'TR', 'SK', 'XWW', 'DK', 'XEU', 'CZ', 'SE', 'NZ', 'KZ', 'NO', 'XYU', 'AT', 'VE', 'CSHH', 'SI', 'SUHH', 'IN', 'NL', 'LT', 'HR', 'TW', 'CN', 'CO', 'IR', 'SG', 'BE', 'EC', 'IE', 'VN', 'PH', 'DZ', 'CH', 'XWG', 'BF', 'HK', 'XSA', 'EE', 'IS', 'PR', 'DDDE', 'IL', 'EG', 'XKO', 'CL', 'JM', 'KR', 'PE', 'BY', 'GE', 'BA', 'DO', 'TH', 'AE', 'ZA', 'PA', 'LV', 'TJ', 'XSI', 'MY', 'UZ', 'AZ', 'ID', 'PK', 'BD', 'CU', 'AL', 'BO', 'XAS', 'NG', 'YUCS', 'GT', 'PY', 'SV', 'CR', 'KP', 'BUMM', 'MM', 'XPI', 'BJ', 'CM', 'KG', 'MA', 'GL', 'MN', 'LI', 'LU', 'MZ', 'MK', 'BM', 'MD', 'ME', 'LB', 'IQ', 'TM', 'TN', 'HT', 'AM', 'CI', 'LK', 'NP', 'QA', 'SY', 'TO', 'CG', 'SN', 'GH', 'JO', 'NE', 'GN', 'VDVN', 'TD', 'SO', 'SD', 'MC', 'TT', 'GA', 'BS', 'LY', 'AO', 'KH', 'MR', 'AF', 'MG', 'ML', 'GY', 'CY', 'ET', 'GU', 'SR', 'MT', 'TG', 'PG', 'MU', 'BI', 'CF', 'NI', 'ZW', 'ZM', 'GW', 'DJ', 'RW', 'TZ', 'GI', 'LA', 'SC', 'GP', 'XAU', 'FO', 'PS', 'ZRCD', 'MO', 'AW', 'KW', 'CV', 'SL', 'SM', 'CD', 'BT', 'LS', 'HN', 'KE', 'MQ', 'AD', 'ER', 'SA', 'CSXX', 'IM', 'XKV', 'BH', 'BB', 'BZ', 'UG', 'AG', 'NU', 'OM', 'BW', 'LR', 'AN', 'MV', 'YE', 'GM', 'KY', 'NC', 'DM', 'TL', 'MP', 'VA', 'GQ', 'FJ', 'SZ', 'TC', 'RE', 'EH', 'PF', 'VG', 'LC', 'MW', 'BN', 'ST', 'KM', 'FM', 'AI', 'GD', 'VI', 'SB', 'GF', 'AQ', 'MH', 'CW', 'WS', 'VC', 'AS', 'XNA', 'MS', 'VU', 'SH', 'KI', 'KN', 'CC', 'TV', 'CK', 'PW', 'NR', 'JE']
language (string) - the language of the title. Unique values:['ja', 'en', 'sv', 'bg', 'tr', 'ru', 'es', 'sr', 'cs', 'fr', 'hi', 'cmn', 'sk', 'fa', 'ca', 'qbn', 'nl', 'pt', 'uz', 'uk', 'qbp', 'ar', 'rn', 'bs', 'ga', 'de', 'yue', 'th', 'yi', 'ka', 'hr', 'sl', 'he', 'it', 'tg', 'kk', 'da', 'el', 'fi', 'be', 'gsw', 'eu', 'gl', 'az', 'ms', 'pl', 'id', 'mr', 'qbo', 'mi', 'la', 'ta', 'lt', 'lv', 'af', 'hy', 'ur', 'bn', 'te', 'ro', 'kn', 'ml', 'mk', 'tl', 'cy', 'et', 'gd', 'qal', 'gu', 'lb', 'zu', 'xh', 'eka', 'ko', 'tk', 'ky', 'wo', 'zh', 'hu', 'no', 'is', 'sq', 'vi', 'pa', 'sd', 'ps', 'ku', 'roa', 'hil', 'tn', 'rm', 'su', 'jv', 'st', 'prs', 'jsl', 'fro', 'haw', 'mn', 'am', 'ne', 'qac', 'lo', 'my', 'myv', 'br', 'iu', 'cr']
 types (array) - Enumerated set of attributes for this alternative title. One or more of the following: ['original', 'imdbDisplay', 'alternative', 'festival', 'dvd', 'working', 'tv', 'video', 'imdbDisplay tv', 'alternative tv', 'imdbDisplay working', 'imdbDisplay festival', 'working tv', 'imdbDisplay video', 'dvd alternative', 'tv video', 'imdbDisplay dvd', 'working video', 'working festival', 'dvd video', 'alternative festival', 'alternative video', 'working alternative']
attributes (array) - Additional terms to describe this alternative title, not enumerated
isOriginalTitle (boolean) – 0: not original title; 1: original title
/* 
3 example rows:
SELECT * FROM title_akas LIMIT 3;
  titleid ordering      title region language       types    attributes isOriginalTitle 
tt0000001        1 Carmencita     \N       \N    original            \N               1 
tt0000001        2 Carmencita     DE       \N          \N literal title               0 
tt0000001        3 Carmencita     US       \N imdbDisplay            \N               0 
*/

CREATE TABLE title_episode (
                tconst TEXT,
                parentTconst TEXT,
                seasonNumber INTEGER,
                episodeNumber INTEGER,
                FOREIGN KEY (tconst) REFERENCES title_basics (tconst),
                FOREIGN KEY (parentTconst) REFERENCES title_basics (tconst)
            )
tconst (string) - alphanumeric identifier of episode
parentTconst (string) - alphanumeric identifier of the parent TV Series
seasonNumber (integer) – season number the episode belongs to
episodeNumber (integer) – episode number of the tconst in the TV series
/* 
3 example rows:
SELECT * FROM title_episode LIMIT 3;
   tconst parentTconst seasonNumber episodeNumber 
tt0031458   tt32857063           \N            \N 
tt0041951    tt0041038            1             9 
tt0042816    tt0989125            1            17 
*/

CREATE TABLE title_crew (
                tconst TEXT,
                directors TEXT,
                writers TEXT,
                FOREIGN KEY (tconst) REFERENCES title_basics (tconst)
            )
tconst (string) - alphanumeric unique identifier of the title
directors (array of nconsts) - director(s) of the given title
writers (array of nconsts) – writer(s) of the given title
/* 
3 example rows:
SELECT * FROM title_crew LIMIT 3;
   tconst directors   writers 
tt0000001 nm0005690        \N 
tt0000002 nm0721526        \N 
tt0000003 nm0721526 nm0721526 
*/

CREATE TABLE title_principals (
                tconst TEXT,
                ordering INTEGER,
                nconst TEXT,
                category TEXT,
                job TEXT,
                characters TEXT,
                FOREIGN KEY (tconst) REFERENCES title_basics (tconst),
                FOREIGN KEY (nconst) REFERENCES name_basics (nconst)
            )
tconst (string) - alphanumeric unique identifier of the title
ordering (integer) – a number to uniquely identify rows for a given titleId
nconst (string) - alphanumeric unique identifier of the name/person
category (string) - the category of job that person was in. Unique values: ['self', 'director', 'producer', 'cinematographer', 'composer', 'writer', 'editor', 'actor', 'actress', 'production_designer', 'archive_footage', 'casting_director', 'archive_sound']
job (string) - the specific job title if applicable, else '\N'
characters (string) - the name of the character played if applicable, else '\N'
/* 
3 example rows:
SELECT * FROM title_principals LIMIT 3;
   tconst ordering    nconst category      job characters 
tt0000001        1 nm1588970     self       \N   ["Self"] 
tt0000001        2 nm0005690 director       \N         \N 
tt0000001        3 nm0005690 producer producer         \N 
*/

CREATE TABLE title_ratings (
                tconst TEXT,
                averageRating REAL,
                numVotes INTEGER,
                FOREIGN KEY (tconst) REFERENCES title_basics (tconst)
            )
tconst (string) - alphanumeric unique identifier of the title
averageRating – weighted average of all the individual user ratings
numVotes - number of votes the title has received
/* 
3 example rows:
SELECT * FROM title_ratings LIMIT 3;
   tconst averageRating numVotes 
tt0000001           5.7     2133 
tt0000002           5.5      289 
tt0000003           6.4     2169 
*/

The query is as follows:
Find the most 'consistent' TV series (titleType = tvSeries) that contains atleast 10 seasons and where the difference between the highest and lowest-rated episodes is minimal. Each season of the TV show can be considered as a separate show. Return the TV show title, season number(int), and the difference between the highest and lowest-rated episodes.

Important: Null values in any field are represented in the data as '\N' (and not as null).

Write a python code that takes the database_path as an input and prints a list of tuples with the desired information. Its type signature is:
def compute_result(database_path: str]) -> List[Tuple]:

The Python code must use sqlite3 to execute the SQL queries, and the outputs can be stored as dataframes.
Please use standard Python libraries for transformations on the dataframe. 
Don't write anything apart from the Python function; use Python comments if needed.
Make sure your code is robust, for instance, some values in the dataframe may be missing -- the code should handle that without crashing.
Assume indentation level as 0 in your code. Your code will be called as print(compute_result(database_path)) to show the output to the user.
If the user question asks for one instance matching a criteria, your code should output all such instances.
Important: make sure your code ONLY outputs the EXACT columns desired in the user instruction. It should NOT output any additional information, even if relevant. It should NOT round up any numerical information, unless especially required for the question.
Important: the output should be returned as a list of tuples without any column names or descriptions. 
Write the final python code enclosed in ```python ```. Provide the full and final code in a single code block. Do not provide any intermediate steps, refinements, or additional explanations.

\end{lstlisting}

\section{Example Text2SQLCode$_{multi}$ Prompt for \texttt{IMDb}}
\subsection{Decomposition}
\begin{lstlisting}
You are a helpful assistant to a database engineer. The user has provided a complex instruction. As a first step, we wish to break it into a series of steps.
You current job is to take this user instruction and create a step-by-step execution plan for achieving it. The raw data is fetched from an SQL database.

We have access to a Text2SQL model, which takes as input a textual instruction and converts it into an SQL query. But, it may or may not be possible to just write a single SQL query for the complex user instruction. This often happens when there is a nested for loop or several complicated conditionals or arithmetic operations or string manipulation needed to express user intent programmatically. This may also happen if we are unsure of SQL syntax for some function, but can execute the same in Python code.
Your job is to generate a step-by-step breakdown for satisfying this instruction. Clearly annotate which step will be done by a Python interpreter and which step by Text2SQL query engine. 
If the instruction is simple enough to be solved by Text2SQL, the decomposition should be only one step. If you require multiple Text2SQL queries, make them independent of each other.

Positive example:
User question: Return the most common payment method used for transactions.
Let's think step by step. We have a table of transactions that has payment_method as a column. We can simply count the frequency of each payment method and take top 1. This question can be handled by SQL easily, no need for Python.
Decomposition:
Text2SQL: Return the most common payment method used for transactions. Output the payment method.

Positive example:
User question: return the date of the monthly high price of AAPL stock between Jan 1st, 2024 and June 30, 2024
Lets's think step by step. This question is complex since we need to first group the time-series data by month, and then compute the max separately for each group. We can fetch raw data for AAPL between Jan 1st and June 30 using SQL and compute aggregates using Python.
Decomposition:
Text2SQL: get daily price of AAPL stock between Jan 1st, 2024 and June 30, 2024
Python: for each month (January-June), find the highest price within that month. and its corresponding date.
Python: for each month, output the corresponding date

Positive example:
User question: How many products are never sold with total value higher than 5?'
Lets's think step by step. Since we are looking for products that are never sold at this price, we can first look for products that do have one occurrence of being sold higher than 5, and exclude them from the set of all products. This will require two independent SQL queries, so it is OK.
Decomposition:
Text2SQL: get list of all products
Text2SQL: get list of all products that are sold with total value higher than 5
Python: find the products that are in list 1 but not in list 2
Python: count the number of such products
Python: output the products

Negative example:
User question: How many products are sold by Zara at a price higher than 5?
Bad Decomposition:
Text2SQL: get list of all products that are sold by Zara
Text2SQL: get prices of all products identified in the previous step
Python: find the products have prices higher than 5
Python: count the number of such products
Python: output the calculated count
This is not a valid decomposition because the second Text2SQL query depends on the output of the previous Text2SQL query. However, all Text2SQL queries must be independently executable and not dependent on each other.

Negative example:
User question: return the date of the monthly high price of AAPL stock between Jan 1st, 2024 and June 30, 2024
Lets's think step by step. This question is complex since we need to first group the time-series data by month, and then compute the max separately for each group. We can fetch raw data for AAPL between Jan 1st and June 30 using SQL and compute aggregates using Python.
Decomposition:
Text2SQL: get daily price of AAPL stock between Jan 1st, 2024 and June 30, 2024
Python: for each month (January-June), find the highest price within that month. and its corresponding date.
Python: for each month, output the highest price and the corresponding date
This is not a valid decomposition because it outputs additional information (the highest price in a month), which is not requested in the original user question. Only the requested information should be outputted.

There is a similar problem, you need to help with. This is how the database was created.
CREATE TABLE name_basics (
                nconst TEXT PRIMARY KEY,
                primaryName TEXT,
                birthYear INTEGER,
                deathYear INTEGER,
                primaryProfession TEXT,
                knownForTitles TEXT
            )
nconst (string) - alphanumeric unique identifier of the name/person
primaryName (string)– name by which the person is most often credited
birthYear – in YYYY format
deathYear – in YYYY format if applicable, else '\N'
primaryProfession (array of strings)– the top-3 professions of the person. One or more of the following: ['accountant', 'actor', 'actress', ..., 'writer']
knownForTitles (array of tconsts) – titles the person is known for

CREATE TABLE title_basics (
                tconst TEXT PRIMARY KEY,
                titleType TEXT,
                primaryTitle TEXT,
                originalTitle TEXT,
                isAdult INTEGER,
                startYear INTEGER,
                endYear INTEGER,
                runtimeMinutes INTEGER,
                genres TEXT
            )
tconst (string) - alphanumeric unique identifier of the title
titleType (string) – the type/format of the title. Unique values:['short', 'movie', 'tvShort', 'tvMovie', 'tvEpisode', 'tvSeries', 'tvMiniSeries', 'tvSpecial', 'video', 'videoGame', 'tvPilot']
primaryTitle (string) – the more popular title / the title used by the filmmakers on promotional materials at the point of release
originalTitle (string) - original title, in the original language
isAdult (boolean) - 0: non-adult title; 1: adult title
startYear (YYYY) – represents the release year of a title. In the case of TV Series, it is the series start year
endYear (YYYY) – TV Series end year. '\N' for all other title types
runtimeMinutes – primary runtime of the title, in minutes
genres (string array) – includes up to three genres associated with the title. One or more of the following:['Action', 'Adult', 'Adventure', 'Animation', 'Biography', 'Comedy', 'Crime', 'Documentary', 'Drama', 'Family', 'Fantasy', 'Film-Noir', 'Game-Show', 'History', 'Horror', 'Music', 'Musical', 'Mystery', 'News', 'Reality-TV', 'Romance', 'Sci-Fi', 'Short', 'Sport', 'Talk-Show', 'Thriller', 'War', 'Western']

CREATE TABLE title_akas (
                titleid TEXT,
                ordering INTEGER,
                title TEXT,
                region TEXT,
                language TEXT,
                types TEXT,
                attributes TEXT,
                isOriginalTitle INTEGER,
                FOREIGN KEY (titleid) REFERENCES title_basics (tconst)
            )
titleId (string) - a tconst, an alphanumeric unique identifier of the title
ordering (integer) – a number to uniquely identify rows for a given titleId
title (string) – the localized title
region (string) - the region for this version of the title. Unique values:['DE', 'US', 'HU', 'GR', 'RU', 'UA', 'JP', 'RO', 'FR', 'GB', 'CA', 'PT', 'MX', 'AU', 'IT', 'ES', 'FI', 'UY', 'AR', 'PL', 'BG', 'RS', 'BR', 'TR', 'SK', 'XWW', 'DK', 'XEU', 'CZ', 'SE', 'NZ', 'KZ', 'NO', 'XYU', 'AT', 'VE', 'CSHH', 'SI', 'SUHH', 'IN', 'NL', 'LT', 'HR', 'TW', 'CN', 'CO', 'IR', 'SG', 'BE', 'EC', 'IE', 'VN', 'PH', 'DZ', 'CH', 'XWG', 'BF', 'HK', 'XSA', 'EE', 'IS', 'PR', 'DDDE', 'IL', 'EG', 'XKO', 'CL', 'JM', 'KR', 'PE', 'BY', 'GE', 'BA', 'DO', 'TH', 'AE', 'ZA', 'PA', 'LV', 'TJ', 'XSI', 'MY', 'UZ', 'AZ', 'ID', 'PK', 'BD', 'CU', 'AL', 'BO', 'XAS', 'NG', 'YUCS', 'GT', 'PY', 'SV', 'CR', 'KP', 'BUMM', 'MM', 'XPI', 'BJ', 'CM', 'KG', 'MA', 'GL', 'MN', 'LI', 'LU', 'MZ', 'MK', 'BM', 'MD', 'ME', 'LB', 'IQ', 'TM', 'TN', 'HT', 'AM', 'CI', 'LK', 'NP', 'QA', 'SY', 'TO', 'CG', 'SN', 'GH', 'JO', 'NE', 'GN', 'VDVN', 'TD', 'SO', 'SD', 'MC', 'TT', 'GA', 'BS', 'LY', 'AO', 'KH', 'MR', 'AF', 'MG', 'ML', 'GY', 'CY', 'ET', 'GU', 'SR', 'MT', 'TG', 'PG', 'MU', 'BI', 'CF', 'NI', 'ZW', 'ZM', 'GW', 'DJ', 'RW', 'TZ', 'GI', 'LA', 'SC', 'GP', 'XAU', 'FO', 'PS', 'ZRCD', 'MO', 'AW', 'KW', 'CV', 'SL', 'SM', 'CD', 'BT', 'LS', 'HN', 'KE', 'MQ', 'AD', 'ER', 'SA', 'CSXX', 'IM', 'XKV', 'BH', 'BB', 'BZ', 'UG', 'AG', 'NU', 'OM', 'BW', 'LR', 'AN', 'MV', 'YE', 'GM', 'KY', 'NC', 'DM', 'TL', 'MP', 'VA', 'GQ', 'FJ', 'SZ', 'TC', 'RE', 'EH', 'PF', 'VG', 'LC', 'MW', 'BN', 'ST', 'KM', 'FM', 'AI', 'GD', 'VI', 'SB', 'GF', 'AQ', 'MH', 'CW', 'WS', 'VC', 'AS', 'XNA', 'MS', 'VU', 'SH', 'KI', 'KN', 'CC', 'TV', 'CK', 'PW', 'NR', 'JE']
language (string) - the language of the title. Unique values:['ja', 'en', 'sv', 'bg', 'tr', 'ru', 'es', 'sr', 'cs', 'fr', 'hi', 'cmn', 'sk', 'fa', 'ca', 'qbn', 'nl', 'pt', 'uz', 'uk', 'qbp', 'ar', 'rn', 'bs', 'ga', 'de', 'yue', 'th', 'yi', 'ka', 'hr', 'sl', 'he', 'it', 'tg', 'kk', 'da', 'el', 'fi', 'be', 'gsw', 'eu', 'gl', 'az', 'ms', 'pl', 'id', 'mr', 'qbo', 'mi', 'la', 'ta', 'lt', 'lv', 'af', 'hy', 'ur', 'bn', 'te', 'ro', 'kn', 'ml', 'mk', 'tl', 'cy', 'et', 'gd', 'qal', 'gu', 'lb', 'zu', 'xh', 'eka', 'ko', 'tk', 'ky', 'wo', 'zh', 'hu', 'no', 'is', 'sq', 'vi', 'pa', 'sd', 'ps', 'ku', 'roa', 'hil', 'tn', 'rm', 'su', 'jv', 'st', 'prs', 'jsl', 'fro', 'haw', 'mn', 'am', 'ne', 'qac', 'lo', 'my', 'myv', 'br', 'iu', 'cr']
types (array) - Enumerated set of attributes for this alternative title. One or more of the following: ['original', 'imdbDisplay', 'alternative', 'festival', 'dvd', 'working', 'tv', 'video', 'imdbDisplay tv', 'alternative tv', 'imdbDisplay working', 'imdbDisplay festival', 'working tv', 'imdbDisplay video', 'dvd alternative', 'tv video', 'imdbDisplay dvd', 'working video', 'working festival', 'dvd video', 'alternative festival', 'alternative video', 'working alternative']
attributes (array) - Additional terms to describe this alternative title, not enumerated
isOriginalTitle (boolean) – 0: not original title; 1: original title

CREATE TABLE title_episode (
                tconst TEXT,
                parentTconst TEXT,
                seasonNumber INTEGER,
                episodeNumber INTEGER,
                FOREIGN KEY (tconst) REFERENCES title_basics (tconst),
                FOREIGN KEY (parentTconst) REFERENCES title_basics (tconst)
            )
tconst (string) - alphanumeric identifier of episode
parentTconst (string) - alphanumeric identifier of the parent TV Series
seasonNumber (integer) - season number the episode belongs to
episodeNumber (integer) - episode number of the tconst in the TV series

CREATE TABLE title_crew (
                tconst TEXT,
                directors TEXT,
                writers TEXT,
                FOREIGN KEY (tconst) REFERENCES title_basics (tconst)
            )
tconst (string) - alphanumeric unique identifier of the title
directors (array of nconsts) - director(s) of the given title
writers (array of nconsts) – writer(s) of the given title

CREATE TABLE title_principals (
                tconst TEXT,
                ordering INTEGER,
                nconst TEXT,
                category TEXT,
                job TEXT,
                characters TEXT,
                FOREIGN KEY (tconst) REFERENCES title_basics (tconst),
                FOREIGN KEY (nconst) REFERENCES name_basics (nconst)
            )
tconst (string) - alphanumeric unique identifier of the title
ordering (integer) – a number to uniquely identify rows for a given titleId
nconst (string) - alphanumeric unique identifier of the name/person
category (string) - the category of job that person was in. Unique values: ['self', 'director', 'producer', 'cinematographer', 'composer', 'writer', 'editor', 'actor', 'actress', 'production_designer', 'archive_footage', 'casting_director', 'archive_sound']
job (string) - the specific job title if applicable, else '\N'
characters (string) - the name of the character played if applicable, else '\N'

CREATE TABLE title_ratings (
                tconst TEXT,
                averageRating REAL,
                numVotes INTEGER,
                FOREIGN KEY (tconst) REFERENCES title_basics (tconst)
            )
tconst (string) - alphanumeric unique identifier of the title
averageRating – weighted average of all the individual user ratings
numVotes - number of votes the title has received

User question: Find the most 'consistent' TV series (titleType = tvSeries) that contains atleast 10 seasons and where the difference between the highest and lowest-rated episodes is minimal. Each season of the TV show can be considered as a separate show. Return the TV show title, season number(int), and the difference between the highest and lowest-rated episodes.
For this example above, write a sequence of steps, each in a different line. Write only one or at most two steps for Text2SQL.
Note that for Text2SQL steps, you should NOT write an SQL query directly. Instead, you should write a prompt for a downstream Text2SQL system.
IMPORTANT -- if you write multiple Text2SQL steps, make them completely and independently executable. One should NOT use the output of other as input.
Finally, recall that if the instruction can be written as a single SQL easily, do not use Python at all and just copy the original question in Text2SQL.
After your chain-of-thought deliberation, start a new line with the word "Decomposition:".
After that begin each step with either "Text2SQL: " or "Python: ".
Do not write any other extra lines.
Let's think step by step.

\end{lstlisting}

\subsection{Python}
\begin{lstlisting}
You are a helpful assistant to a database engineer, helping a user find the best code for their instruction. The user has provided a complex instruction and an LLM has generated a decomposition for it into atomic steps.
Some of those steps require fetching data from an SQL database. That data has already been fetched and is available in Pandas dataframes.

You current job is to take this user instruction and its suggested decomposition, along with the Pandas dataframes and write a Python code that produces the final answer as desired in the user instruction.

Your code should be directly executable by a Python interpreter (you may assume that the dataframes are available to the interpreter already). So, any explanation or your thought process should be written in Python comments.

User instruction: Find the most 'consistent' TV series (titleType = tvSeries) that contains atleast 10 seasons and where the difference between the highest and lowest-rated episodes is minimal. Each season of the TV show can be considered as a separate show. Return the TV show title, season number(int), and the difference between the highest and lowest-rated episodes.
Decomposition:
Use SQL to get all episodes (with their ratings) for TV series that have at least 10 seasons. But note: we need to know which series have at least 10 seasons. We can compute the number of seasons per series in SQL.
However, to avoid dependency between SQL queries, we can first get all episodes of all TV series along with their ratings and season numbers. Then, in Python, we can group by parent series and season to compute the range of ratings for each season. Then, we filter for series with at least 10 seasons and find the season with the smallest range.
First, get all TV series (titleType='tvSeries') and their number of seasons (from title_episode). But we need to filter those with at least 10 seasons.
However, if we do this in one SQL, it might be complex. But to avoid dependent SQL queries, we can get all episodes with their ratings, along with the series title and season number.
Text2SQL: For all TV series (titleType = 'tvSeries'), return each episode's parent series tconst, the season number, the episode's average rating, and the primary title of the parent series.
Python: Group the data by parent series and season. For each season, compute the range (max rating min rating). Then, for each series, count the number of distinct seasons. Filter to keep only series with at least 10 distinct seasons. Then, among these, find the season with the smallest range. Output the series title, season number, and the range.

Notice that '\N' represents null value in any field.

The list (listOfDFs) has 1 dataframe corresponding to the Text2SQL query in the decomposition above.
listOfDFs[0] has the columns as: [parentTconst      object
seasonNumber      object
averageRating    float64
primaryTitle      object]
Some sample data in listOfDFs[0] is follows:
       parentTconst seasonNumber  averageRating              primaryTitle
0        tt32857063           \N            6.9     Teatro lírico español
1         tt0041038            1            7.6           The Lone Ranger
2         tt0989125            1            7.6  BBC Sunday-Night Theatre
...             ...          ...            ...                       ...
429768    tt0985991            3            7.8              Horrid Henry
429769    tt0985991            4            6.9              Horrid Henry
429770    tt0985991            4            7.9              Horrid Henry


Write a Python function (compute_result) that takes as input the list of dataframes (listOfDFs) as described above and returns a list of tuples with the desired information. Its type signature is:
def compute_result(listOfDFs: List[DataFrame]) -> List[Tuple]:

The Python function must only use standard Python libraries.
Don't write anything apart from the Python function; use Python comments if needed.
Make sure your code is robust, for instance, some values in the dataframe may be missing -- the code should handle that without crashing.
Assume indentation level as 0 in your code. Your code will be called as print(compute_results(listOfDFs)) to show the output to the user.
If the user question asks for one instance matching a criteria, your code should output all such instances.
Important: make sure your code ONLY outputs the EXACT columns desired in the user instruction. It should NOT output any additional information, even if relevant. It should NOT round up any numerical information, unless especially required for the question.
Important: the output should be returned as a list of tuples without any column names or descriptions. 
Write the python code enclosed in ```python ```
\end{lstlisting}

\section{Example Text2SQL Prompt for \texttt{Olist}}
\begin{lstlisting}
You are a helpful assistant to a database engineer. The user has provided a complex instruction. Your job is to write a valid SQL for it.

This is how the database was created.
CREATE TABLE olist_geolocation_dataset (
                geolocation_zip_code_prefix TEXT,
                geolocation_lat REAL,
                geolocation_lng REAL,
                geolocation_city TEXT,
                geolocation_STATE TEXT
            )
This table has information Brazilian zip codes and its lat/lng coordinates.
geolocation_zip_code_prefix (string) - first 5 digits of zip code
geolocation_lat (float) - latitude
geolocation_lng (float) - longitude
geolocation_city (string) - city name
geolocation_state (string) - state
/* 
3 example rows:
SELECT * FROM olist_geolocation_dataset LIMIT 3;
geolocation_zip_code_prefix     geolocation_lat    geolocation_lng geolocation_city geolocation_STATE 
                      01037  -23.54562128115268 -46.63929204800168        sao paulo                SP 
                      01046 -23.546081127035535 -46.64482029837157        sao paulo                SP 
                      01046  -23.54612896641469 -46.64295148361138        sao paulo                SP 
*/

CREATE TABLE olist_customers_dataset (
                customer_id TEXT PRIMARY KEY,
                customer_unique_id TEXT NOT NULL,
                customer_zip_code_prefix TEXT,
                customer_city TEXT,
                customer_state TEXT
            )
This table has information about the customer and its location. Use it to identify unique customers in the orders dataset and to find the orders delivery location.
customer_id (string) - key to the orders dataset. Each order has a unique customer_id.
customer_unique_id (string) - unique identifier of a customer.
customer_zip_code_prefix (string) - first five digits of customer zip code
customer_city (string) - customer city name
customer_state (string) - customer state
/* 
3 example rows:
SELECT * FROM olist_customers_dataset LIMIT 3;
                     customer_id               customer_unique_id customer_zip_code_prefix         customer_city customer_state 
06b8999e2fba1a1fbc88172c00ba8bc7 861eff4711a542e4b93843c6dd7febb0                    14409                franca             SP 
18955e83d337fd6b2def6b18a428ac77 290c77bc529b7ac935b93aa66c333dc3                    09790 sao bernardo do campo             SP 
4e7b3e00288586ebd08712fdd0374a03 060e732b5b29e8181a18229c7b0b2b5e                    01151             sao paulo             SP 
*/

CREATE TABLE olist_sellers_dataset (
                seller_id TEXT PRIMARY KEY,
                seller_zip_code_prefix TEXT,
                seller_city TEXT,
                seller_state TEXT
            )
This table includes data about the sellers that fulfilled orders made at Olist. 
seller_id (string) - seller unique identifier
seller_zip_code_prefix (string) - first 5 digits of seller zip code
seller_city (string) - seller city name
seller_state (string) - seller state
/* 
3 example rows:
SELECT * FROM olist_sellers_dataset LIMIT 3;
                       seller_id seller_zip_code_prefix    seller_city seller_state 
3442f8959a84dea7ee197c632cb2df15                  13023       campinas           SP 
d1b65fc7debc3361ea86b5f14c68d2e2                  13844     mogi guacu           SP 
ce3ad9de960102d0677a81f5d0bb7b2d                  20031 rio de janeiro           RJ 
*/

CREATE TABLE olist_products_dataset (
                product_id TEXT PRIMARY KEY,
                product_category_name TEXT,
                product_name_lenght INTEGER,
                product_description_lenght INTEGER,
                product_photos_qty INTEGER,
                product_weight_g INTEGER,
                product_length_cm INTEGER,
                product_height_cm INTEGER,
                product_width_cm INTEGER
            )
This table includes data about the products sold by Olist.
product_id (string) - unique product identifier
product_category_name (string) - root category of product, in Portuguese.
product_name_lenght (int) - number of characters extracted from the product name.
product_description_lenght (int) - number of characters extracted from the product description.
product_photos_qty (int) - number of product published photos
product_weight_g (int) - product weight measured in grams.
product_length_cm (int) - product length measured in centimeters.
product_height_cm (int) - product height measured in centimeters.
product_width_cm (int) - product width measured in centimeters.
/* 
3 example rows:
SELECT * FROM olist_products_dataset LIMIT 3;
                      product_id product_category_name product_name_lenght product_description_lenght product_photos_qty product_weight_g product_length_cm product_height_cm product_width_cm 
1e9e8ef04dbcff4541ed26657ea517e5            perfumaria                  40                        287                  1              225                16                10               14 
3aa071139cb16b67ca9e5dea641aaa2f                 artes                  44                        276                  1             1000                30                18               20 
96bd76ec8810374ed1b65e291975717f         esporte_lazer                  46                        250                  1              154                18                 9               15 
*/

CREATE TABLE olist_orders_dataset (
                order_id TEXT PRIMARY KEY,
                customer_id TEXT NOT NULL,
                order_status TEXT,
                order_purchase_timestamp TEXT,
                order_approved_at TEXT,
                order_delivered_carrier_date TEXT,
                order_delivered_customer_date TEXT,
                order_estimated_delivery_date TEXT,
                FOREIGN KEY (customer_id) REFERENCES olist_customers_dataset (customer_id)
            )
This is the core table. From each order you might find all other information.
order_id (string) - unique identifier of the order.
customer_id (string) - key to the customer dataset. Each order has a unique customer_id.
order_status (string) - Reference to the order status (Value can be one of: delivered, invoiced, shipped, processing, unavailable, canceled, created, and approved).
order_purchase_timestamp (date) - Shows the purchase timestamp, format in YYYY-MM-DD HH:MM:SS
order_approved_at (date) - Shows the payment approval timestamp, format in YYYY-MM-DD HH:MM:SS
order_delivered_carrier_date (date) - Shows the order posting timestamp. When it was handled to the logistic partner, format in YYYY-MM-DD HH:MM:SS
order_delivered_customer_date (date) - Shows the actual order delivery date to the customer, format in YYYY-MM-DD HH:MM:SS
order_estimated_delivery_date (date) - Shows the estimated delivery date that was informed to customer at the purchase moment, format in YYYY-MM-DD HH:MM:SS
/* 
3 example rows:
SELECT * FROM olist_orders_dataset LIMIT 3;
                        order_id                      customer_id order_status order_purchase_timestamp   order_approved_at order_delivered_carrier_date order_delivered_customer_date order_estimated_delivery_date 
e481f51cbdc54678b7cc49136f2d6af7 9ef432eb6251297304e76186b10a928d    delivered      2017-10-02 10:56:33 2017-10-02 11:07:15          2017-10-04 19:55:00           2017-10-10 21:25:13           2017-10-18 00:00:00 
53cdb2fc8bc7dce0b6741e2150273451 b0830fb4747a6c6d20dea0b8c802d7ef    delivered      2018-07-24 20:41:37 2018-07-26 03:24:27          2018-07-26 14:31:00           2018-08-07 15:27:45           2018-08-13 00:00:00 
47770eb9100c2d0c44946d9cf07ec65d 41ce2a54c0b03bf3443c3d931a367089    delivered      2018-08-08 08:38:49 2018-08-08 08:55:23          2018-08-08 13:50:00           2018-08-17 18:06:29           2018-09-04 00:00:00 
*/

CREATE TABLE olist_order_payments_dataset (
                order_id TEXT,
                payment_sequential INTEGER NOT NULL,
                payment_type TEXT NOT NULL,
                payment_installments INTEGER NOT NULL,
                payment_value REAL NOT NULL,
                FOREIGN KEY (order_id) REFERENCES olist_orders_dataset (order_id)
            )
This table includes data about the orders payment options.
order_id (string) - unique identifier of an order.
payment_sequential (int) - a customer may pay an order with more than one payment method. If he does so, a sequence will be created to accommodate all payments.
payment_type (string) - method of payment chosen by the customer (Value can be one of credit_card, boleto, voucher, debit_card, not_defined).
payment_installments (int) - number of installments chosen by the customer.
payment_value (float) - transaction value.
/* 
3 example rows:
SELECT * FROM olist_order_payments_dataset LIMIT 3;
                        order_id payment_sequential payment_type payment_installments payment_value 
b81ef226f3fe1789b1e8b2acac839d17                  1  credit_card                    8         99.33 
a9810da82917af2d9aefd1278f1dcfa0                  1  credit_card                    1         24.39 
25e8ea4e93396b6fa0d3dd708e76c1bd                  1  credit_card                    1         65.71 
*/

CREATE TABLE olist_order_items_dataset (
                order_id TEXT,
                order_item_id INTEGER NOT NULL,
                product_id TEXT NOT NULL,
                seller_id TEXT NOT NULL,
                shipping_limit_date TEXT,
                price REAL NOT NULL,
                freight_value REAL,
                FOREIGN KEY (order_id) REFERENCES olist_orders_dataset (order_id),
                FOREIGN KEY (product_id) REFERENCES olist_products_dataset (product_id)
                FOREIGN KEY (seller_id) REFERENCES olist_sellers_dataset (seller_id)
            )
This table includes data about the items purchased within each order.
order_id (string) - order unique identifier
order_item_id (int) - sequential number identifying number of items included in the same order.
product_id (string) - product unique identifier
seller_id (string) - seller unique identifier
shipping_limit_date (date) - Shows the seller shipping limit date for handling the order over to the logistic partner, format in YYYY-MM-DD HH:MM:SS
price (float) - item price
freight_value (float) - item freight value item (if an order has more than one item the freight value is already splitted between items)
/* 
3 example rows:
order_id    order_item_id   product_id  seller_id   shipping_limit_date price   freight_value
00143d0f86d6fbd9f9b38ab440ac16f5    1   e95ee6822b66ac6058e2e4aff656071a    a17f621c590ea0fab3d5d883e1630ec6    2017-10-20 16:07:52 21.33   15.10
00143d0f86d6fbd9f9b38ab440ac16f5    2   e95ee6822b66ac6058e2e4aff656071a    a17f621c590ea0fab3d5d883e1630ec6    2017-10-20 16:07:52 21.33   15.10
00143d0f86d6fbd9f9b38ab440ac16f5    3   e95ee6822b66ac6058e2e4aff656071a    a17f621c590ea0fab3d5d883e1630ec6    2017-10-20 16:07:52 21.33   15.10
*/

For the example above, the calculation for order_id 00143d0f86d6fbd9f9b38ab440ac16f5 will be:
The total order_item value is: 21.33 * 3 = 63.99
The total freight value is: 15.10 * 3 = 45.30
The total order value (product + freight) is: 45.30 + 63.99 = 109.29

CREATE TABLE olist_order_reviews_dataset (
                review_id TEXT,
                order_id TEXT NOT NULL,
                review_score INTEGER NOT NULL,
                review_comment_title TEXT,
                review_comment_message TEXT,
                review_creation_date TEXT,
                review_answer_timestamp TEXT,
                FOREIGN KEY (order_id) REFERENCES olist_orders_dataset (order_id)
            )
This table includes data about the reviews made by the customers.
review_id (string) - unique review identifier
order_id (string) - unique order identifier
review_score (int) - Note ranging from 1 to 5 given by the customer on a satisfaction survey.
review_comment_title (string) - Comment title from the review left by the customer, in Portuguese.
review_comment_message (string) - Comment message from the review left by the customer, in Portuguese.
review_creation_date (date) - Shows the date in which the satisfaction survey was sent to the customer, format in YYYY-MM-DD HH:MM:SS
review_answer_timestamp (date) - Shows satisfaction survey answer timestamp, format in YYYY-MM-DD HH:MM:SS 
/* 
3 example rows:
SELECT * FROM olist_order_reviews_dataset LIMIT 3;
                       review_id                         order_id review_score review_comment_title review_comment_message review_creation_date review_answer_timestamp 
7bc2406110b926393aa56f80a40eba40 73fc7af87114b39712e6da79b0a377eb            4                                              2018-01-18 00:00:00     2018-01-18 21:46:59 
80e641a11e56f04c1ad469d5645fdfde a548910a1c6147796b98fdf73dbeba33            5                                              2018-03-10 00:00:00     2018-03-11 03:05:13 
228ce5500dc1d8e020d8d1322874b6f0 f9e4b658b201a9f2ecdecbb34bed034b            5                                              2018-02-17 00:00:00     2018-02-18 14:36:24 
*/


CREATE TABLE product_category_name_translation (
                product_category_name TEXT PRIMARY KEY,
                product_category_name_english TEXT
            )
Translates the product_category_name to english.
product_category_name (string) - category name in Portuguese
product_category_name_english (string) - category name in English
/* 
3 example rows:
SELECT * FROM product_category_name_translation LIMIT 3;
 product_category_name product_category_name_english 
          beleza_saude                 health_beauty 
informatica_acessorios         computers_accessories 
            automotivo                          auto 
*/

-- Using valid SQLite, answer the following questions for the tables provided above. Write the SQL query in a separate line enclosed in ```sql ```
-- Make sure your code ONLY outputs the EXACT columns desired in the user instruction. It should NOT output any additional information, even if relevant. It should NOT round up any numerical information, unless especially required for the question.
-- Finally, if multiple entities match question's criteria, output all of them, even if question asks for one.
-- User question: Does the number of products in an order affect the overall review score of the order? Return the Pearson correlation coefficient and pvalue rounded to the second decimal place.
The final SQL is: Let's think step by step.

\end{lstlisting}

\section{Example Text2SQLCode$_{single}$ Prompt for \texttt{Olist}}
\begin{lstlisting}
You are a highly capable AI assistant specializing in database engineering and code generation. 
Your task is to process a user's natural language query in the following steps:
1) Decompose the query into SQL and Python steps
2) Write python code that uses SQL for data fetching operations and python for computing the final answer from the fetched data.


--- Decomposition Stage ---

We have access to a Text2SQL model, which takes as input a textual instruction and converts it into an SQL query. But, it may or may not be possible to just write a single SQL query for the complex user instruction. This often happens when there is a nested for loop or several complicated conditionals or arithmetic operations or string manipulation needed to express user intent programmatically. This may also happen if we are unsure of SQL syntax for some function, but can execute the same in Python code.
Your job is to generate a step-by-step breakdown for satisfying this instruction. Clearly annotate which step will be done by a Python interpreter and which step by Text2SQL query engine. 
If the instruction is simple enough to be solved by Text2SQL, the decomposition should be only one step. If you require multiple Text2SQL queries, make them independent of each other.

Examples: 

User question: Return the most common payment method used for transactions.
Let's think step by step. We have a table of transactions that has payment_method as a column. We can simply count the frequency of each payment method and take top 1. This question can be handled by SQL easily, no need for Python.
Decomposition:
Text2SQL: Return the most common payment method used for transactions. Output the payment method.

User question: return the date of the monthly high price of AAPL stock between Jan 1st, 2024 and June 30, 2024
Lets's think step by step. This question is complex since we need to first group the time-series data by month, and then compute the max separately for each group. We can fetch raw data for AAPL between Jan 1st and June 30 using SQL and compute aggregates using Python.
Decomposition:
Text2SQL: get daily price of AAPL stock between Jan 1st, 2024 and June 30, 2024
Python: for each month (January-June), find the highest price within that month. and its corresponding date.
Python: for each month, output the corresponding date

--- Code Stage --

Now write python code that contains execution of SQL queries for the Text2SQL steps and python code for computing the final result.

This is how the database was created.
CREATE TABLE olist_geolocation_dataset (
                geolocation_zip_code_prefix TEXT,
                geolocation_lat REAL,
                geolocation_lng REAL,
                geolocation_city TEXT,
                geolocation_STATE TEXT
            )
This table has information Brazilian zip codes and its lat/lng coordinates.
geolocation_zip_code_prefix (string) - first 5 digits of zip code
geolocation_lat (float) - latitude
geolocation_lng (float) - longitude
geolocation_city (string) - city name
geolocation_state (string) - state
/* 
3 example rows:
SELECT * FROM olist_geolocation_dataset LIMIT 3;
geolocation_zip_code_prefix     geolocation_lat    geolocation_lng geolocation_city geolocation_STATE 
                      01037  -23.54562128115268 -46.63929204800168        sao paulo                SP 
                      01046 -23.546081127035535 -46.64482029837157        sao paulo                SP 
                      01046  -23.54612896641469 -46.64295148361138        sao paulo                SP 
*/

CREATE TABLE olist_customers_dataset (
                customer_id TEXT PRIMARY KEY,
                customer_unique_id TEXT NOT NULL,
                customer_zip_code_prefix TEXT,
                customer_city TEXT,
                customer_state TEXT
            )
This table has information about the customer and its location. Use it to identify unique customers in the orders dataset and to find the orders delivery location.
customer_id (string) - key to the orders dataset. Each order has a unique customer_id.
customer_unique_id (string) - unique identifier of a customer.
customer_zip_code_prefix (string) - first five digits of customer zip code
customer_city (string) - customer city name
customer_state (string) - customer state
/* 
3 example rows:
SELECT * FROM olist_customers_dataset LIMIT 3;
                     customer_id               customer_unique_id customer_zip_code_prefix         customer_city customer_state 
06b8999e2fba1a1fbc88172c00ba8bc7 861eff4711a542e4b93843c6dd7febb0                    14409                franca             SP 
18955e83d337fd6b2def6b18a428ac77 290c77bc529b7ac935b93aa66c333dc3                    09790 sao bernardo do campo             SP 
4e7b3e00288586ebd08712fdd0374a03 060e732b5b29e8181a18229c7b0b2b5e                    01151             sao paulo             SP 
*/

CREATE TABLE olist_sellers_dataset (
                seller_id TEXT PRIMARY KEY,
                seller_zip_code_prefix TEXT,
                seller_city TEXT,
                seller_state TEXT
            )
This table includes data about the sellers that fulfilled orders made at Olist. 
seller_id (string) - seller unique identifier
seller_zip_code_prefix (string) - first 5 digits of seller zip code
seller_city (string) - seller city name
seller_state (string) - seller state
/* 
3 example rows:
SELECT * FROM olist_sellers_dataset LIMIT 3;
                       seller_id seller_zip_code_prefix    seller_city seller_state 
3442f8959a84dea7ee197c632cb2df15                  13023       campinas           SP 
d1b65fc7debc3361ea86b5f14c68d2e2                  13844     mogi guacu           SP 
ce3ad9de960102d0677a81f5d0bb7b2d                  20031 rio de janeiro           RJ 
*/

CREATE TABLE olist_products_dataset (
                product_id TEXT PRIMARY KEY,
                product_category_name TEXT,
                product_name_lenght INTEGER,
                product_description_lenght INTEGER,
                product_photos_qty INTEGER,
                product_weight_g INTEGER,
                product_length_cm INTEGER,
                product_height_cm INTEGER,
                product_width_cm INTEGER
            )
This table includes data about the products sold by Olist.
product_id (string) - unique product identifier
product_category_name (string) - root category of product, in Portuguese.
product_name_lenght (int) - number of characters extracted from the product name.
product_description_lenght (int) - number of characters extracted from the product description.
product_photos_qty (int) - number of product published photos
product_weight_g (int) - product weight measured in grams.
product_length_cm (int) - product length measured in centimeters.
product_height_cm (int) - product height measured in centimeters.
product_width_cm (int) - product width measured in centimeters.
/* 
3 example rows:
SELECT * FROM olist_products_dataset LIMIT 3;
                      product_id product_category_name product_name_lenght product_description_lenght product_photos_qty product_weight_g product_length_cm product_height_cm product_width_cm 
1e9e8ef04dbcff4541ed26657ea517e5            perfumaria                  40                        287                  1              225                16                10               14 
3aa071139cb16b67ca9e5dea641aaa2f                 artes                  44                        276                  1             1000                30                18               20 
96bd76ec8810374ed1b65e291975717f         esporte_lazer                  46                        250                  1              154                18                 9               15 
*/

CREATE TABLE olist_orders_dataset (
                order_id TEXT PRIMARY KEY,
                customer_id TEXT NOT NULL,
                order_status TEXT,
                order_purchase_timestamp TEXT,
                order_approved_at TEXT,
                order_delivered_carrier_date TEXT,
                order_delivered_customer_date TEXT,
                order_estimated_delivery_date TEXT,
                FOREIGN KEY (customer_id) REFERENCES olist_customers_dataset (customer_id)
            )
This is the core table. From each order you might find all other information.
order_id (string) - unique identifier of the order.
customer_id (string) - key to the customer dataset. Each order has a unique customer_id.
order_status (string) - Reference to the order status (Value can be one of: delivered, invoiced, shipped, processing, unavailable, canceled, created, and approved).
order_purchase_timestamp (date) - Shows the purchase timestamp, format in YYYY-MM-DD HH:MM:SS
order_approved_at (date) - Shows the payment approval timestamp, format in YYYY-MM-DD HH:MM:SS
order_delivered_carrier_date (date) - Shows the order posting timestamp. When it was handled to the logistic partner, format in YYYY-MM-DD HH:MM:SS
order_delivered_customer_date (date) - Shows the actual order delivery date to the customer, format in YYYY-MM-DD HH:MM:SS
order_estimated_delivery_date (date) - Shows the estimated delivery date that was informed to customer at the purchase moment, format in YYYY-MM-DD HH:MM:SS
/* 
3 example rows:
SELECT * FROM olist_orders_dataset LIMIT 3;
                        order_id                      customer_id order_status order_purchase_timestamp   order_approved_at order_delivered_carrier_date order_delivered_customer_date order_estimated_delivery_date 
e481f51cbdc54678b7cc49136f2d6af7 9ef432eb6251297304e76186b10a928d    delivered      2017-10-02 10:56:33 2017-10-02 11:07:15          2017-10-04 19:55:00           2017-10-10 21:25:13           2017-10-18 00:00:00 
53cdb2fc8bc7dce0b6741e2150273451 b0830fb4747a6c6d20dea0b8c802d7ef    delivered      2018-07-24 20:41:37 2018-07-26 03:24:27          2018-07-26 14:31:00           2018-08-07 15:27:45           2018-08-13 00:00:00 
47770eb9100c2d0c44946d9cf07ec65d 41ce2a54c0b03bf3443c3d931a367089    delivered      2018-08-08 08:38:49 2018-08-08 08:55:23          2018-08-08 13:50:00           2018-08-17 18:06:29           2018-09-04 00:00:00 
*/

CREATE TABLE olist_order_payments_dataset (
                order_id TEXT,
                payment_sequential INTEGER NOT NULL,
                payment_type TEXT NOT NULL,
                payment_installments INTEGER NOT NULL,
                payment_value REAL NOT NULL,
                FOREIGN KEY (order_id) REFERENCES olist_orders_dataset (order_id)
            )
This table includes data about the orders payment options.
order_id (string) - unique identifier of an order.
payment_sequential (int) - a customer may pay an order with more than one payment method. If he does so, a sequence will be created to accommodate all payments.
payment_type (string) - method of payment chosen by the customer (Value can be one of credit_card, boleto, voucher, debit_card, not_defined).
payment_installments (int) - number of installments chosen by the customer.
payment_value (float) - transaction value.
/* 
3 example rows:
SELECT * FROM olist_order_payments_dataset LIMIT 3;
                        order_id payment_sequential payment_type payment_installments payment_value 
b81ef226f3fe1789b1e8b2acac839d17                  1  credit_card                    8         99.33 
a9810da82917af2d9aefd1278f1dcfa0                  1  credit_card                    1         24.39 
25e8ea4e93396b6fa0d3dd708e76c1bd                  1  credit_card                    1         65.71 
*/

CREATE TABLE olist_order_items_dataset (
                order_id TEXT,
                order_item_id INTEGER NOT NULL,
                product_id TEXT NOT NULL,
                seller_id TEXT NOT NULL,
                shipping_limit_date TEXT,
                price REAL NOT NULL,
                freight_value REAL,
                FOREIGN KEY (order_id) REFERENCES olist_orders_dataset (order_id),
                FOREIGN KEY (product_id) REFERENCES olist_products_dataset (product_id)
                FOREIGN KEY (seller_id) REFERENCES olist_sellers_dataset (seller_id)
            )
This table includes data about the items purchased within each order.
order_id (string) - order unique identifier
order_item_id (int) - sequential number identifying number of items included in the same order.
product_id (string) - product unique identifier
seller_id (string) - seller unique identifier
shipping_limit_date (date) - Shows the seller shipping limit date for handling the order over to the logistic partner, format in YYYY-MM-DD HH:MM:SS
price (float) - item price
freight_value (float) - item freight value item (if an order has more than one item the freight value is already splitted between items)
/* 
3 example rows:
order_id    order_item_id   product_id  seller_id   shipping_limit_date price   freight_value
00143d0f86d6fbd9f9b38ab440ac16f5    1   e95ee6822b66ac6058e2e4aff656071a    a17f621c590ea0fab3d5d883e1630ec6    2017-10-20 16:07:52 21.33   15.10
00143d0f86d6fbd9f9b38ab440ac16f5    2   e95ee6822b66ac6058e2e4aff656071a    a17f621c590ea0fab3d5d883e1630ec6    2017-10-20 16:07:52 21.33   15.10
00143d0f86d6fbd9f9b38ab440ac16f5    3   e95ee6822b66ac6058e2e4aff656071a    a17f621c590ea0fab3d5d883e1630ec6    2017-10-20 16:07:52 21.33   15.10
*/

For the example above, the calculation for order_id 00143d0f86d6fbd9f9b38ab440ac16f5 will be:
The total order_item value is: 21.33 * 3 = 63.99
The total freight value is: 15.10 * 3 = 45.30
The total order value (product + freight) is: 45.30 + 63.99 = 109.29

CREATE TABLE olist_order_reviews_dataset (
                review_id TEXT,
                order_id TEXT NOT NULL,
                review_score INTEGER NOT NULL,
                review_comment_title TEXT,
                review_comment_message TEXT,
                review_creation_date TEXT,
                review_answer_timestamp TEXT,
                FOREIGN KEY (order_id) REFERENCES olist_orders_dataset (order_id)
            )
This table includes data about the reviews made by the customers.
review_id (string) - unique review identifier
order_id (string) - unique order identifier
review_score (int) - Note ranging from 1 to 5 given by the customer on a satisfaction survey.
review_comment_title (string) - Comment title from the review left by the customer, in Portuguese.
review_comment_message (string) - Comment message from the review left by the customer, in Portuguese.
review_creation_date (date) - Shows the date in which the satisfaction survey was sent to the customer, format in YYYY-MM-DD HH:MM:SS
review_answer_timestamp (date) - Shows satisfaction survey answer timestamp, format in YYYY-MM-DD HH:MM:SS 
/* 
3 example rows:
SELECT * FROM olist_order_reviews_dataset LIMIT 3;
                       review_id                         order_id review_score review_comment_title review_comment_message review_creation_date review_answer_timestamp 
7bc2406110b926393aa56f80a40eba40 73fc7af87114b39712e6da79b0a377eb            4                                              2018-01-18 00:00:00     2018-01-18 21:46:59 
80e641a11e56f04c1ad469d5645fdfde a548910a1c6147796b98fdf73dbeba33            5                                              2018-03-10 00:00:00     2018-03-11 03:05:13 
228ce5500dc1d8e020d8d1322874b6f0 f9e4b658b201a9f2ecdecbb34bed034b            5                                              2018-02-17 00:00:00     2018-02-18 14:36:24 
*/


CREATE TABLE product_category_name_translation (
                product_category_name TEXT PRIMARY KEY,
                product_category_name_english TEXT
            )
Translates the product_category_name to english.
product_category_name (string) - category name in Portuguese
product_category_name_english (string) - category name in English
/* 
3 example rows:
SELECT * FROM product_category_name_translation LIMIT 3;
 product_category_name product_category_name_english 
          beleza_saude                 health_beauty 
informatica_acessorios         computers_accessories 
            automotivo                          auto 
*/

The query is as follows:
Does the number of products in an order affect the overall review score of the order? Return the Pearson correlation coefficient and pvalue rounded to the second decimal place

Write a python code that takes the database_path as an input and prints a list of tuples with the desired information. Its type signature is:
def compute_result(database_path: str]) -> List[Tuple]:

The Python code must use sqlite3 to execute the SQL queries, and the outputs can be stored as dataframes.
Please use standard Python libraries for transformations on the dataframe. 
Don't write anything apart from the Python function; use Python comments if needed.
Make sure your code is robust, for instance, some values in the dataframe may be missing -- the code should handle that without crashing.
Assume indentation level as 0 in your code. Your code will be called as print(compute_result(database_path)) to show the output to the user.
If the user question asks for one instance matching a criteria, your code should output all such instances.
Important: make sure your code ONLY outputs the EXACT columns desired in the user instruction. It should NOT output any additional information, even if relevant. It should NOT round up any numerical information, unless especially required for the question.
Important: the output should be returned as a list of tuples without any column names or descriptions. 
Write the final python code enclosed in ```python ```
\end{lstlisting}

\section{Example Text2SQLCode$_{multi}$ Prompt for \texttt{Olist}}
\subsection{Decomposition}
\begin{lstlisting}
You are a helpful assistant to a database engineer. The user has provided a complex instruction. As a first step, we wish to break it into a series of steps.
You current job is to take this user instruction and create a step-by-step execution plan for achieving it. The raw data is fetched from an SQL database.

We have access to a Text2SQL model, which takes as input a textual instruction and converts it into an SQL query. But, it may or may not be possible to just write a single SQL query for the complex user instruction. This often happens when there is a nested for loop or several complicated conditionals or arithmetic operations or string manipulation needed to express user intent programmatically. This may also happen if we are unsure of SQL syntax for some function, but can execute the same in Python code.
Your job is to generate a step-by-step breakdown for satisfying this instruction. Clearly annotate which step will be done by a Python interpreter and which step by Text2SQL query engine.
If the instruction is simple enough to be solved by Text2SQL, the decomposition should be only one step. If you require multiple Text2SQL queries, make them independent of each other.

Positive example:
User question: Return the most common payment method used for transactions.
Let's think step by step. We have a table of transactions that has payment_method as a column. We can simply count the frequency of each payment method and take top 1. This question can be handled by SQL easily, no need for Python.
Decomposition:
Text2SQL: Return the most common payment method used for transactions. Output the payment method.

Positive example:
User question: return the date of the monthly high price of AAPL stock between Jan 1st, 2024 and June 30, 2024
Lets's think step by step. This question is complex since we need to first group the time-series data by month, and then compute the max separately for each group. We can fetch raw data for AAPL between Jan 1st and June 30 using SQL and compute aggregates using Python.
Decomposition:
Text2SQL: get daily price of AAPL stock between Jan 1st, 2024 and June 30, 2024
Python: for each month (January-June), find the highest price within that month. and its corresponding date.
Python: for each month, output the corresponding date

Negative example:
User question: How many products are sold by Zara at a price higher than 5?
Bad Decomposition:
Text2SQL: get list of all products that are sold by Zara
Text2SQL: get prices of all products identified in the previous step
Python: find the products have prices higher than 5
Python: count the number of such products
Python: output the calculated count
This is not a valid decomposition because the second Text2SQL query depends on the output of the previous Text2SQL query. However, all Text2SQL queries must be independently executable and not dependent on each other.

Negative example:
User question: return the date of the monthly high price of AAPL stock between Jan 1st, 2024 and June 30, 2024
Lets's think step by step. This question is complex since we need to first group the time-series data by month, and then compute the max separately for each group. We can fetch raw data for AAPL between Jan 1st and June 30 using SQL and compute aggregates using Python.
Decomposition:
Text2SQL: get daily price of AAPL stock between Jan 1st, 2024 and June 30, 2024
Python: for each month (January-June), find the highest price within that month. and its corresponding date.
Python: for each month, output the highest price and the corresponding date
This is not a valid decomposition because it outputs additional information (the highest price in a month), which is not requested in the original user question. Only the requested information should be outputted.


There is a similar problem, you need to help with. This is how the database was created.

CREATE TABLE olist_geolocation_dataset (
                geolocation_zip_code_prefix TEXT,
                geolocation_lat REAL,
                geolocation_lng REAL,
                geolocation_city TEXT,
                geolocation_STATE TEXT
            )
This table has information Brazilian zip codes and its lat/lng coordinates.
geolocation_zip_code_prefix (string) - first 5 digits of zip code
geolocation_lat (float) - latitude
geolocation_lng (float) - longitude
geolocation_city (string) - city name
geolocation_state (string) - state

CREATE TABLE olist_customers_dataset (
                customer_id TEXT PRIMARY KEY,
                customer_unique_id TEXT NOT NULL,
                customer_zip_code_prefix TEXT,
                customer_city TEXT,
                customer_state TEXT
            )
This table has information about the customer and its location. Use it to identify unique customers in the orders dataset and to find the orders delivery location.
customer_id (string) - key to the orders dataset. Each order has a unique customer_id.
customer_unique_id (string) - unique identifier of a customer.
customer_zip_code_prefix (string) - first five digits of customer zip code
customer_city (string) - customer city name
customer_state (string) - customer state

CREATE TABLE olist_sellers_dataset (
                seller_id TEXT PRIMARY KEY,
                seller_zip_code_prefix TEXT,
                seller_city TEXT,
                seller_state TEXT
            )
This table includes data about the sellers that fulfilled orders made at Olist. 
seller_id (string) - seller unique identifier
seller_zip_code_prefix (string) - first 5 digits of seller zip code
seller_city (string) - seller city name
seller_state (string) - seller state

CREATE TABLE olist_products_dataset (
                product_id TEXT PRIMARY KEY,
                product_category_name TEXT,
                product_name_lenght INTEGER,
                product_description_lenght INTEGER,
                product_photos_qty INTEGER,
                product_weight_g INTEGER,
                product_length_cm INTEGER,
                product_height_cm INTEGER,
                product_width_cm INTEGER
            )
This table includes data about the products sold by Olist.
product_id (string) - unique product identifier
product_category_name (string) - root category of product, in Portuguese.
product_name_lenght (int) - number of characters extracted from the product name.
product_description_lenght (int) - number of characters extracted from the product description.
product_photos_qty (int) - number of product published photos
product_weight_g (int) - product weight measured in grams.
product_length_cm (int) - product length measured in centimeters.
product_height_cm (int) - product height measured in centimeters.
product_width_cm (int) - product width measured in centimeters.

CREATE TABLE olist_orders_dataset (
                order_id TEXT PRIMARY KEY,
                customer_id TEXT NOT NULL,
                order_status TEXT,
                order_purchase_timestamp TEXT,
                order_approved_at TEXT,
                order_delivered_carrier_date TEXT,
                order_delivered_customer_date TEXT,
                order_estimated_delivery_date TEXT,
                FOREIGN KEY (customer_id) REFERENCES olist_customers_dataset (customer_id)
            )
This is the core table. From each order you might find all other information.
order_id (string) - unique identifier of the order.
customer_id (string) - key to the customer dataset. Each order has a unique customer_id.
order_status (string) - Reference to the order status (Value can be one of: delivered, invoiced, shipped, processing, unavailable, canceled, created, and approved).
order_purchase_timestamp (date) - Shows the purchase timestamp, format in YYYY-MM-DD HH:MM:SS
order_approved_at (date) - Shows the payment approval timestamp, format in YYYY-MM-DD HH:MM:SS
order_delivered_carrier_date (date) - Shows the order posting timestamp. When it was handled to the logistic partner, format in YYYY-MM-DD HH:MM:SS
order_delivered_customer_date (date) - Shows the actual order delivery date to the customer, format in YYYY-MM-DD HH:MM:SS
order_estimated_delivery_date (date) - Shows the estimated delivery date that was informed to customer at the purchase moment, format in YYYY-MM-DD HH:MM:SS

CREATE TABLE olist_order_payments_dataset (
                order_id TEXT,
                payment_sequential INTEGER NOT NULL,
                payment_type TEXT NOT NULL,
                payment_installments INTEGER NOT NULL,
                payment_value REAL NOT NULL,
                FOREIGN KEY (order_id) REFERENCES olist_orders_dataset (order_id)
            )
This table includes data about the orders payment options.
order_id (string) - unique identifier of an order.
payment_sequential (int) - a customer may pay an order with more than one payment method. If he does so, a sequence will be created to accommodate all payments.
payment_type (string) - method of payment chosen by the customer (Value can be one of credit_card, boleto, voucher, debit_card, not_defined).
payment_installments (int) - number of installments chosen by the customer.
payment_value (float) - transaction value.

CREATE TABLE olist_order_items_dataset (
                order_id TEXT,
                order_item_id INTEGER NOT NULL,
                product_id TEXT NOT NULL,
                seller_id TEXT NOT NULL,
                shipping_limit_date TEXT,
                price REAL NOT NULL,
                freight_value REAL,
                FOREIGN KEY (order_id) REFERENCES olist_orders_dataset (order_id),
                FOREIGN KEY (product_id) REFERENCES olist_products_dataset (product_id)
                FOREIGN KEY (seller_id) REFERENCES olist_sellers_dataset (seller_id)
            )
olist_order_items_dataset
This table includes data about the items purchased within each order.
order_id (string) - order unique identifier
order_item_id (int) - sequential number identifying number of items included in the same order.
product_id (string) - product unique identifier
seller_id (string) - seller unique identifier
shipping_limit_date (date) - Shows the seller shipping limit date for handling the order over to the logistic partner, format in YYYY-MM-DD HH:MM:SS
price (float) - item price
freight_value (float) - item freight value item (if an order has more than one item the freight value is already splitted between items)
/* 
3 example rows:
order_id    order_item_id   product_id  seller_id   shipping_limit_date price   freight_value
00143d0f86d6fbd9f9b38ab440ac16f5    1   e95ee6822b66ac6058e2e4aff656071a    a17f621c590ea0fab3d5d883e1630ec6    2017-10-20 16:07:52 21.33   15.10
00143d0f86d6fbd9f9b38ab440ac16f5    2   e95ee6822b66ac6058e2e4aff656071a    a17f621c590ea0fab3d5d883e1630ec6    2017-10-20 16:07:52 21.33   15.10
00143d0f86d6fbd9f9b38ab440ac16f5    3   e95ee6822b66ac6058e2e4aff656071a    a17f621c590ea0fab3d5d883e1630ec6    2017-10-20 16:07:52 21.33   15.10
*/

For the example above, the calculation for order_id 00143d0f86d6fbd9f9b38ab440ac16f5 will be:
The total order_item value is: 21.33 * 3 = 63.99
The total freight value is: 15.10 * 3 = 45.30
The total order value (product + freight) is: 45.30 + 63.99 = 109.29

CREATE TABLE olist_order_reviews_dataset (
                review_id TEXT,
                order_id TEXT NOT NULL,
                review_score INTEGER NOT NULL,
                review_comment_title TEXT,
                review_comment_message TEXT,
                review_creation_date TEXT,
                review_answer_timestamp TEXT,
                FOREIGN KEY (order_id) REFERENCES olist_orders_dataset (order_id)
            )
This table includes data about the reviews made by the customers.
review_id (string) - unique review identifier
order_id (string) - unique order identifier
review_score (int) - Note ranging from 1 to 5 given by the customer on a satisfaction survey.
review_comment_title (string) - Comment title from the review left by the customer, in Portuguese.
review_comment_message (string) - Comment message from the review left by the customer, in Portuguese.
review_creation_date (date) - Shows the date in which the satisfaction survey was sent to the customer, format in YYYY-MM-DD HH:MM:SS
review_answer_timestamp (date) - Shows satisfaction survey answer timestamp, format in YYYY-MM-DD HH:MM:SS 


CREATE TABLE product_category_name_translation (
                product_category_name TEXT PRIMARY KEY,
                product_category_name_english TEXT
            )
Translates the product_category_name to english.
product_category_name (string) - category name in Portuguese
product_category_name_english (string) - category name in English

User question: Does the number of products in an order affect the overall review score of the order? Return the Pearson correlation coefficient and pvalue rounded to the second decimal place

For this example above, write a sequence of steps, each in a different line. Write only one or at most two steps for Text2SQL. And keep each step short for ease of processing.
Note that for Text2SQL steps, you should NOT write an SQL query directly. Instead, you should write a short prompt for a downstream Text2SQL system. You can also copy relevant text from user question.
Important: Note that for Text2SQL steps, include in the step all information desired in the question. For example, if name is asked, include name in the Text2SQL step, or else the Python code will not be able to return it.
Finally, recall that if the instruction can be written as a single SQL easily, do not use Python at all and just copy the original question in Text2SQL.

After your chain-of-thought deliberation, start a new line with the word "Decomposition:".
After that begin each step with either "Text2SQL: " or "Python: ".
Do not write any other extra lines.
Let's think step by step.
\end{lstlisting}

\subsection{Python}
\begin{lstlisting}
You are a helpful assistant to a database engineer, helping a user find the best code for their instruction. The user has provided a complex instruction and an LLM has generated a decomposition for it into atomic steps.
Some of those steps require fetching data from an SQL database. That data has already been fetched and is available in Pandas dataframes.

You current job is to take this user instruction and its suggested decomposition, along with the Pandas dataframes and write a Python code that produces the final answer as desired in the user instruction.

Your code should be directly executable by a Python interpreter (you may assume that the dataframes are available to the interpreter already). So, any explanation or your thought process should be written in Python comments.

User instruction: Does the number of products in an order affect the overall review score of the order? Return the Pearson correlation coefficient and pvalue rounded to the second decimal place
Decomposition:
Counts the number of items (order_item_id) for each order in olist_order_items_dataset.
Joins with olist_order_reviews_dataset to get the review_score for that order.
Text2SQL: For each order that has a review, return the order_id, the count of products in the order, and the review score.
Python: Compute the Pearson correlation coefficient and p-value between the number of products and the review score, rounded to two decimal places.


The list (listOfDFs) has 1 dataframe corresponding to the Text2SQL query in the decomposition above.
listOfDFs[0] has the columns as: [order_id         object
product_count     int64
review_score      int64]
Some sample data in listOfDFs[0] is follows:
                               order_id  product_count  review_score
0      00010242fe8c5a6d1ba2dd792cb16214              1             5
1      00018f77f2f0320c557190d7a144bdd3              1             4
2      000229ec398224ef6ca0657da4fc703e              1             5
...                                 ...            ...           ...
98115  fffce4705a9662cd70adb13d4a31832d              1             5
98116  fffe18544ffabc95dfada21779c9644f              1             5
98117  fffe41c64501cc87c801fd61db3f6244              1             5


Write a Python function (compute_result) that takes as input the list of dataframes (listOfDFs) as described above and returns a list of tuples with the desired information. Its type signature is:
def compute_result(listOfDFs: List[DataFrame]) -> List[Tuple]:

The Python function must only use standard Python libraries.
Don't write anything apart from the Python function; use Python comments if needed.
Make sure your code is robust, for instance, some values in the dataframe may be missing -- the code should handle that without crashing.
Assume indentation level as 0 in your code. Your code will be called as print(compute_results(listOfDFs)) to show the output to the user.
Important: make sure your code ONLY outputs the EXACT columns desired in the user instruction. It should NOT output any additional information, even if relevant. It should NOT round up any numerical information, unless especially required for the question.
Important: the output should be returned as a list of tuples without any column names or descriptions.
Finally, if multiple entities match question's criteria, your code should output all of them, even if question asks for one.
Write the python code enclosed in ```python ```

\end{lstlisting}



\end{document}